\documentclass[preprint,prd,aps,superscriptaddress,titlepage,%
nofootinbib,eqsecnum,floatfix]{revtex4}
\usepackage{graphicx}

\begin{document}

\preprint{
  KEK-CP-128
  }

\title{
  $B^0-\bar{B}^0$ mixing in quenched lattice QCD
  }

\newcommand{\Tsukuba}{
  Institute of Physics, 
  University of Tsukuba, 
  Tsukuba, 305-8571, 
  Japan}

\newcommand{\RCCP}{
  Center for Computational Physics, 
  University of Tsukuba, 
  Tsukuba, 305-8577, 
  Japan}

\newcommand{\ICRR}{
  Institute for Cosmic Ray Research, 
  University of Tokyo, 
  Kashiwa, 277-8582, 
  Japan}

\newcommand{\KEK}{
  High Energy Accelerator Research Organization (KEK), 
  Tsukuba, 305-0801, 
  Japan}

\newcommand{\Hiroshima}{
  Department of Physics, 
  Hiroshima University,
  Higashi-Hiroshima, 739-8526, Japan}

\newcommand{\YITP}{
  Yukawa Institute for Theoretical Physics, 
  Kyoto University, 
  Kyoto, 606-8502, Japan}

\author{S.~Aoki}
\affiliation{\Tsukuba}

\author{M.~Fukugita}
\affiliation{\ICRR}

\author{S.~Hashimoto}
\affiliation{\KEK}

\author{K-I.~Ishikawa}
\affiliation{\Tsukuba}
\affiliation{\RCCP}

\author{N.~Ishizuka}
\affiliation{\Tsukuba}
\affiliation{\RCCP}

\author{Y.~Iwasaki}
\affiliation{\Tsukuba}

\author{K.~Kanaya}
\affiliation{\Tsukuba}

\author{T.~Kaneko}
\affiliation{\KEK}

\author{Y.~Kuramashi}
\affiliation{\KEK}

\author{M.~Okawa}
\affiliation{\Hiroshima}

\author{T.~Onogi}
\affiliation{\YITP}

\author{N.~Tsutsui}
\affiliation{\KEK}

\author{A.~Ukawa}
\affiliation{\Tsukuba}
\affiliation{\RCCP}

\author{N.~Yamada}
\affiliation{\KEK}

\author{T.~Yoshi\'{e}}
\affiliation{\Tsukuba}
\affiliation{\RCCP}

\collaboration{JLQCD Collaboration}
\noaffiliation

\date{\today}

\begin{abstract}
  We present our results of lattice calculations of $B$ parameters,
  which parameterize $\Delta B$=2 transition amplitudes together with
  the leptonic decay constant.
  Calculations are made in the quenched approximation at $\beta$=5.7,
  5.9, 6.0 and 6.1, using NRQCD action for heavy quark and the
  $O(a)$-improved Wilson action for light quark.
  The operators are perturbatively renormalized including the
  correction of $O(\alpha_s/(aM)^m)$ ($m\ge$0).
  We examine the scaling behavior of $B$ parameters, and
  discuss the systematic uncertainties based on the results with 
  several different
  truncations of higher order terms in $1/M$ and $\alpha_s$
  expansions.
  We find $B_{B_d}(m_b)=0.84(3)(5)$,
  $B_{B_s}/B_{B_d}=1.020(21)(^{+15}_{-16})(^{+5}_{-0})$ and
  $B_{S_s}(m_b)=0.85(1)(5)(^{+1}_{-0})$ in the quenched
  approximation.
  The errors represent statistical and systematic as well as the
  uncertainty in the determination of strange quark mass.
\end{abstract}

\maketitle

\section{Introduction}
\label{sec:Introduction}
The determination of the Cabibbo-Kobayashi-Maskawa (CKM) matrix
element $|V_{td}|$ plays a crucial role in testing the unitarity
relation of the CKM matrix, since the position of the vertex of the
unitarity triangle would be essentially identified together with the
angle $\phi_1$ of the unitarity triangle.
Now that the angle $\phi_1$ has already been measured experimentally 
by the asymmetric $B$ factories  \cite{Aubert:2001nu,Abe:2001xe} 
and its precision is expected to be improved substantially 
in the near future, the accuracy of $|V_{td}|$ really 
determines the accuracy of the Standard model predictions.
Then, other measurements of the CKM matrix elements, such as the
determination of $|V_{ub}|$ through a measurement of 
$b\rightarrow ul\nu$, may be used for a test of the CKM mechanism
of the quark flavor mixing and the CP violation in the Standard Model.

The CKM matrix element $|V_{td}|$ may be determined using the mass
difference $\Delta M_d$ in the neutral $B$ meson mixing, as it emerges 
through a loop diagram mediated by top quark and $W$ boson, which is
proportional to $|V_{td}V_{tb}^*|^2$.
The precision in the current world average 
($\Delta M_d$ = 0.489 $\pm$ 0.008 ps$^{-1}$ \cite{Hocker:2001jb})
is already as good as 1.6\%.
The constraint on $|V_{td}|$ is, however, limited by the theoretical
uncertainty in the calculation of the hadronic matrix element
$\langle\bar{B}^0|\mathcal{O}_{L_d}|B^0\rangle$ of the $\Delta B=2$
four-quark operator 
$\mathcal{O}_{L_d}=
\bar{b}\gamma_\mu(1-\gamma_5)d\ \bar{b}\gamma_\mu(1-\gamma_5)d$.
It is usually parameterized as $\frac{8}{3}f_B^2 B_B M_B^2$ using the
$B$ meson leptonic decay constant $f_B$ and the $B$ parameter $B_B$.
In the vacuum saturation approximation, which is valid when both $b$
and (anti-)$d$ quarks are nonrelativistic, $B_B$ is normalized to unity.

The best available theoretical method to calculate $f_B$ and $B_B$ is
the numerical simulation of QCD on the lattice, whose current status
is reviewed in \cite{Ryan:2001ej,Kronfeld:2001ss,Yamada:2002_rev}.
For the decay constant $f_B$, several groups investigated the
systematic errors in the lattice calculation, performing the
simulations on several different lattices.
It is found that the error associated with the large $b$ quark mass is 
controlled reasonably well if one uses an effective theory for heavy
quark such as the non-relativistic QCD (NRQCD)
\cite{Thacker:1991bm,Lepage:1992tx} or the Fermilab formalism
\cite{El-Khadra:1997mp}, and the results are insensitive to the
lattice spacing
\cite{Aoki:1998ji,El-Khadra:1998hq,Bernard:1998xi,%
Ishikawa:2000xu,Collins:2001ix}.

For the $B$ parameter $B_B$, on the other hand, most lattice
calculations relying on the effective theory for heavy quark are
limited to the static approximation, in which the $1/m_b$ correction
is neglected \cite{Ewing:1996ih,Christensen:1997sj,Gimenez:1999mw},
and the study of the systematic uncertainty depending on the lattice 
spacing has not been made.
Recently some of us used the NRQCD action, for the first time,
to calculate $B_B$ \cite{Hashimoto:1999ck},
and the $1/m_b$ correction was studied at a fixed lattice spacing
\cite{Hashimoto:2000eh}. 
They also calculated another $B$ parameter $B_S$ 
\cite{Hashimoto:2000eh,Hashimoto:2000yh}
to parameterize the matrix element of the operator
$\mathcal{O}_{S_s}=\bar{b}(1-\gamma_5)s\ \bar{b}(1-\gamma_5)s$,
which appears in the heavy quark expansion of the width difference of
$B_s$ \cite{Beneke:1996gn,Beneke:1999sy,Beneke:2001cu}. 

In this paper we extend the previous studies
\cite{Hashimoto:1999ck,Hashimoto:2000yh,Hashimoto:2000eh} to
investigate the systematic errors in the calculation of the $B$ meson
$B$ parameters. 
Using the same NRQCD action as in
\cite{Hashimoto:1999ck,Hashimoto:2000eh,Hashimoto:2000yh} we calculate
the $B$ parameters at four lattice spacings to estimate the size of
systematic errors depending on the lattice spacing.
In order to minimize other sources of systematic errors, we use the
$O(a)$-improved Wilson quark action \cite{Sheikholeslami:1985ij} for
light quarks with the improvement coefficient $c_{\mathrm{SW}}$
calculated at the one-loop level
\cite{Wohlert:1987rf,Naik:1993ux,Luscher:1996vw} 
and nonperturbatively \cite{Luscher:1997ug}.

Since NRQCD is an effective theory valid for heavy quark and the
action is constructed by an expansion in inverse heavy quark mass,
there is a potential source of systematic error due to the
truncation of $1/m_b$ expansion.
Furthermore, in order to match the effective theory to the full theory
one has to use perturbation theory, and errors from higher order
corrections should also be taken into account.
We introduce a method to estimate these systematic errors by treating
the neglected higher order terms in different ways.
It turned out that the error estimated in this way is quite consistent
with a naive order counting assuming typical sizes for the expansion
parameters.

Due to the systematic errors discussed above, it is not
straightforward to obtain an accuracy better than 10--15\% for
$f_B\sqrt{B_B}$ which is relevant for the determination of $|V_{td}|$.
Alternatively, one could use the ratio $\Delta M_s/\Delta M_d$, once
the mass difference in the $B_s-\bar{B}_s$ mixing is measured.
(The current experimental bound is 
$\Delta M_s > 13.1$~ps$^{-1}$ at 95\% CL \cite{PDG2002}.)
It is proportional to $\xi^2 |V_{ts}/V_{td}|^2$, where $\xi$ is a
ratio to describe SU(3) flavor breaking of the hadronic matrix
elements given by
\begin{equation}
  \label{eq:xi}
  \xi = \frac{f_{B_s}\sqrt{B_{B_s}}}{f_B\sqrt{B_B}}.
\end{equation}
Since the bulk of the systematic errors in the calculations of $f_B$
and $B_B$ cancels in this ratio, one may achieve much better
accuracy, as stressed in \cite{Bernard:1998dg}.
The largest remaining uncertainty comes from the chiral extrapolation
of lattice data, which is also discussed in this paper.

The $B$ meson $B$ parameters have also been calculated using the 
conventional relativistic actions for heavy quark
\cite{Becirevic:2001nv,Lellouch:2001tw}.
Since the lattice spacing in the present simulations is not small
enough compared to the Compton wave length of the $b$ quark, one has
to extrapolate the results obtained around the charm quark mass to the
bottom quark mass, which is a significant source of systematic
uncertainty.
In fact, the extrapolation with the linear form in $1/M$
does not seem to agree with explicit calculations in the static
limit \cite{Hashimoto:2000bk,Bernard:2001ki,Gimenez:2001jj}.
Therefore one may use the static result to constrain the heavy quark
extrapolation in the infinite mass limit \cite{Becirevic:2001xt}.
We present a comparison of our result with these previous
calculations. 

This paper is organized as follows.
In the next section, we summarize some phenomenological formulas 
for the mass and width difference in the $B^0-\bar{B}^0$ mixing.
The lattice action and operators we employ in this work are
defined in Section~\ref{sec:NRQCD_action_and_operators}, where the
method to extract the continuum $B$ parameters from lattice matrix
elements is also described.
Simulation details and results are given in
Section~\ref{sec:Lattice_simulations} and
\ref{sec:Simulation_results}, respectively.
In Section \ref{sec:Physics_results} we present our results for the
$B$ parameters and their systematic uncertainties are discussed.
Using these results we also predict the mass and width difference of 
$B_s$ meson system.
Calculation of the SU(3) breaking ratio $\xi$ is briefly discussed in
Section~\ref{sec:SU(3)_breaking_ratio_xi}, and our conclusions are
given in Section~\ref{sec:Conclusions}.

Preliminary reports of this work have already been presented in
\cite{Yamada:2001ym,Hashimoto:2001zq,Yamada:2001ik,Yamada:2001xp}.

\section{Phenomenological formulas}
\label{sec:Phenomenological_formulas}

In this section we summarize the phenomenological formulas which
involves the $B$ meson $B$ parameters.
We also present some notations which will be used throughout the
paper.

\subsection{Mass difference}
In the Standard Model the mass difference in the neutral
$B_q^0-\bar{B}_q^0$ mesons ($q$ denotes $d$ or $s$) is 
given by
\begin{equation}
  \Delta M_q =
  |V_{tb}^*V_{tq}|^2
  \frac{G_F^2 m_W^2}{16 \pi^2 M_{B_q}}
  S_0(x_t) \eta_{2B}
  \left[ \alpha_s(\mu_b) \right]^{-\gamma_0/2\beta_0}
  \left[ 1 + \frac{\alpha_s(\mu_b)}{4\pi}J_5 \right]
  \langle\bar{B}_q^0|\mathcal{O}_{L_q}(\mu_b) | B_q^0\rangle.
  \label{eq:dm_formula}
\end{equation}
$S_0(x_t)$ ($x_t=m_t^2/m_W^2$) is the Inami-Lim function
\cite{Inami:1981fz} and $\eta_{2B}$ is the short distance QCD
correction \cite{Buras:1990fn}, whose full expression is found in 
\cite{Buchalla:1996vs}.

The four-quark operator $\mathcal{O}_{L_q}(\mu_b)$ is defined as 
\begin{equation}
  \mathcal{O}_{L_q}(\mu_b) =
  \bar{b}\gamma_\mu (1-\gamma_5) q\,
  \bar{b}\gamma_\mu (1-\gamma_5) q,
  \label{eq:def_of_OL}
\end{equation}
which depends on the renormalization scale $\mu_b$ if it is defined in
the continuum renormalization schemes, such as the naive dimensional
regularization (NDR) with the modified minimal subtraction
($\overline{\mathrm{MS}}$) scheme. 
The scale dependence cancels with the prefactor
$\left[ \alpha_s(\mu_b) \right]^{-\gamma_0/2\beta_0}
\left[ 1 + \frac{\alpha_s(\mu_b)}{4\pi}J_5 \right]$
such that the physical mass difference is scale independent.
In the NDR-$\overline{\mathrm{MS}}$ scheme the anomalous dimensions
are written as
\begin{equation}
  J_{n_f} = \frac{\gamma_0\beta_1}{2\beta_0^2}
          - \frac{\gamma_1}{2\beta_0},
  \label{eq:J_5}
\end{equation}
and
\begin{equation}
  \begin{array}{c@{\hspace{3em}}c}
    \displaystyle \beta_0 = 11 - \frac{2}{3} n_f, &
    \displaystyle \beta_1 = 102 - \frac{38}{3} n_f, \\[2ex]
    \displaystyle \gamma_0 = 4, &
    \displaystyle \gamma_1 = -7 + \frac{4}{9} n_f.
  \end{array}
  \label{eq:anomalous_dimensions}
\end{equation}
The renormalization scale $\mu_b$ is usually taken at the $b$ quark
mass $m_b$.

The $B$ parameter $B_{B_q}$ is defined through
\begin{equation}
  \langle\bar{B}_q^0|\mathcal{O}_{L_q}(\mu_b) | B_q^0\rangle
  = \frac{8}{3}B_{B_q}(\mu_b)f_{B_q}^2M_{B_q}^2,
  \label{eq:para_BB}
\end{equation}
and the scale-independent $\hat{B}_{B_q}$ is given by
\begin{equation}
  \hat{B}_{B_q} =
  \left[ \alpha_s(\mu_b) \right]^{-\gamma_0/2\beta_0}
  \left[ 1 + \frac{\alpha_s(\mu_b)}{4\pi}J_5 \right]
  B_{B_q}(\mu_b).
  \label{eq:para_BBhat}
\end{equation}
The number of flavors $n_f$ is 5.
To evaluate this expression we use the strong coupling constant 
$\alpha_s(\mu_b)$ defined in the $\overline{\mathrm{MS}}$ scheme with 
$\Lambda_{\overline{\mathrm{MS}}}^{(5)}$ = 225~MeV, which 
corresponds to $\alpha_s^{(5)}(4.8~\mbox{GeV})$ = 0.216.

The bulk of the theoretical uncertainties cancels in the ratio 
\begin{equation}
  \label{eq:dm_ratio}
  \frac{\Delta M_s}{\Delta M_d} =
  \left|\frac{V_{ts}}{V_{td}}\right|^2
  \frac{M_{B_s}}{M_{B_d}} \,
  \xi^2,
\end{equation}
where $\xi$ describes the SU(3) flavor breaking of the matrix element 
$\langle\bar{B}_q^0|\mathcal{O}_{L_q}(\mu_b) | B_q^0\rangle$
as defined in (\ref{eq:xi}).
If one assumes the unitarity relation among the CKM matrix elements
$|V_{ts}|\simeq|V_{cb}|$, (\ref{eq:dm_ratio}) may be used to determine
$|V_{td}|$.

\subsection{Width difference}

Using the heavy quark expansion, the width difference in the neutral 
$B_s^0-\bar{B}_s^0$ mixing is calculated as 
\cite{Beneke:1996gn,Beneke:1999sy}
\begin{equation}
  \Delta\Gamma_{B_s} 
  = - 2 \frac{1}{2M_{B_s}}
  \langle \bar{B}_q|
  \,\mathrm{Im}\,
  i\!\int\!d^4x\,
  T\mathcal{H}_{\mathrm{eff}}(x)\mathcal{H}_{\mathrm{eff}}(0) 
  |B_q\rangle,
  \label{eq:width_difference_formula_1}
\end{equation}
where $\mathcal{H}_{\mathrm{eff}}$ is the $\Delta B$ = 1 weak
transition Hamiltonian. 
The main contribution comes from the transition 
$b\bar{s}\rightarrow c\bar{c}$ followed by 
$c\bar{c}\rightarrow \bar{b}s$,
and other contributions mediated by penguin operators are also
considered.

The operator product expansion (OPE) may be used to
approximate the transition operator
$i\!\int\!d^4x\,T\mathcal{H}_{\mathrm{eff}}(x)\mathcal{H}_{\mathrm{eff}}(0)$,
which gives an $1/m_b$ expansion.
At the leading order in $1/m_b$ the $\Delta B$ = 2 four-quark operator
$\mathcal{O}_{L_s}$ defined in (\ref{eq:def_of_OL}) and another
operator 
\begin{equation}
  \mathcal{O}_{S_s} =
  \bar{b} (1-\gamma_5) s\, \bar{b} (1-\gamma_5) s,
  \label{eq:def_of_OS}
\end{equation}
appear.
Then, the following formula
\begin{eqnarray}
  \left(\frac{\Delta\Gamma}{\Gamma}\right)_{B_s}
  & = & 
  \frac{16\pi^2 B(B_s\rightarrow Xe\nu)}{
    g(z)\tilde{\eta}_{QCD}}
  \frac{f_{B_s}^2 M_{B_s}}{m_b^3} |V_{cs}|^2
  \nonumber\\
  & & \times
  \left( G(z)   \frac{8}{3} B_{B_s}(m_b)
       + G_S(z) \frac{5}{3} \frac{B_{S_s}(m_b)}{\mathcal{R}(m_b)^2}
       + \sqrt{1-4z}\,\delta_{1/m}
  \right)
  \label{eq:width_difference_formula_2}
\end{eqnarray}
is obtained at the next-to-leading order \cite{Beneke:1999sy},
where $m_b$ is the pole mass of $b$ quark.
The width difference is normalized by the total decay width of $B_s$
meson $\Gamma_{B_s}$, which is written in terms of the semileptonic
decay branching ratio $B(B_s\rightarrow Xe\nu)$ on the right hand side
in order to remove an uncertainty in the value of $|V_{cb}|$.
The phase space factor 
$g(z)=1-8z+8z^3-z^4-12z^2\ln z$, where $z=m_c^2/m_b^2$, and
the short distance QCD correction\footnote{
  Following the treatment in \cite{Beneke:1999sy},
  the approximate form of \cite{Kim:1989ac} is used for
  $\tilde{\eta}_{QCD}$.
}
\begin{equation}
  \tilde{\eta}_{QCD}
  =
  1-\frac{2\alpha_s(m_b)}{3\pi}
  \left[
    (\pi^2-\frac{31}{4})(1-\sqrt{z})^2+\frac{3}{2}
  \right]
\end{equation}
are known factors, and the functions $G(z)$ and $G_S(z)$ describe the
next-to-leading order QCD corrections \cite{Beneke:1999sy} appearing
in the calculation of the operator product expansion.
The correction term $\delta_{1/m}$ denotes the next-to-leading order
contribution in the $1/m_b$ expansion, which is estimated in
\cite{Beneke:1996gn} using the factorization approximation. 

The $B$ parameter $B_{S_s}$ is defined through
\begin{equation}
  \langle\bar{B}_s^0| \mathcal{O}_{S_s}(\mu_b) | B_s^0\rangle
  = -\frac{5}{3} f_{B_s}^2 
  \frac{B_{S_s}(\mu_b)}{\mathcal{R}(\mu_b)^2} M_{B_s}^2,
  \label{eq:para_BS}
\end{equation}
where
\begin{equation}
  \mathcal{R}(\mu_b)=\frac{\bar{m}_b(\mu_b)+\bar{m}_s(\mu_b)}{M_{B_s}} 
\end{equation}
is the ratio of matrix elements of heavy-light axial vector current and
pseudoscalar density and $\bar{m}(\mu_b)$ represents a
$\overline{\rm MS}$ quark mass.

In the following analysis, the scale $\mu_b$ is set to the pole mass
of $b$ quark, $m_b$ = 4.8 GeV, 
for which $G(z)$=0.03 and $G_S(z)$=0.937.
With input parameters 
$z=0.085$,
$|V_{cs}|$ = $1-\frac{\lambda^2}{2}$=0.976,
$M_{B_s}$ = 5.37~GeV, 
$B(B_s\rightarrow Xe\nu)$ = 0.107,
we obtain
\begin{equation}
  \left(\frac{\Delta\Gamma}{\Gamma}\right)_{B_s}
  = \left(\frac{f_{B_s}}{230~\mathrm{MeV}} \right)^2
  \left[\ 0.007 B_{B_s}(m_b)
        + 0.207 B_{S_s}(m_b)
        - 0.077\
  \right].
  \label{eq:width_difference_formula_3}
\end{equation}
For the central value of the decay constant, we choose a
recent world average of unquenched lattice calculations
$f_{B_s}$ = 230(30)~MeV \cite{Ryan:2001ej,Yamada:2002_rev}.
The uncertainties associated with these input parameters are discussed 
in Section \ref{sec:Physics_results}.

\section{NRQCD action and operators}
\label{sec:NRQCD_action_and_operators}

In this section we describe the lattice NRQCD action and operators
used in our calculations.
The perturbative matching of the lattice operators to the continuum
ones is summarized.

\subsection{NRQCD action}
\label{subsec:NRQCD_action}

To treat the heavy quark on a lattice with moderate lattice spacing
$a$, the idea of the heavy quark effective theory (HQET) 
\cite{Eichten:1990zv,Georgi:1990um,Grinstein:1990mj}
is useful, as it allows us to describe the heavy quark of mass $M$
without introducing large systematic errors scaling as a (positive)
power of $aM$. 
In HQET the Lagrangian is organized as an expansion in inverse powers
of $M$, and the terms beyond some fixed order are truncated.
Since the physical expansion parameter is $\Lambda_{\mathrm{QCD}}/M$
with $\Lambda_{\mathrm{QCD}}\simeq$ 300--500~MeV, one may typically
achieve the accuracy of order few percent for $B$ mesons at the
next-to-leading order, \textit{i.e.} including terms of order $1/M$.

Though NRQCD was originally introduced in the continuum
\cite{Caswell:1986ui,Bodwin:1995jh} and on the lattice 
\cite{Thacker:1991bm,Lepage:1992tx} to describe the quarkonium
systems such as charmonium and bottomonium,
for which the expansion parameter is a velocity of heavy quark rather
than $\Lambda_{\mathrm{QCD}}/M$,
the formulation can also be used for the lattice study of heavy-light
mesons as first demonstrated in \cite{Hashimoto:1994nd}.
At the next-to-leading order in $\Lambda_{\mathrm{QCD}}/M$, the
Lagrangian in the continuum Euclidean space-time is written
as 
\begin{equation}
  \label{eq:NRQCD_continuum}
  \mathcal{L}_{\mathrm{NRQCD}}^{cont} =
  Q^{\dagger} \left[
    D_0 + \frac{\mathbf{D}^2}{2M}
    + g \frac{\mathbf{\sigma}\cdot\mathbf{B}}{2M}
  \right]
  Q
  +
  \chi^{\dagger} \left[
    D_0 - \frac{\mathbf{D}^2}{2M}
    - g \frac{\mathbf{\sigma}\cdot\mathbf{B}}{2M}
  \right]
  \chi,
\end{equation}
for heavy quark field $Q$ and heavy anti-quark field $\chi$.
Both are represented by a two-component non-relativistic spinor.
The derivatives $D_0$ and $\mathbf{D}$ are temporal and
spatial covariant derivatives respectively.
The leading order term $D_0$ represents a heavy quark as a
static color source.
The leading correction terms of order $\Lambda_{\mathrm{QCD}}/M$ are 
the non-relativistic kinetic term $\mathbf{D}^2/2M$ and
the spin-(chromo)magnetic interaction term 
$\mathbf{\sigma}\cdot\mathbf{B}/2M$, 
where $\mathbf{B}$ denotes the chromomagnetic field strength.
In the usual HQET approach, only the leading terms are kept 
in the effective Lagrangian and corrections of order
$\Lambda_{\mathrm{QCD}}/M$ are treated as operator insertions.
Alternatively, in our lattice calculation we include the correction 
terms in the Lagrangian (\ref{eq:NRQCD_continuum}) and
evaluate the matrix elements with the heavy quark propagator
including the effect of order $\Lambda_{\mathrm{QCD}}/M$.

On the lattice we use a discretized version of the Lagrangian
(\ref{eq:NRQCD_continuum}), whose explicit form is written as
\begin{equation}
  \label{eq:NRQCD_lattice}
  S_{\mathrm{NRQCD}}= 
  \sum_{x,y} Q^{\dagger}(x)( \delta_{x,y} - K_{Q}(x,y) ) Q(y) +
  \sum_{x,y} \chi^{\dagger}(x)( \delta_{x,y} - K_{\chi}(x.y) ) \chi(y).
\end{equation}
The kernel to describe the time evolution of (anti-)heavy quark is
given by
\begin{eqnarray}
  \label{eq:evolution_kernel_Q}
  K_{Q}(x,y) 
  & \equiv &
  \left[ 
    \left( 1-\frac{a H_{0}}{2 n} \right)^{n}
    \left( 1-\frac{a \delta H}{2} \right)
    \delta^{(-)}_{4}{U^{\dagger}_{4}}
    \left( 1-\frac{a \delta H}{2} \right)
    \left( 1-\frac{a H_{0}}{2 n} \right)^{n}
  \right](x,y), \\
  \label{eq:evolution_kernel_chi}
  K_{\chi}(x,y) 
  & \equiv &
  \left[ 
    \left( 1-\frac{a H_{0}}{2 n} \right)^{n}
    \left( 1-\frac{a \delta H}{2} \right)
    \delta^{(+)}_{4}{U_{4}}
    \left( 1-\frac{a \delta H}{2} \right)
    \left( 1-\frac{a H_{0}}{2 n} \right)^{n}
  \right](x,y),
\end{eqnarray}
where $n$ denotes a stabilization parameter introduced in
order to remove the instability arising from unphysical
momentum modes in the evolution equation
\cite{Thacker:1991bm,Lepage:1992tx}.
The operator $\delta^{(\pm)}_4$ is defined as 
$\delta^{(\pm)}_4(x,y) \equiv \delta_{x_4\pm 1,y_4}
 \delta_{\mathbf{x},\mathbf{y}}$,
and $H_{0}$ and $\delta H$ are lattice Hamiltonians defined
by 
\begin{eqnarray}
  \label{eq:kinetic_term}
  H_{0}      
  & \equiv &
  -\frac{\mathbf{\Delta}^{(2)}}{2 aM_{0}},
  \\
  \label{eq:spin-magnetic}
  \delta{H}
  & \equiv &
  -c_{B} \frac{g}{2aM_{0}} \mathbf{\sigma}\cdot\mathbf{B},
\end{eqnarray}
where 
$\mathbf{\Delta}^{(2)}\equiv\sum_{i=1}^{3}\Delta^{(2)}_i$ 
is the Laplacian defined on the lattice with
$\Delta^{(2)}_i\equiv\Delta^{(+)}_i\Delta^{(-)}_i$,
$\Delta^{(+)}_i$ and $\Delta^{(-)}_i$ being forward and backward
covariant derivatives in the $i$-th direction.
In (\ref{eq:spin-magnetic}) the chromomagnetic field operator
$\mathbf{B}$ is the usual clover-leaf type lattice field strength
\cite{Lepage:1992tx}. 
In these definitions, the lattice operators
$\Delta^{(2)}$ and $\mathbf{B}$ are dimensionless,
\textit{i.e.} appropriate powers of $a$ are understood.
The space-time indices $x$ and $y$ are implicit in these
expressions. 
The bare heavy quark mass $M_0$ is distinguished from the
renormalized one $M$.

At the tree level, the lattice action (\ref{eq:NRQCD_lattice})
describes the continuum NRQCD (\ref{eq:NRQCD_continuum}) in the limit
of vanishing lattice spacing $a$.
(We may identify $M_0$ with $M$ and take the tree level value $c_B=1$.)
The leading discretization error for the spatial derivative is of
order $(a\Lambda_{\mathrm{QCD}})^2\Lambda_{\mathrm{QCD}}/M$.
Since the temporal derivative is discretized asymmetrically, the
leading error appears at order $aD_0^2$, whose typical size is 
estimated as $a\Lambda_{\mathrm{QCD}}^3/M^2$ using the equation
of motion. 
The gauge potential part is automatically improved, as it is
exponentiated into the temporal link variable $U_4$.

In the presence of radiative corrections, the heavy quark mass $M_0$ 
and the chromomagnetic coupling $c_B$ have to be tuned in such a way
that the continuum values are reproduced at each value of the strong
coupling constant $\alpha_s$.
Furthermore, the radiative corrections generate many other terms which
do not exist in the continuum Lagrangian (\ref{eq:NRQCD_continuum}),
because NRQCD is not a renormalizable field theory.
In general these terms appear with some factor of form 
$\alpha_s^k/(aM_0)^m$ with positive integers $k$ and $m$
($k\ge 1$ and $m\ge 2$).\footnote{%
  There are also the lowest dimension operators $Q^\dagger Q$ and
  $\chi^\dagger \chi$, but they only give the energy shift and do not 
  contribute to the dynamics of heavy quark.
}
Therefore, NRQCD should be considered as an effective theory valid for
small $1/(aM)$ up to higher order terms in $1/(aM)$.

Perturbation theory can be used to calculate the renormalization of
the parameters.
For example, the one-loop calculations of energy shift and mass
renormalization were carried out for lattice NRQCD by Davies and
Thacker \cite{Davies:1992py} and by Morningstar
\cite{Morningstar:1993de}, and by ourselves
\cite{Hashimoto:1999ck,Ishikawa:2000xu,Hashimoto:2000eh} for
the above particular form of the NRQCD action.\footnote{%
  We note that the evolution kernels
  (\ref{eq:evolution_kernel_Q}) and
  (\ref{eq:evolution_kernel_chi}) are slightly different 
  from the definition used, for example, in
  \cite{Morningstar:1993de}, where the $(1-aH_0/2n)^n$ terms 
  appear inside of the $(1-a\delta H/2)$ terms.
}
To improve the perturbative expansion we utilize the tadpole
improvement \cite{Lepage:1993xa}, namely, all the gauge links in the
action (\ref{eq:NRQCD_lattice}) are divided by its mean field value
$u_0$ determined from the plaquette expectation value
$u_0 \equiv (\langle\mathrm{Tr}U_P\rangle/3)^{1/4}$,
where some counter terms are introduced in perturbative calculations.
The one-loop tuning of the coupling constant $c_B$ for the
spin-(chromo)magnetic interaction term (\ref{eq:spin-magnetic}) has
not yet been performed, so we take the tree level value $c_B$ = 1
after making the tadpole improvement.

The relativistic four-component Dirac spinor field $b$ is related to
the two-component non-relativistic field $Q$ and $\chi$ appearing in
the NRQCD action (\ref{eq:NRQCD_lattice}) via the
Foldy-Wouthuysen-Tani (FWT) transformation  
\begin{equation}
  \label{eq:FWT_transformation}
  b = 
  \left(
    1 -
    \frac{\mathbf{\gamma}\cdot\mathbf{\Delta}^{(\pm)}}{2aM_0}
  \right)
  h,
\end{equation}
where
\begin{equation}
  h \equiv
  \left( 
    \begin{array}{c}
      Q \\ \chi^{\dagger}
    \end{array}
  \right).
\end{equation}
The symbol $\mathbf{\Delta}^{(\pm)}$ denotes a symmetric covariant
differentiation operator 
$\Delta^{(\pm)}_i\equiv\Delta^{(+)}_i\Delta^{(-)}_i$.

\subsection{Bilinear operators}
\label{subsec:Bilinear_operators}
The heavy-light axial-vector current 
$\mathcal{A}_\mu=\bar{b}\gamma_\mu\gamma_5q$
and pseudoscalar density $\mathcal{P}=\bar{b}\gamma_5q$ appear in the 
definition of the $B$ parameters through the vacuum saturation 
approximation.
We use the calligraphic symbols $\mathcal{A}_\mu$ and $\mathcal{P}$ to 
denote the currents defined in the continuum full theory.
Since the pseudoscalar density diverges in the continuum, it is
renormalized with the $\overline{\mathrm{MS}}$ scheme at a scale
$\mu$. 
On the other hand, the axial-vector current does not need
renormalization, because it is partially conserved in the continuum
full theory.

The corresponding lattice operators are 
\begin{eqnarray}
  \label{eq:lattice_currents}
  J^{(0)}_\Gamma & = & \bar{b}\Gamma q,
  \\
  \label{eq:lattice_current_J1}
  J^{(1)}_\Gamma & = & \frac{-1}{2aM_0}
  \bar{b}
  \left( \mathbf{\gamma}\cdot\loarrow{\mathbf{\Delta}}^{(\pm)}
  \right) \Gamma q,
  \\
  \label{eq:lattice_current_J2}
  J^{(2)}_\Gamma & = & \frac{-1}{2aM_0}
  \bar{b} \Gamma
  \left( \mathbf{\gamma}\cdot\mathbf{\Delta}^{(\pm)}
  \right) q,
\end{eqnarray}
where $J_\Gamma$ is $A_4$ for $\Gamma=\gamma_4\gamma_5$ or
$P$ for $\Gamma=\gamma_5$.
The light quark field $q$ is described by the $O(a)$-improved
Wilson quark action \cite{Sheikholeslami:1985ij}.
We apply the tadpole improvement \cite{Lepage:1993xa}
for the light quark field using the critical hopping parameter
$\kappa_c$ to define the mean link variable $u_0=1/8\kappa_c$, 
so that we normalize the light quark field with a factor
$\sqrt{1-\frac{3\kappa}{4\kappa_c}}$.
The heavy quark field $b$ is defined in (\ref{eq:FWT_transformation}).

The one-loop matching between the continuum and lattice operators is
written as
\begin{equation}
  \label{eq:current_matching}
  \mathcal{J}_\Gamma =
       \left[ 1 + \frac{\alpha_s}{4\pi} \rho^{(0)}_\Gamma \right]
                                                J^{(0)}_\Gamma
     + \frac{\alpha_s}{4\pi} \rho^{(1)}_\Gamma  J^{(1)}_\Gamma
     + \frac{\alpha_s}{4\pi} \rho^{(2)}_\Gamma  J^{(2)}_\Gamma,
\end{equation}
with one-loop coefficients $\rho^{(i)}_\Gamma$.
The coefficient $\rho^{(0)}_\Gamma$ is written as
\begin{eqnarray}
  \label{eq:rho0_A4}
  \rho^{(0)}_A &=&
  2\ln(a^2 M^2) + \zeta_A,
  \\
  \label{eq:rho0_P}
  \rho^{(0)}_P &=&
  \frac{9}{2}\ln(\mu^2/M^2) + \frac{3}{2}\ln(a^2M^2)
  + \zeta_P,
\end{eqnarray}
for $\Gamma=\gamma_4\gamma_5$ and $\gamma_5$, respectively.
In the static limit the numerical constants are
$\zeta_A=-16.55$ \cite{Ishikawa:1999rv,Golden:1991dx,Borrelli:1992fy}
and $\zeta_P=-11.21$ \cite{Ishikawa:1999rv}.
For the NRQCD action (\ref{eq:NRQCD_lattice}) with a finite heavy
quark mass $M_0$ the numerical values for $\zeta_A$ and $\zeta_P$ are
available in Table~III of \cite{Hashimoto:2000eh}.\footnote{%
  The same quantity was previously calculated in
  \cite{Davies:1993ec,Morningstar:1998ep}, but for a slightly
  different NRQCD action. 
}

In the static limit, while the second term of
(\ref{eq:current_matching}) vanishes, the third term remains finite
and describes the $O(\alpha_s a\Lambda_{\rm QCD})$ improvement, and
its coefficient $\rho^{(2)}_\Gamma$ is
$\rho^{(2)}_A/2aM_0=\rho^{(2)}_P/2aM_0=13.01$ \cite{Ishikawa:1999rv}.
Away from the static limit, these terms give contributions of the
$O(\alpha_s a\Lambda_{\rm QCD})$ and $O(\alpha_s\Lambda_{\rm QCD}/M)$,
and the one-loop coefficients are calculated only for the axial vector
current $\Gamma=\gamma_4\gamma_5$ for our choice of the NRQCD action
\cite{Ishikawa:2000xu}.\footnote{%
  Note that a different notation is used in \cite{Ishikawa:2000xu}.
  Similar calculation was previously made by Morningstar and
  Shigemitsu \cite{Morningstar:1998yx}.
}

\subsection{$\Delta B$=2 operators}
\label{subsec:DeltaB=2_operators}
We assume that the continuum four-quark operators $\mathcal{O}_L(\mu)$ 
and $\mathcal{O}_S(\mu)$ are renormalized in the
$\overline{\mathrm{MS}}$ scheme with totally anti-commuting
$\gamma_5$. 
In the renormalization of $\mathcal{O}_S(\mu)$, the subtraction of 
evanescent operators is made with the definition given by
Eqs.~(13)-(15) of \cite{Beneke:1999sy}.

For matching of these four-quark operators, the following lattice 
operators are involved at the lowest dimension:
\begin{eqnarray}
  \label{eq:four_quark_operators}
  O_L &=& \bar{b}\gamma_\mu(1-\gamma_5)q\
          \bar{b}\gamma_\mu(1-\gamma_5)q,
  \\
  O_R &=& \bar{b}\gamma_\mu(1+\gamma_5)q\
          \bar{b}\gamma_\mu(1+\gamma_5)q,
  \\
  O_S &=& \bar{b}(1-\gamma_5)q\ \bar{b}(1-\gamma_5)q,
  \\
  O_N &=& 2\ \bar{b}\gamma_\mu(1-\gamma_5)q\
             \bar{b}\gamma_\mu(1+\gamma_5)q
        + 4\ \bar{b}(1-\gamma_5)q\ \bar{b}(1+\gamma_5)q,
  \\
  O_M &=& 2\ \bar{b}\gamma_\mu(1-\gamma_5)q\
             \bar{b}\gamma_\mu(1+\gamma_5)q
        - 4\ \bar{b}(1-\gamma_5)q\ \bar{b}(1+\gamma_5)q,
  \\
  O_P &=& 2\ \bar{b}\gamma_\mu(1-\gamma_5)q\
             \bar{b}\gamma_\mu(1+\gamma_5)q
       + 12\ \bar{b}(1-\gamma_5)q\ \bar{b}(1+\gamma_5)q,
  \\
  O_T &=& 5\ \bar{b}\gamma_\mu(1-\gamma_5)q\
             \bar{b}\gamma_\mu(1+\gamma_5)q
       - 34\ \bar{b}(1-\gamma_5)q\ \bar{b}(1+\gamma_5)q.
  \label{eq:four_quark_operators_end}
\end{eqnarray}
As in the bilinear operator case, one has to introduce higher
dimensional operators to remove $O(a)$ errors at the one-loop level.
For the four-quark operators, however, the necessary one-loop
calculations to remove the $O(\alpha_s a\Lambda_{\rm QCD})$ error are
made only in the static limit \cite{Ishikawa:1999rv,Hashimoto:2000yh}.
The higher dimensional operators appearing in that limit are 
\begin{eqnarray}
  \label{eq:four-quark_operators_higher-dim}
  O_{LD} & = & 
  \bar{b}\gamma_\mu(1-\gamma_5)q\
  \bar{b}\gamma_\mu(1-\gamma_5)
  (a\mathbf{\gamma}\cdot\mathbf{\Delta}^{(\pm)})q,
  \\
  O_{ND} & = &
  2\ \bar{b}\gamma_\mu(1-\gamma_5)q\
     \bar{b}\gamma_\mu(1+\gamma_5)
     (a\mathbf{\gamma}\cdot\mathbf{\Delta}^{(\pm)})q
  \nonumber\\
  & &
  + 4\ \bar{b}(1-\gamma_5)q\ 
       \bar{b}(1+\gamma_5)
       (a\mathbf{\gamma}\cdot\mathbf{\Delta}^{(\pm)})q,
  \\
  O_{SD} & = &
  \bar{b}(1-\gamma_5)q\
  \bar{b}(1-\gamma_5)
  (a\mathbf{\gamma}\cdot\mathbf{\Delta}^{(\pm)})q,
  \\
  O_{PD} & = &
  2\ \bar{b}\gamma_\mu(1-\gamma_5)q\
     \bar{b}\gamma_\mu(1+\gamma_5)
     (a\mathbf{\gamma}\cdot\mathbf{\Delta}^{(\pm)})q
  \nonumber\\
  & &
  + 12\ \bar{b}(1-\gamma_5)q\ 
       \bar{b}(1+\gamma_5)
       (a\mathbf{\gamma}\cdot\mathbf{\Delta}^{(\pm)})q.
\end{eqnarray}

The one-loop matching is written as follows:
\begin{eqnarray}
  \mathcal{O}_L(\mu)
  & = & O_L
  + \frac{\alpha_s}{4\pi}\rho_{L,L} O_L
  + \frac{\alpha_s}{4\pi}\zeta_{L,S} O_S
  \nonumber\\
  & &
  + \frac{\alpha_s}{4\pi}\zeta_{L,R} O_R
  + \frac{\alpha_s}{4\pi}\zeta_{L,N} O_N
  + \frac{\alpha_s}{4\pi}\zeta_{L,M} O_M
  \nonumber\\
  & &
  + \frac{\alpha_s}{4\pi}\zeta_{L,LD} O_{LD}
  + \frac{\alpha_s}{4\pi}\zeta_{L,ND} O_{ND},
  \label{eq:matching_L}
  \\
  \mathcal{O}_S(\mu)
  & = & O_S
  + \frac{\alpha_s}{4\pi}\rho_{S,S} O_S
  + \frac{\alpha_s}{4\pi}\rho_{S,L} O_L 
  \nonumber\\
  & & 
  + \frac{\alpha_s}{4\pi}\zeta_{S,R} O_R
  + \frac{\alpha_s}{4\pi}\zeta_{S,P} O_P
  + \frac{\alpha_s}{4\pi}\zeta_{S,T} O_T
  \nonumber\\
  & & 
  + \frac{\alpha_s}{4\pi}\zeta_{S,SD} O_{SD}
  + \frac{\alpha_s}{4\pi}\zeta_{S,LD} O_{LD}
  + \frac{\alpha_s}{4\pi}\zeta_{S,PD} O_{PD},
  \label{eq:matching_S}
\end{eqnarray}
where the coefficients $\rho_{L,L}$, $\rho_{S,S}$, and $\rho_{S,L}$
contain the physical scale $\mu$ and $m_b$ as follows.
\begin{eqnarray}
  \label{eq:rho_LL}
  \rho_{L,L} & = &
  - 2 \ln\left(\frac{\mu^2}{M^2}\right)
  + 4 \ln(a^2M^2) + \zeta_{L,L},
  \\
  \label{eq:rho_SS}
  \rho_{S,S} & = &
  \frac{16}{3}\ln\left(\frac{\mu^2}{M^2}\right)
  + \frac{4}{3} \ln(a^2M^2) + \zeta_{S,S},
  \\
  \label{eq:rho_SL}
  \rho_{S,L} & = &
  \frac{1}{3}\ln\left(\frac{\mu^2}{M^2}\right)
  - \frac{2}{3}\ln(a^2M^2) + \zeta_{S,L}.
\end{eqnarray}
The numerical results for the one-loop coefficients 
$\zeta_{L,X}$ ($X=L$, $S$, $R$, $N$, and $M$) and
$\zeta_{S,X}$ ($X=S$, $L$, $R$, $P$, and $T$) are given in
Table~VI and VIII of \cite{Hashimoto:2000eh}.
In the static limit, $\zeta_{L,M}$ and $\zeta_{S,T}$ vanish, and
others agree with the previous calculations
\cite{Flynn:1991qz,Borrelli:1992fy,%
DiPierro:1998ty,Gimenez:1999mw,Ishikawa:1999rv}.\footnote{%
  A numerical error in \cite{Borrelli:1992fy} was later corrected in
  \cite{DiPierro:1998ty,Gimenez:1999mw,Ishikawa:1999rv}.
}

The last lines in (\ref{eq:matching_L}) and (\ref{eq:matching_S}) are
added to remove the error of $O(\alpha_s a\Lambda_{\rm QCD})$, but
their coefficients are known only in the static limit.
Their values are
$\zeta_{L,LD}=-17.20$,
$\zeta_{L,ND}= -9.20$,
$\zeta_{S,SD}= -6.88$,
$\zeta_{S,LD}=  2.58$ and
$\zeta_{S,PD}=  1.15$
\cite{Ishikawa:1999rv,Hashimoto:2000yh}.

\subsection{Truncation of expansions}
\label{subsec:Truncation_of_expansions}
As in the matching of the NRQCD action discussed in
Section~\ref{subsec:NRQCD_action}, we have to truncate the $1/M$ and
the perturbative expansions in the matching of the bilinear and
four-quark operators.
The $1/M$ expansion is truncated at $O(1/M)$, which is consistent with
our choice of the NRQCD action (\ref{eq:NRQCD_continuum}), 
and the perturbative corrections of order $\alpha_s^2$
and higher are neglected. 

In addition, there are mixed corrections of
$O(\alpha_s\Lambda_{\rm QCD}/M)$. 
In the matching of the bilinear operators the matching coefficients
for the mixed corrections are available and such corrections were
actually included in \cite{Ishikawa:2000xu} by combining with higher
dimensional operators as shown in (\ref{eq:current_matching}).
For the four-quark operators, however, the mixing with higher
dimensional operators at the one-loop level has not been 
calculated yet\footnote{%
  Except for the static limit, where the mixing terms describe
  the correction of order $\alpha_s a$ rather than $\alpha_s/M$.
}.
Thus, in this paper, the mixed corrections are not considered in both
of the bilinear and four-quark operators.
This means that, for the bilinear operators, only the first term of
(\ref{eq:current_matching}) is taken, thus the matching becomes
multiplicative in this approximation.

At this level of accuracy, it is arbitrary to apply the FWT
transformation to a heavy quark field,
if the heavy quark field forms an operator appearing in the one-loop
corrections.
Namely, in (\ref{eq:matching_L}) and (\ref{eq:matching_S}),
we may replace all the four-quark operators $O_X$ except for that in
the first term by $O'_X$,
where
\begin{equation}
  \label{eq:four_quark_operators_unrotated}
  O'_L = \bar{h}\gamma_\mu(1-\gamma_5)q\
         \bar{h}\gamma_\mu(1-\gamma_5)q,
\end{equation}
and so on, and the heavy quark field $h$ is not rotated by the FWT
transformation (\ref{eq:FWT_transformation}). 
Therefore $O'_X$ differ from $O_X$ at $O(\Lambda_{\rm QCD}/M)$.
In the naive order counting analysis both choices are equivalent up to
unknown corrections of $O(\alpha_s\Lambda_{\rm QCD}/M)$.

In the calculation of $B$ parameters through the ratios
\begin{eqnarray}
  \label{eq:BB_from_ratio}
  B_B(\mu_b) & = &
  \frac{
    \langle\bar{B}|\mathcal{O}_L(\mu_b)|B\rangle
  }{
    \frac{8}{3}
    \langle\bar{B}|\mathcal{A}_\mu|0\rangle
    \langle 0|\mathcal{A}_\mu|B\rangle
  },
  \\
  \label{eq:BS_from_ratio}
  B_S(\mu_b) & = &
  \frac{
    \langle\bar{B}|\mathcal{O}_S(\mu_b)|B\rangle
  }{
    \frac{5}{3}
    \langle\bar{B}|\mathcal{P}(\mu_b)|0\rangle
    \langle 0|\mathcal{P}(\mu_b)|B\rangle
  },
\end{eqnarray}
the perturbative and $1/M$ expansions may be truncated in several
different ways.
A natural choice to match the ratios (\ref{eq:BB_from_ratio}) and
(\ref{eq:BS_from_ratio}) is to write the numerator and denominator as
they stand:
\begin{eqnarray}
  \label{eq:BB_from_ratio_I}
  B_B^{\mathrm{(I)}}(\mu_b) & = &
  \frac{
    \left[ 1 + \frac{\alpha_s}{4\pi}\rho_{L,L} \right] B_{B,L}^{lat}
    + \sum_{X=S,R,N,M}
    \frac{\alpha_s}{4\pi}\zeta_{L,X} B_{B,X}^{lat}
  }{
    \left[ 1 + \frac{\alpha_s}{4\pi}\rho^{(0)}_A \right]^2
  },
  \\
  \label{eq:BS_from_ratio_I}
  B_S^{\mathrm{(I)}}(\mu_b) & = &
  \frac{
    \left[ 1 + \frac{\alpha_s}{4\pi}\rho_{S,S} \right] B_{S,S}^{lat}
    + \frac{\alpha_s}{4\pi}\rho_{S,L} B_{S,L}^{lat}
    + \sum_{X=R,P,T}
    \frac{\alpha_s}{4\pi}\zeta_{S,X} B_{S,X}^{lat}
  }{
    \left[ 1 + \frac{\alpha_s}{4\pi}\rho^{(0)}_P \right]^2
  }.
\end{eqnarray}
A roman numeral, (I) in this case, as a superscript of $B_B$ or
$B_S$ distinguishes the method to truncate the expansion.
$B_{B,X}^{lat}$ and $B_{S,X}^{lat}$ are $B$ parameters defined with
the lattice operators as
\begin{eqnarray}
  \label{eq:BB^lat}
  B_{B,X}^{lat} & = &
  \frac{
    \langle\bar{B}|O_X|B\rangle
  }{
    \frac{8}{3}
    \langle\bar{B}|A_4^{(0)}|0\rangle
    \langle 0|A_4^{(0)}|B\rangle
  },
  \\
  \label{eq:BS^lat}
  B_{S,X}^{lat} & = &
  \frac{
    \langle\bar{B}|O_X|B\rangle
  }{
    \frac{5}{3}
    \langle\bar{B}|P^{(0)}|0\rangle
    \langle 0|P^{(0)}|B\rangle
  },
\end{eqnarray}
which are directly measured in the numerical simulation from a ratio
of correlation functions as we describe in the next section.

Alternatively, one may linearize the perturbative expansion as
\begin{eqnarray}
  \label{eq:BB_from_ratio_II}
  B_B^{\mathrm{(II)}}(\mu_b) & = &
  \left[ 1 + \frac{\alpha_s}{4\pi}(\rho_{L,L}-2\rho_A^{(0)}) 
  \right] B_{B,L}^{lat}
  + \sum_{X=S,R,N,M}
  \frac{\alpha_s}{4\pi}\zeta_{L,X} B_{B,X}^{lat},
  \\
  \label{eq:BS_from_ratio_II}
  B_S^{\mathrm{(II)}}(\mu_b) & = &
  \left[ 1 + \frac{\alpha_s}{4\pi}(\rho_{S,S}-2\rho_P^{(0)}) 
  \right] B_{S,S}^{lat}
  + \frac{\alpha_s}{4\pi}\rho_{S,L} B_{S,L}^{lat}
  + \sum_{X=R,P,T}
  \frac{\alpha_s}{4\pi}\zeta_{S,X} B_{S,X}^{lat}.
\end{eqnarray}
Formally they are different from the method I by order
$\alpha_s^2$, which is not known.
We expect, however, that perturbative expansion behaves better for
the method II, because the contributions from factorized diagrams to the
four-quark operators are the same as those of the corresponding bilinear
operators, so that the radiative corrections partly cancel in the
combination 
$\rho_{L,L}-2\rho_A^{(0)}$ or $\rho_{S,S}-2\rho_P^{(0)}$.

For each method I or II, we also consider the variation of replacing
$O_X$ and ${J}^{(0)}_\Gamma$ by $O'_X$ and ${J'}^{(0)}_\Gamma$,
respectively, as discussed above, and define the methods as I' and
II', where ${J'}^{(0)}_\Gamma$ are defined similarly to
(\ref{eq:four_quark_operators_unrotated}).
The difference of the method I' (II') from I (II) is of order
$\alpha_s\Lambda_{\rm QCD}/M$. 

Since the level of accuracy of these four methods is equivalent in the
naive order counting argument, they can be used to estimate possible
systematic errors due to the truncation of expansions.

\section{Lattice simulations}
\label{sec:Lattice_simulations}

\subsection{Simulation sets}

We have performed numerical simulations at four $\beta$ values. 
For three of them ($\beta$=6.1, 5.9, and 5.7), 
which we call the simulation set \textit{A}, the $O(a)$-improvement 
coefficient $c_{\mathrm{SW}}$ in the light quark action is determined
using the one-loop expression
$c_{\mathrm{SW}}=(1/P^{3/4})[1+0.199\alpha_V(1/a)]$.
The one-loop coefficient is calculated in
\cite{Wohlert:1987rf,Naik:1993ux,Luscher:1996vw}, and 
we apply the tadpole improvement \cite{Lepage:1993xa} with the
plaquette expectation value to define the mean link variable.
For the last simulation ($\beta$=6.0), which we denote as the 
simulation set \textit{B}, the non-perturbative value is used for 
$c_{\mathrm{SW}}$ \cite{Luscher:1997ug}.
Therefore, as far as the light quark sector is concerned, the
discretization error is minimized in the set \textit{B}, for which the
leading error is of $O(a^2)$, while the effect of $O(\alpha_s^2 a)$ is
remaining in the set \textit{A}.
For the quantities involving heavy quarks, however, both sets of
simulations give the same order of accuracy, since the heavy quark
action and operators are not improved at the same level.

Simulation parameters are summarized in
Table~\ref{tab:simulation_parameters}.
The parameters for the simulation set \textit{A} are almost the same
as in our previous work for the leptonic decay constant
\cite{Ishikawa:2000xu}, except that
the number of statistical ensembles is increased in this work to
obtain stable signals for three-point functions.
The set \textit{B} is our new simulation set, which is intended for
comparison with our recent unquenched simulations
\cite{Aoki:2001yi,Aoki:2002yr}, and its $\beta$ value, $\beta$ = 6.0,
is chosen so that the inverse lattice spacing becomes about 2~GeV.
In this paper, we present only the quenched results leaving the
unquenched calculations for future publications.

For both simulation sets, \textit{A} and \textit{B},
we take the standard plaquette gauge action, and the configuration
generation and gauge fixing are made as in \cite{Ishikawa:2000xu}.
Four values of
the light quark hopping parameter are chosen for each $\beta$ as given
in Table~\ref{tab:simulation_parameters}.
They correspond to the light quark mass $m_q$ covering the range 
$m_s/2<m_q<2m_s$, where $m_s$ denotes the physical strange quark 
mass. 
The hopping parameter corresponding to the strange quark mass is
determined using the $K$ or $\phi$ meson masses as input, and will be
denoted as $\kappa_{s1}$ and $\kappa_{s2}$ respectively.

The heavy quark mass in our simulation ranges from $2m_b/3$ to $4m_b$.
The smallest heavy quark mass in the lattice unit $aM_0$ is limited
around unity due to the possibly large systematic error in the
matching calculation as discussed in
Section~\ref{sec:NRQCD_action_and_operators}.
The limit in the heaviest side is set by the exponentially growing
statistical error \cite{Hashimoto:1994nd}.

The lattice spacing $a$ is determined through the string tension (for
the set \textit{A}) or the rho meson mass (for the set \textit{B}).
For the simulation set \textit{A} it is confirmed that both
determinations are in good agreement (3.5\% variation depending on
$\beta$) \cite{Aoki:1998ji}.
Therefore, in effect the lattice spacing is set using the rho meson
mass for both data sets.

We use the simulation set \textit{B} to obtain our central value and
the other to investigate the systematic errors depending on the
lattice spacing. 
The primary reason is that the discretization error is minimized by
the non-perturbative $O(a)$ improvement for the set \textit{B}.
The set \textit{B} is also advantageous since we have larger
statistics and hence the numerical results are more stable.

\subsection{Correlation functions}

The method to calculate two- and three-point functions mostly follows
that of \cite{Hashimoto:1999ck}.
We put a local source at the origin of the lattice and solve for the
light quark propagator.
The heavy quark and anti-quark propagators are obtained from the same
local source by solving the evolution kernels
(\ref{eq:evolution_kernel_Q}) and 
(\ref{eq:evolution_kernel_chi}), respectively.

Three point functions are constructed as
\begin{equation}
  \label{eq:three_point_function}
  C_X^{(3)}(t_1,t_2)
  = \sum_{\vec{x}_1}\sum_{\vec{x}_2}
  \langle 0 |
  {\cal T}
  \left[ 
    {A_4^S}^\dag(t_1,\vec{x}_1) O_X(0,\vec{0})
    {A_4^S}^\dag(t_2,\vec{x}_2) 
  \right]
  |0 \rangle,
\end{equation}
where $O_X$ is one of the four-quark operators defined in
(\ref{eq:four_quark_operators})--(\ref{eq:four_quark_operators_end}).
We take $t_1>0$ and $t_2<0$ so that a $\bar{B}$ meson propagates in
the positive direction in time and a $B$ meson propagates in the
opposite direction.
We also measure two point functions
\begin{eqnarray}
  \label{eq:two_point_function_A}
  C_A^{(2)}(t)
  & = & 
  \sum_{\vec{x}}
  \langle 0 |
  {\cal T}
  \left[ {A_4^S}^{\dag}(t,\vec{x}) A_4^{(0)}(0,\vec{0}) \right]
  | 0 \rangle,
  \\
  \label{eq:two_point_function_P}
  C_P^{(2)}(t)
  & = & 
  \sum_{\vec{x}}
  \langle 0 |
  {\cal T}
  \left[ {A_4^S}^{\dag}(t,\vec{x}) P^{(0)}(0,\vec{0}) \right]
  | 0 \rangle,
\end{eqnarray}
for positive and negative values of $t$.

A smeared current $A_4^S$, defined as
\begin{equation}
  A_4^{S}(t,\vec{x}) =
  \sum_{\vec{y}} \phi(\vec{y}) 
  \bar{b}(t,\vec{x}+\vec{y}) \gamma_4\gamma_5 q(t,\vec{x}),
\end{equation}
is used to enhance the overlap with the
ground state $B$ meson.
We measure the smearing function $\phi(\vec{x})$ for each set of heavy
and light quark masses with a limited number of gauge configurations
before starting the main simulation. 

We extract the lattice $B$ parameters $B_{B,X}^{lat}$
(\ref{eq:BB^lat}) and $B_{S,X}^{lat}$ (\ref{eq:BS^lat}) from the
following ratios
\begin{eqnarray}
  \label{eq:BB^lat_from_ratio}
  R_{B,X}(t_1,t_2) \equiv
  \frac{C_X^{(3)}(t_1,t_2)}{
    \frac{8}{3} C_A^{(2)}(t_1) C_A^{(2)}(t_2)
  }
  & \stackrel{|t_i|\gg 1}{\longrightarrow} &
  B_{B,X}^{lat},
  \\
  \label{eq:BS^lat_from_ratio}
  R_{S,X}(t_1,t_2) \equiv
  \frac{C_X^{(3)}(t_1,t_2)}{
    \frac{5}{3} C_P^{(2)}(t_1) C_P^{(2)}(t_2)
  }
  & \stackrel{|t_i|\gg 1}{\longrightarrow} &
  B_{S,X}^{lat},
\end{eqnarray}
for large enough $|t_i|$ ($i=1,2$).

\subsection{Meson masses}
In order to calculate the heavy-light meson masses precisely, we also
calculate two-point functions with the smeared source and local sink,
\begin{eqnarray}
  \label{eq:two_point_function_A^LS}
  C_A^{(2)LS}(t)
  & = & 
  \sum_{\vec{x}}
  \langle 0 |
  {\cal T}
  \left[ A_4^{(0)\dag}(t,\vec{x}) A_4^S(0,\vec{0}) \right]
  | 0 \rangle,
  \\
  \label{eq:two_point_function_P^LS}
  C_P^{(2)LS}(t)
  & = & 
  \sum_{\vec{x}}
  \langle 0 |
  {\cal T}
  \left[ A_4^{(0)\dag}(t,\vec{x}) P^S(0,\vec{0}) \right]
  | 0 \rangle,
\end{eqnarray}
for which the statistical signal is much better than those with
the local source and smeared sink.
The heavy-light meson mass is, then, obtained by adding the binding
energy $E_{\mathrm{bin}}$ measured from the two-point functions and bare
quark mass $aM_0$.
Including one-loop corrections we use
\begin{equation}
  aM_P = Z_m aM_0 + E_{\mathrm{bin}} - \delta m,
\end{equation}
where perturbative corrections $Z_m$ and $\delta m$ are given as
\begin{eqnarray}
  Z_m      &=& 1 + \alpha_s B,\\
  \delta m &=& \alpha_s A,
\end{eqnarray}
and $A$ and $B$ for each bare quark mass are given in Table. I of
\cite{Ishikawa:2000xu}.

\section{Simulation results}
\label{sec:Simulation_results}

\subsection{Ratio of correlation functions}
\label{Ratio_of_correlation_functions}
\begin{figure}[p]
  \centering
  \includegraphics*[width=8cm,clip=true]{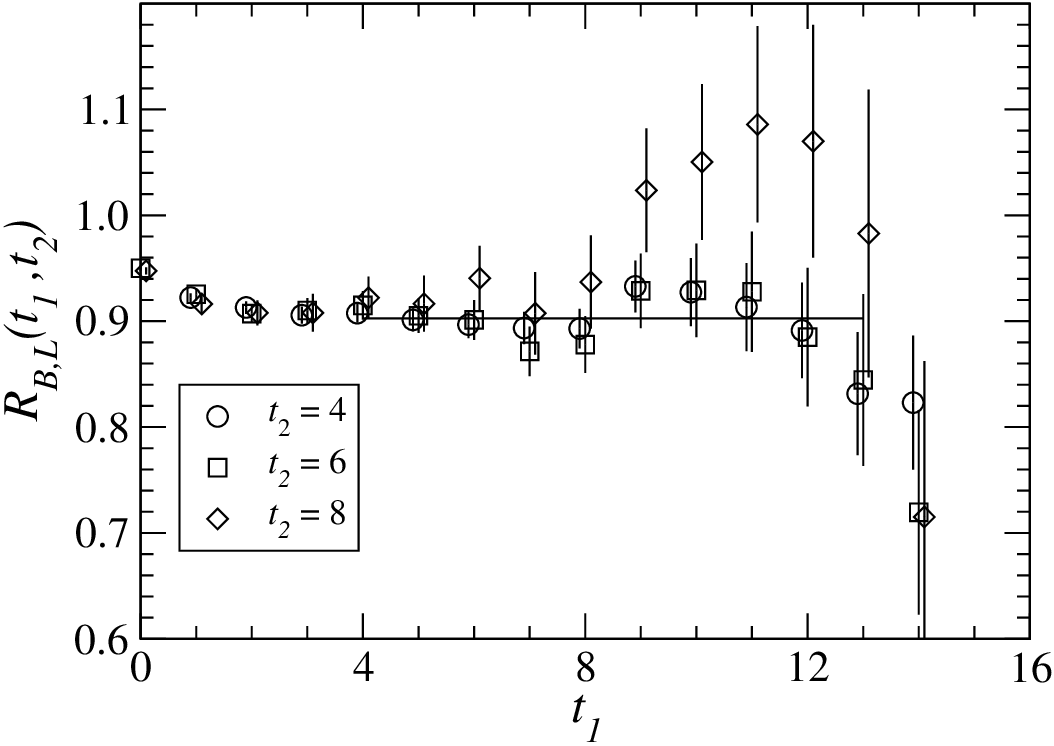}
  \\[2mm]
  \includegraphics*[width=8cm,clip=true]{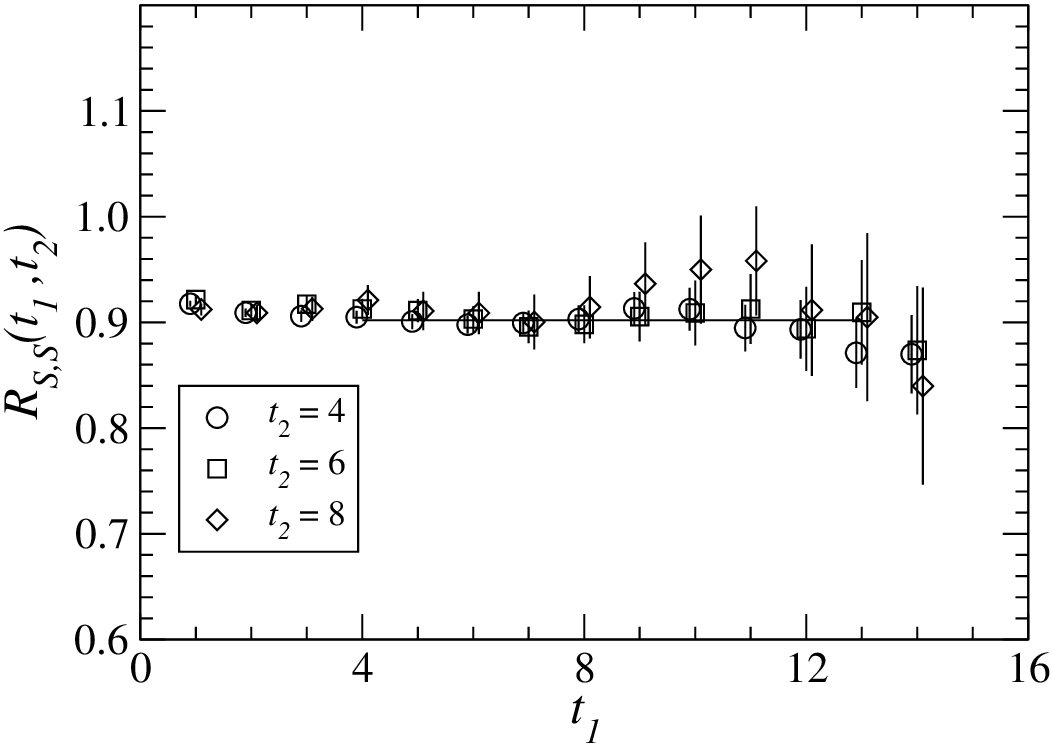}
  \\[2mm]
  \includegraphics*[width=8cm,clip=true]{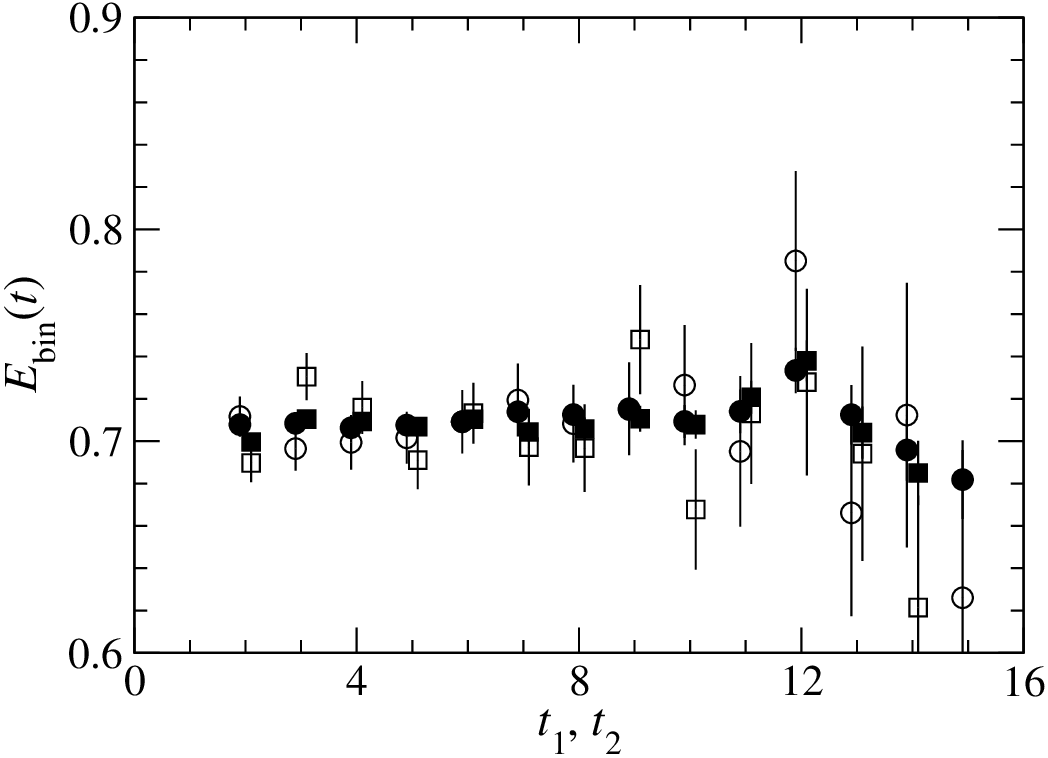}
  \caption{
    $R_{B,L}(t_1,t_2)$ (top) and $R_{S,S}(t_1,t_2)$ (middle) at
    $\beta$ = 5.7, $aM_0$ = 3.8 and $\kappa$ = 0.13920.
    Horizontal line represents a fit with a range $t_1 = [4,13]$ for a
    fixed $t_2 = 4$. 
    The bottom plot shows an effective mass for two point functions
    $C_A^{(2)}(t)$ (open symbols) and $C_A^{(2)LS}(t)$ (filled
    symbols).
    Circles and squares represent data points for positive and
    negative $t$ respectively.
    }
  \label{fig:R_b57}
\end{figure}

\begin{figure}[p]
  \centering
  \includegraphics*[width=8cm,clip=true]{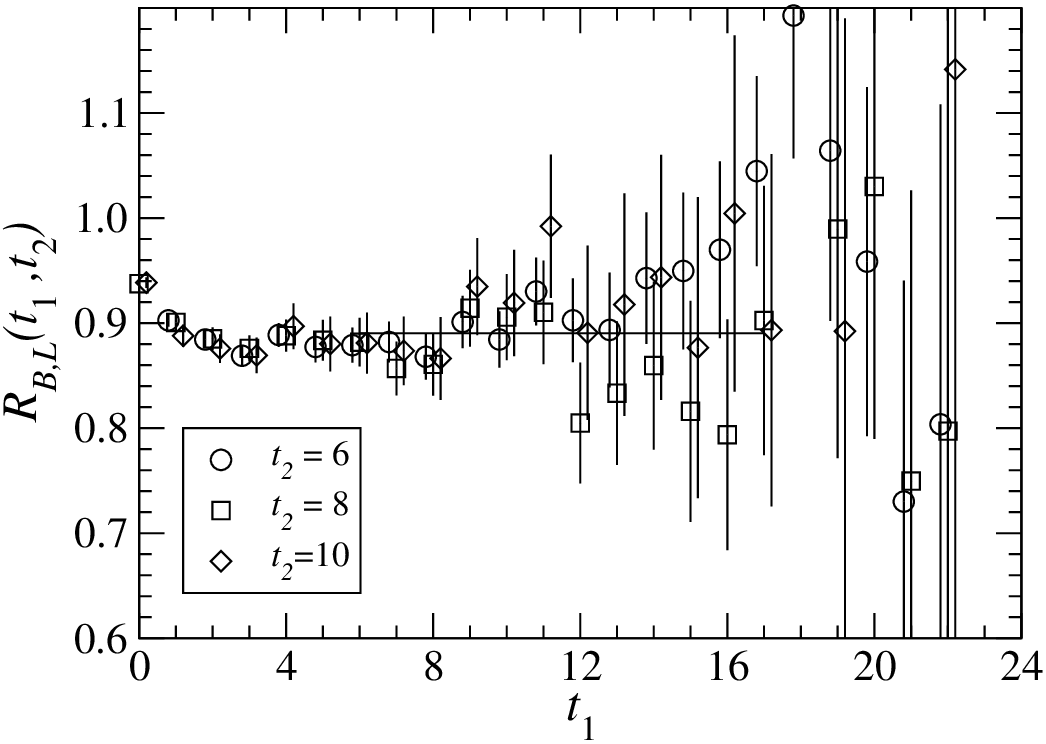}
  \\[2mm]
  \includegraphics*[width=8cm,clip=true]{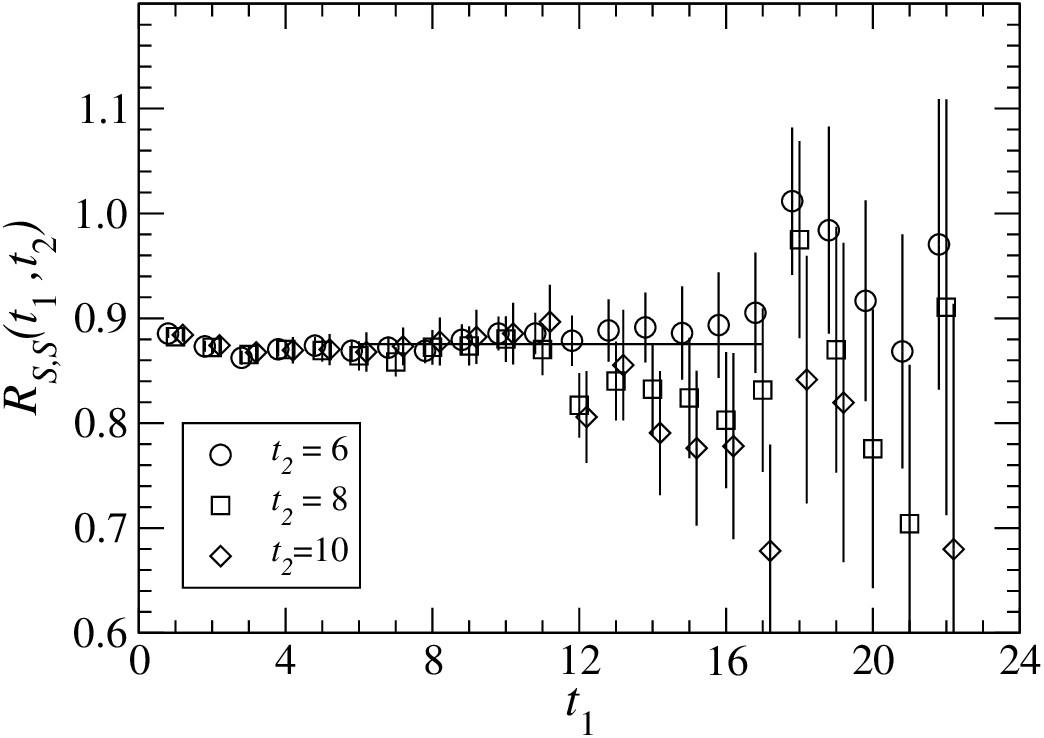}
  \\[2mm]
  \includegraphics*[width=8cm,clip=true]{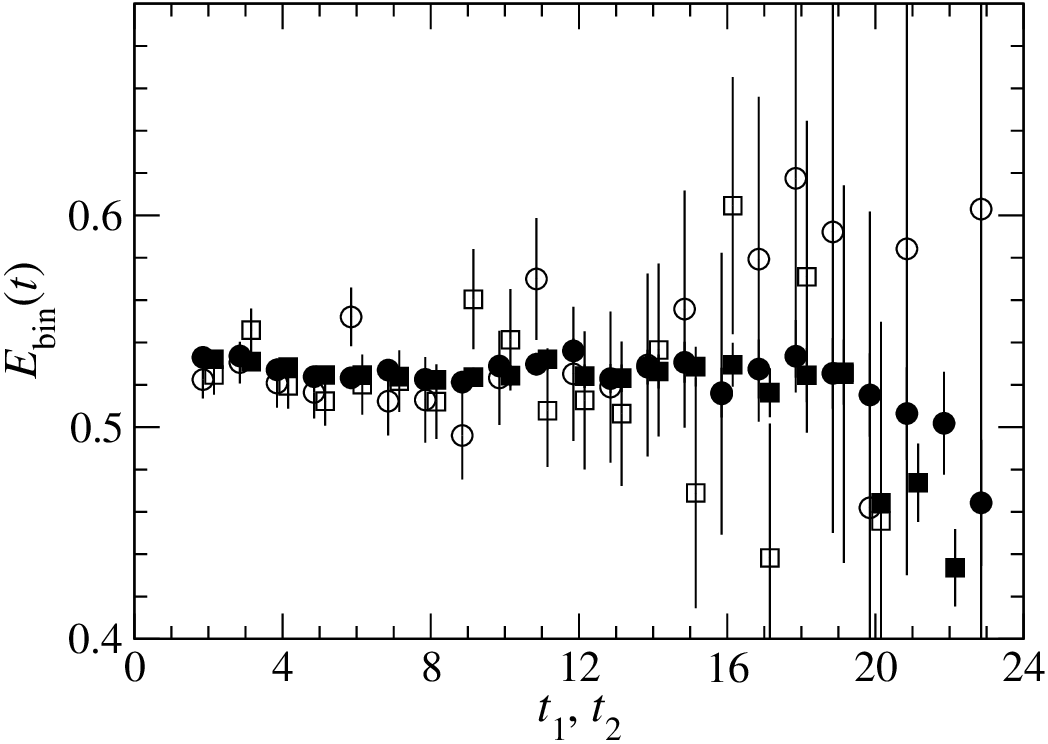}
  \caption{
    Same as Fig.~\ref{fig:R_b57}, but for
    $\beta$ = 5.9, $aM_0$ = 2.1 and $\kappa$ = 0.13816.
    Horizontal line represents a fit with a range $t_1 = [6,20]$ for a
    fixed $t_2 = 6$. 
    }
  \label{fig:R_b59}
\end{figure}

\begin{figure}[p]
  \centering
  \includegraphics*[width=8cm,clip=true]{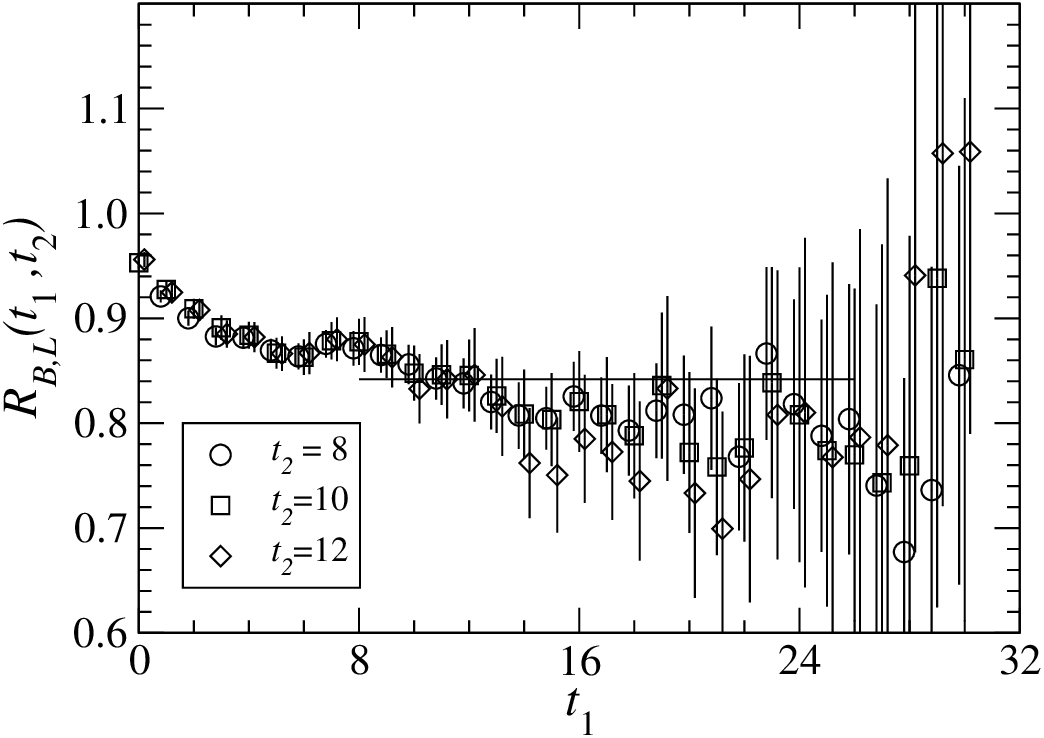}
  \\[2mm]
  \includegraphics*[width=8cm,clip=true]{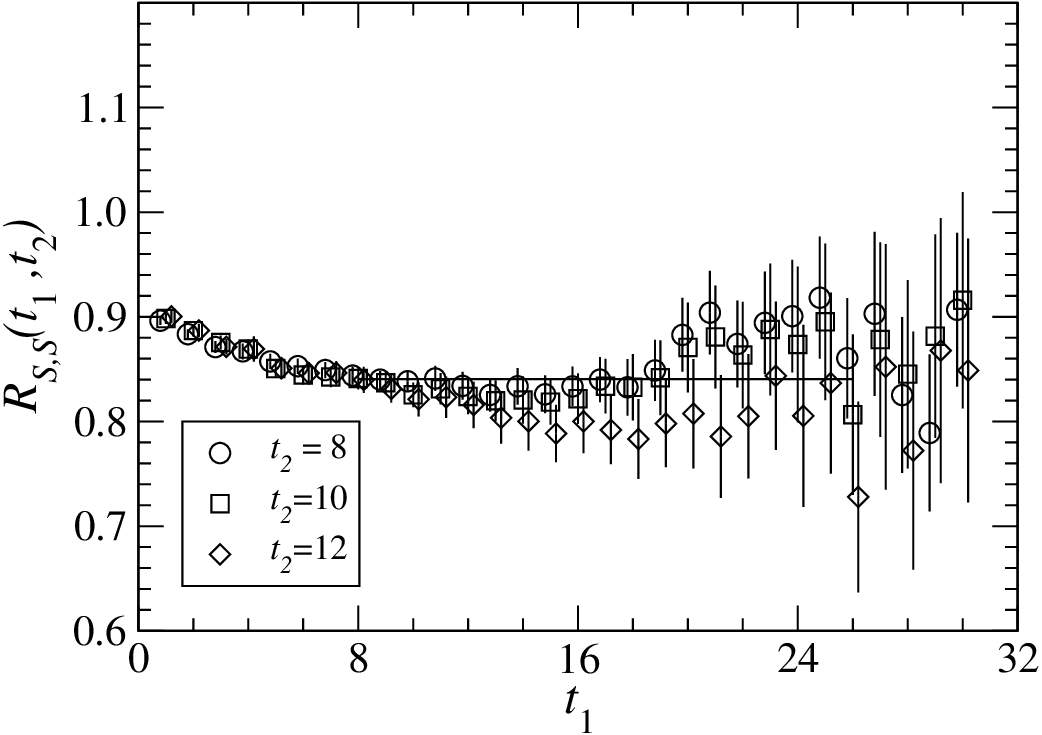}
  \\[2mm]
  \includegraphics*[width=8cm,clip=true]{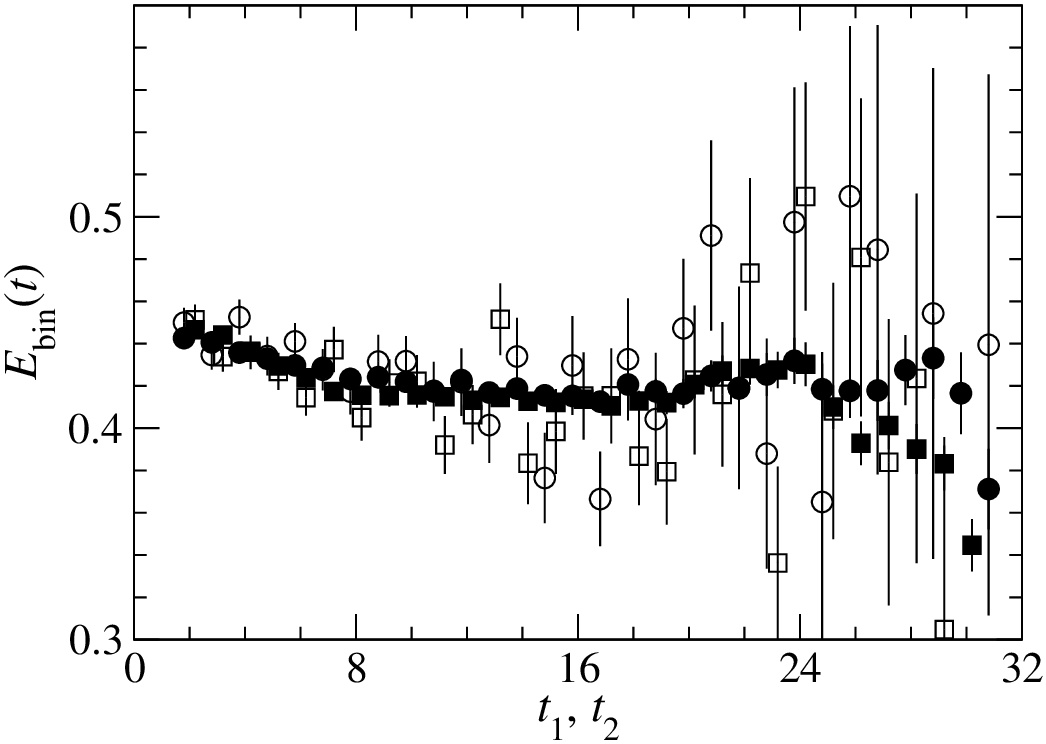}
  \caption{
    Same as Fig.~\ref{fig:R_b57}, but for
    $\beta$ = 6.1, $aM_0$ = 1.5 and $\kappa$ = 0.13716.
    Horizontal line represents a fit with a range $t_1 = [8,28]$ for a
    fixed $t_2 = 8$. 
    }
  \label{fig:R_b61}
\end{figure}

\begin{figure}[p]
  \centering
  \includegraphics*[width=8cm,clip=true]{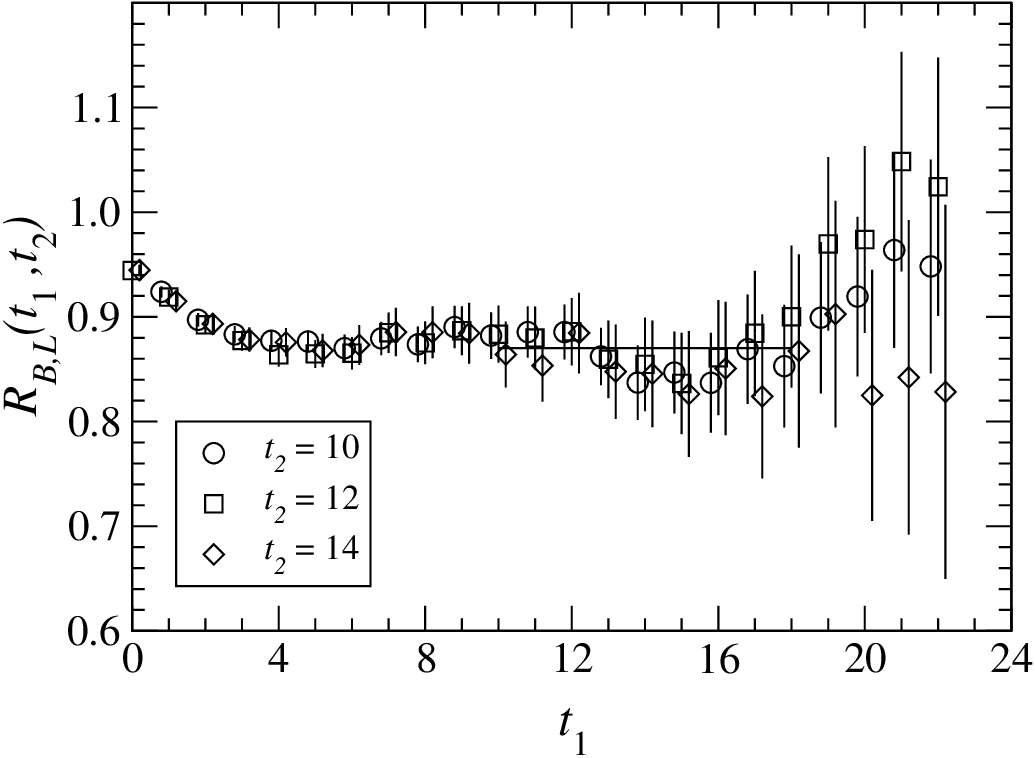}
  \\[2mm]
  \includegraphics*[width=8cm,clip=true]{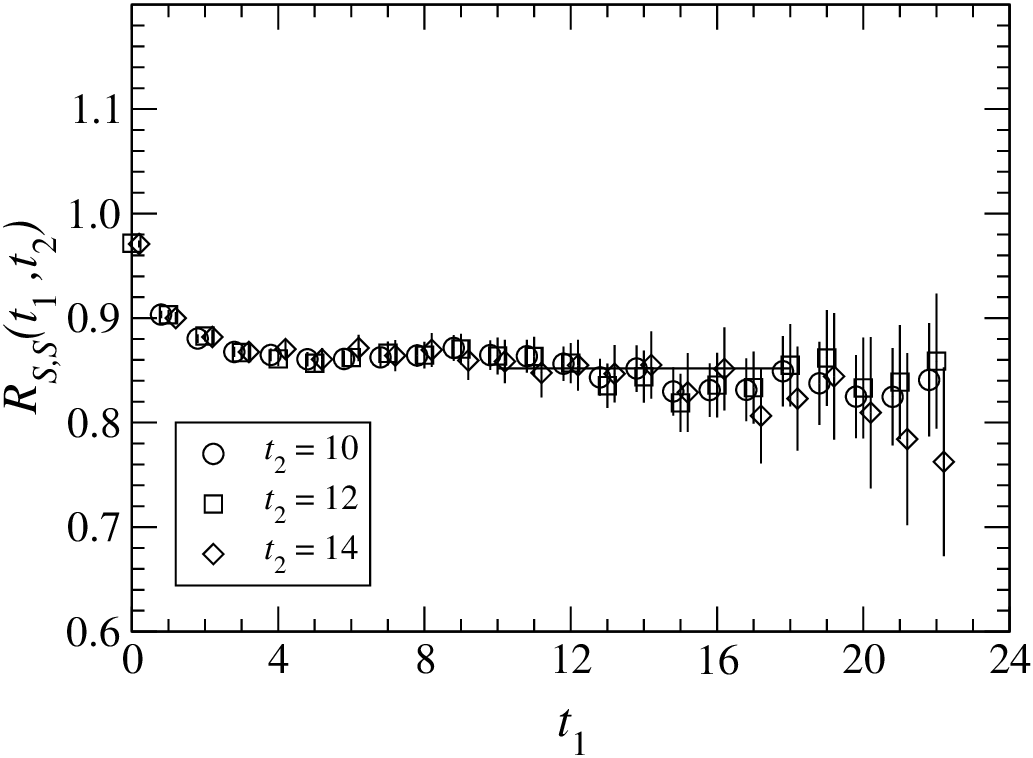}
  \\[2mm]
  \includegraphics*[width=8cm,clip=true]{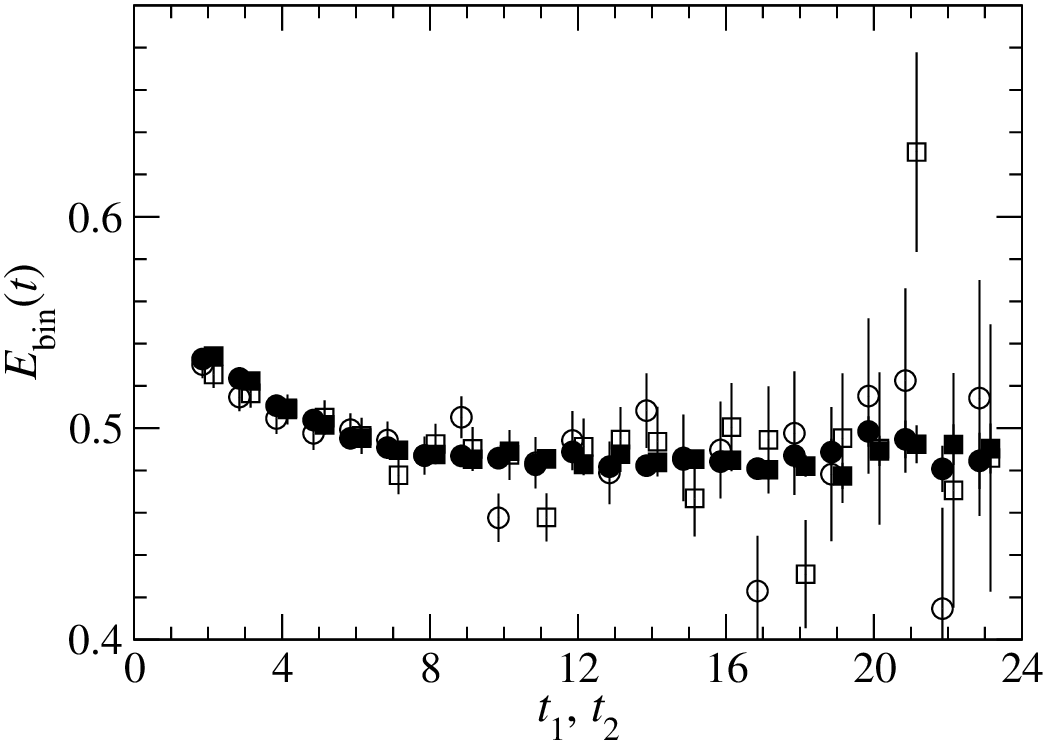}
  \caption{
    Same as Fig.~\ref{fig:R_b57}, but for 
    $\beta$ = 6.0, $aM_0$ = 2.1 and $\kappa$ = 0.13432.
    Horizontal line represents a fit with a range $t_1 = [10,18]$ for a
    fixed $t_2 = 10$. 
    }
  \label{fig:R_b60}
\end{figure}

We first extract the $B$ parameters defined on the lattice 
$B_{B,X}^{lat}$ and $B_{S,X}^{lat}$, which are obtained from the
asymptotic behavior of the ratios
$R_{B,X}(t_1,t_2)$ and $R_{S,X}(t_1,t_2)$ as
(\ref{eq:BB^lat_from_ratio})--(\ref{eq:BS^lat_from_ratio}).
In Figures \ref{fig:R_b57}--\ref{fig:R_b60} (top and middle panels in
each Figure) we plot these ratios as a function of $t_1$ for some
fixed values of $t_2$.
For illustration we show the operators giving leading contributions,
\textit{i.e.} $B_{B,L}^{lat}$ and $B_{S,S}^{lat}$,
for the heavy quark mass closest to the physical $b$ quark mass and
the lightest quark mass (largest $\kappa$ value).
We obtain similar plots for other mass parameters, but the statistical 
signal becomes much noisier for larger heavy quark mass.

The range of $t$ ($t_1$ and $t_2$) to be included in the fit of the
ratios has to be chosen such that the effect of excited states is
negligible. 
We identify the plateau seen in the plots of $R_{B,X}(t_1,t_2)$
and $R_{S,X}(t_1,t_2)$ as the region where the ground state
contribution dominates.
To be more conservative, we also check that the plateau is reached for
the effective mass plot of two-point functions $C_A^{(2)}(t)$ and
$C_A^{(2)LS}(t)$, which are calculated for the same smearing function
as used in the calculation of three-point functions.
The plots are shown in the bottom panel of Figures
\ref{fig:R_b57}--\ref{fig:R_b60}.

In the fit of the ratios we take a range of $t$ as wide as
possible in order to avoid possible contamination from the statistical 
fluctuation \cite{Aoki:1996bb}.
The fit is done for a fixed value of $t_2=t_{1min}$ and changing $t_1$
in the range $[t_{1min},t_{1max}]$.
The value of $[t_{1min},t_{1max}]$ is listed in
Table~\ref{tab:simulation_parameters}.
In order to quantify the possible effect from excited state
contamination, we also carried out a fit with larger values of
$t_{1min}$ (= $t_2$).
Since the statistical error grows rapidly as $t_{1min}$ is taken
larger, the maximum change for $t_{1min}$ is chosen to keep the statistical
error smaller than 8--10\%.
The effect for $B_B$ is found to be 1\% or less except for $\beta$ =
6.1 where it is at most 3\%.
For $B_S$ it brings a 1.5--3\% effect except for $\beta$=5.7 where it
is negligible.
The variations among different choices of the fit range are taken into
account in the final results.

\subsection{Chiral extrapolation}
\label{subsec:Chiral_extrapolation}
\begin{figure}[p]
  \centering
  \includegraphics*[width=8cm,clip=true]{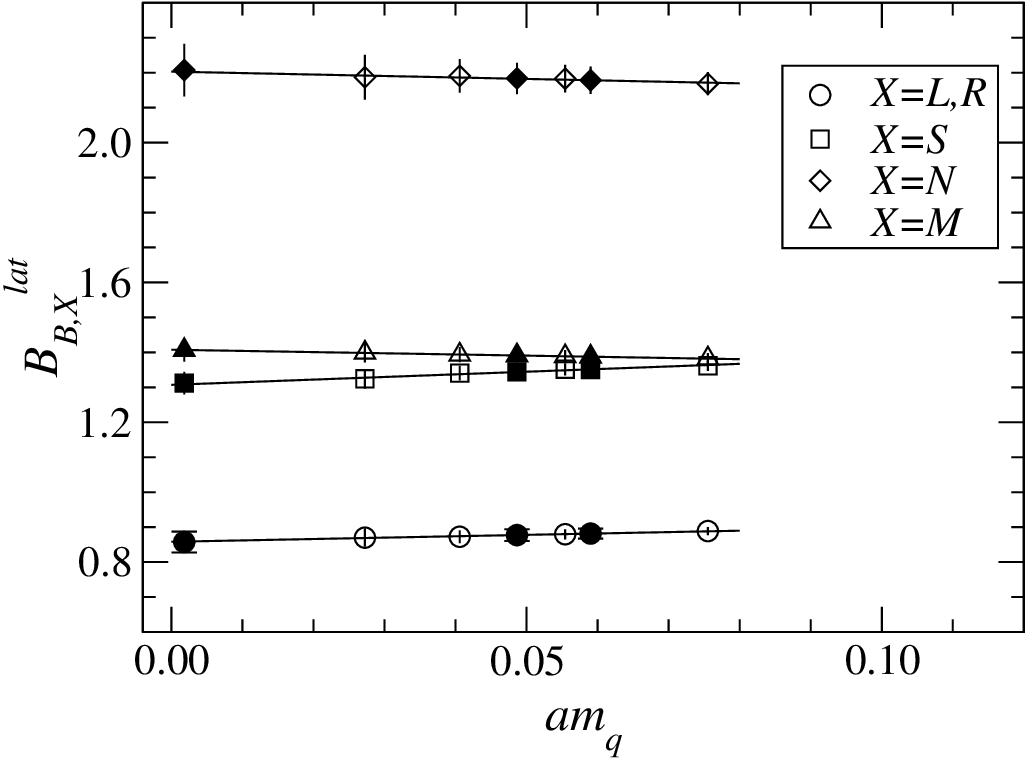}
  \\[2mm]
  \includegraphics*[width=8cm,clip=true]{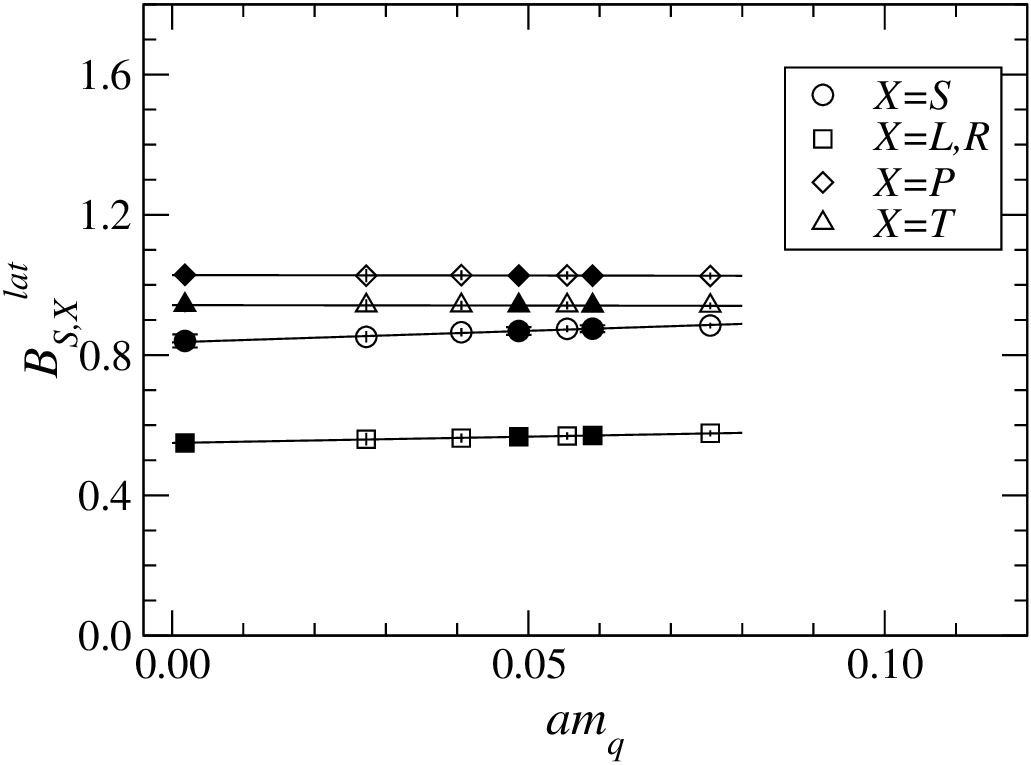}
  \caption{
    Chiral extrapolation of $B_{B,X}^{lat}$ (top panel) and 
    $B_{S,X}^{lat}$ (bottom panel)
    at $\beta$ = 6.0 and $aM_0$ = 2.1.
    Data are normalized by their vacuum saturation approximation
    (VSA). It is
    $B_{B,L}^{\mathrm{(VSA)}} = 1$,
    $B_{B,S}^{\mathrm{(VSA)}} = -5/8$,
    $B_{B,N}^{\mathrm{(VSA)}} = 1$,
    $B_{B,M}^{\mathrm{(VSA)}} = -6$
    for $B_{B,X}^{lat}$, and
    $B_{S,S}^{\mathrm{(VSA)}} = 1$,
    $B_{S,L}^{\mathrm{(VSA)}} = -8/5$,
    $B_{S,P}^{\mathrm{(VSA)}} = -64/5$,
    $B_{S,T}^{\mathrm{(VSA)}} = 288/5$
    for $B_{S,X}^{lat}$.
    In VSA the correction of order $1/M$ is neglected.
    }
  \label{fig:chiral_extrapolation}
\end{figure}

The results of $B_{B,X}^{lat}$ and $B_{S,X}^{lat}$ are insensitive to
the light quark mass.
An example is shown in Figure~\ref{fig:chiral_extrapolation}, where
the data at $\beta$ = 6.0 are plotted as a function of
$am_q \equiv 
\frac{1}{2}\left(\frac{1}{\kappa}-\frac{1}{\kappa_c}\right)$.
The $B$ parameters for all relevant operators are plotted:
$X$ = $L$, $R$, $S$, $N$ and $M$ for $B_{B,X}^{lat}$, and 
$X$ = $S$, $L$, $R$, $P$, $T$ for $B_{S,X}^{lat}$.
Each $B$ parameter is normalized by its vacuum saturation
approximation.
We averaged the matrix elements with $X$ = $L$ and $R$, as they should
be equal in the infinite number of statistics by parity symmetry.
In later sections, the averaged matrix elements are denoted by 
$X$ = $LR$.

We extrapolate these $B$ parameters to the chiral limit of light quark
assuming a linear function in $am_q$.
In most cases the chiral extrapolation changes the value of $B$
parameters from the lightest measured data by about 1\% or less.
Therefore, the chiral extrapolation is extremely stable and the
associated systematic error is negligible.
To confirm this observation we also tried a quadratic
extrapolation for some parameter sets, for which we find that the
results are consistent with the linear extrapolation within the
statistical error. 

In chiral perturbation theory for heavy-light mesons, the 
logarithmic dependence such as $m_q\ln m_q$ is predicted for $B_B$
\cite{Booth:1995hx,Sharpe:1996qp}.
In the quenched approximation the chiral limit is even divergent as
$\ln m_q$.
The more divergent term $\ln m_q$ has a factor $1-3g^2$ as its
coefficient, and
the $B^*B\pi$ coupling $g$ is evaluated in the range 0.2--0.7 using 
$D^*\rightarrow D\pi$ decay \cite{Stewart:1998ke},
$D^*$ decay width \cite{Anastassov:2002cw}, 
quark models \cite{Casalbuoni:1997pg}, 
and quenched lattice calculations
\cite{deDivitiis:1998kj,Aoki:2001rd,Abada:2002xe}.
It means that this divergent logarithm is relatively unimportant because
of its small coefficient $1-3g^2$ = 0.2(7).
It is, however, difficult to resolve such logarithmic dependences from
the data taken in the range of our light quark masses.
In this work, therefore, we do not further consider them, leaving the
study of the chiral logarithm including the effect of unquenching for
future publications.

Results of the linear extrapolation are summarized in
Table~\ref{tab:lattice_B_parameters_57B}--\ref{tab:lattice_B_parameters_60S},
where we list the values of $B_{B,X}^{lat}$ and $B_{S,X}^{lat}$ 
at each $\beta$ and $aM_0$.
The value of $\kappa$ corresponding to the physical $u$ or
$d$ quark mass, which we denote $\kappa_{ud}$, is very close to the
critical value $\kappa_c$. 
The value of $\kappa_s$ corresponding to the strange quark mass
depends on the input quantity.
We list the results at $\kappa_{s1}$, for which the $K$ meson mass is
used as input, and at $\kappa_{s2}$, for which $\phi$ meson mass is
used.

\subsection{$1/M_P$ dependence}
\label{subsec:1/M_P_dependence}
\begin{figure}[p]
  \centering
  \includegraphics*[width=8cm,clip=true]{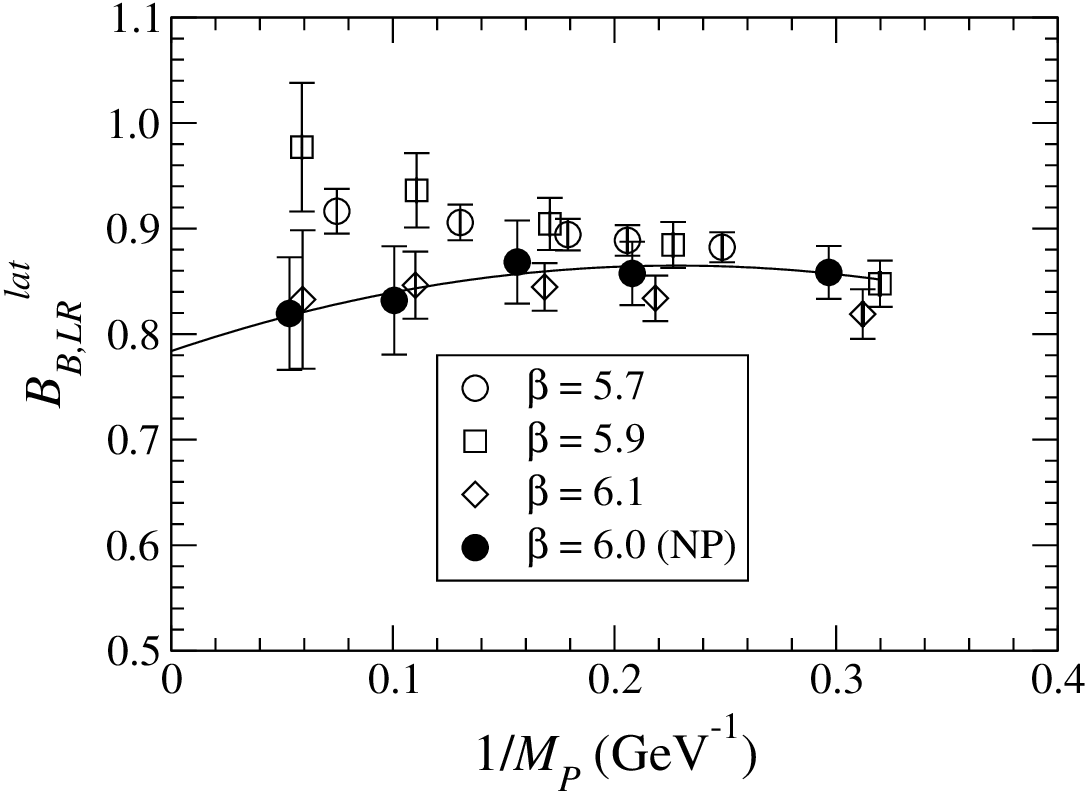}
  \includegraphics*[width=8cm,clip=true]{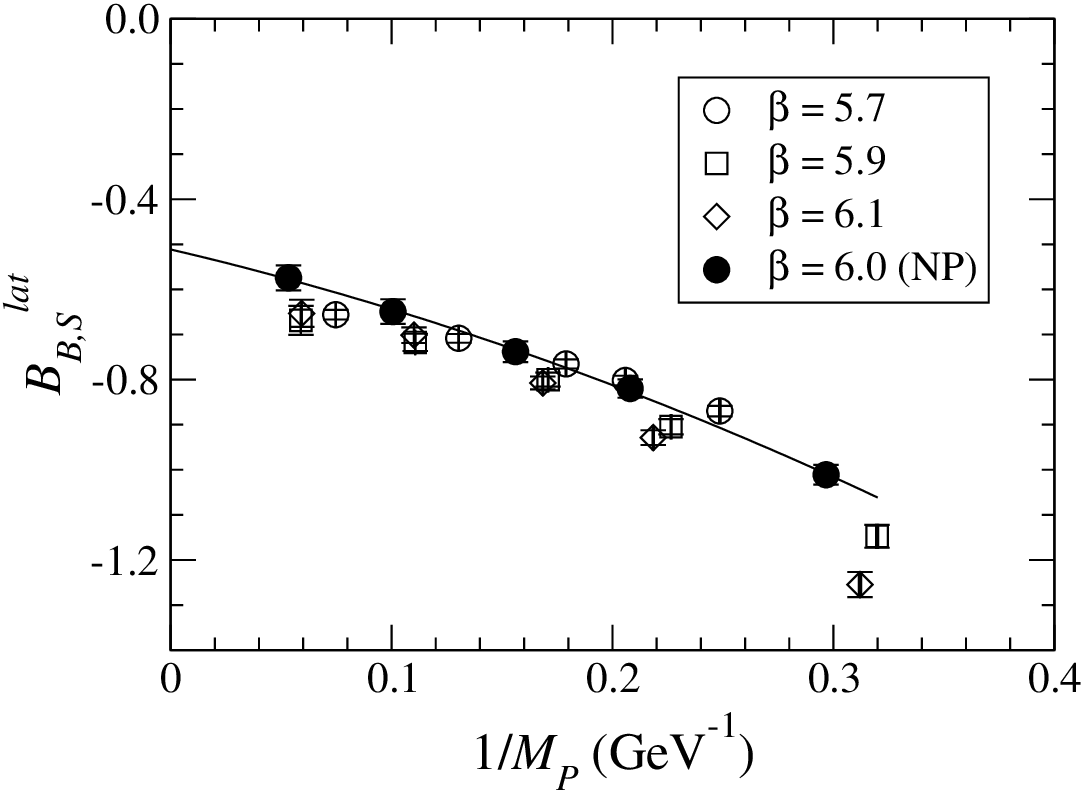}\\
  \includegraphics*[width=8cm,clip=true]{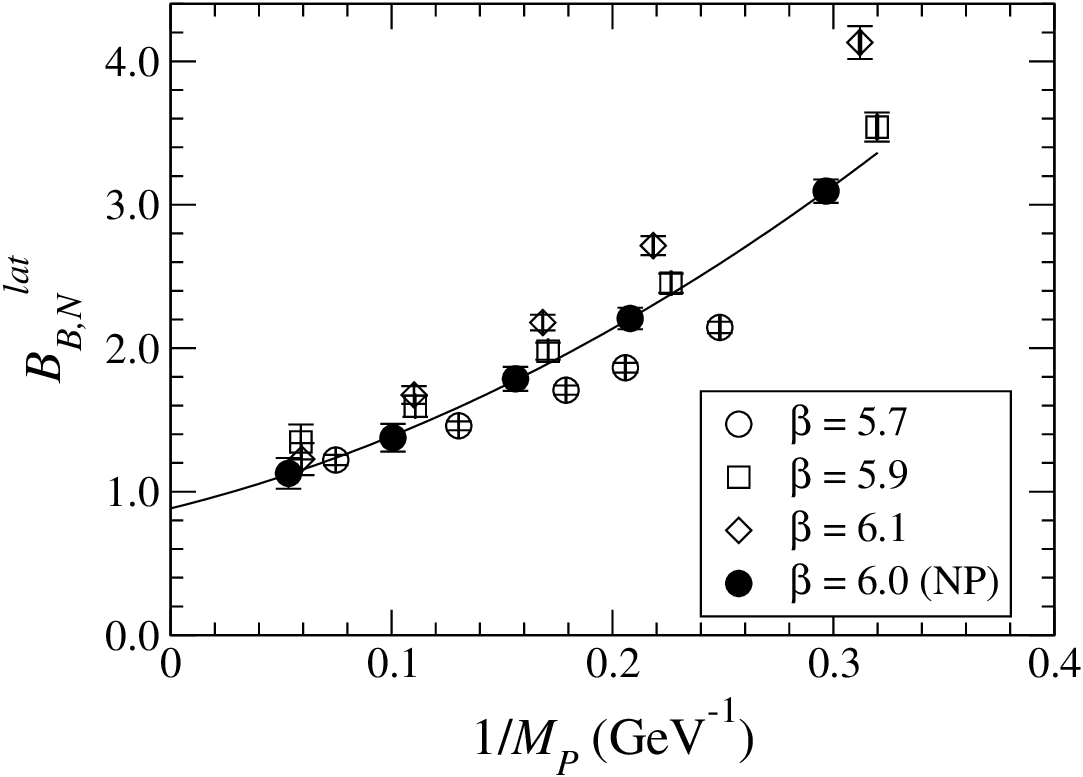}
  \includegraphics*[width=8cm,clip=true]{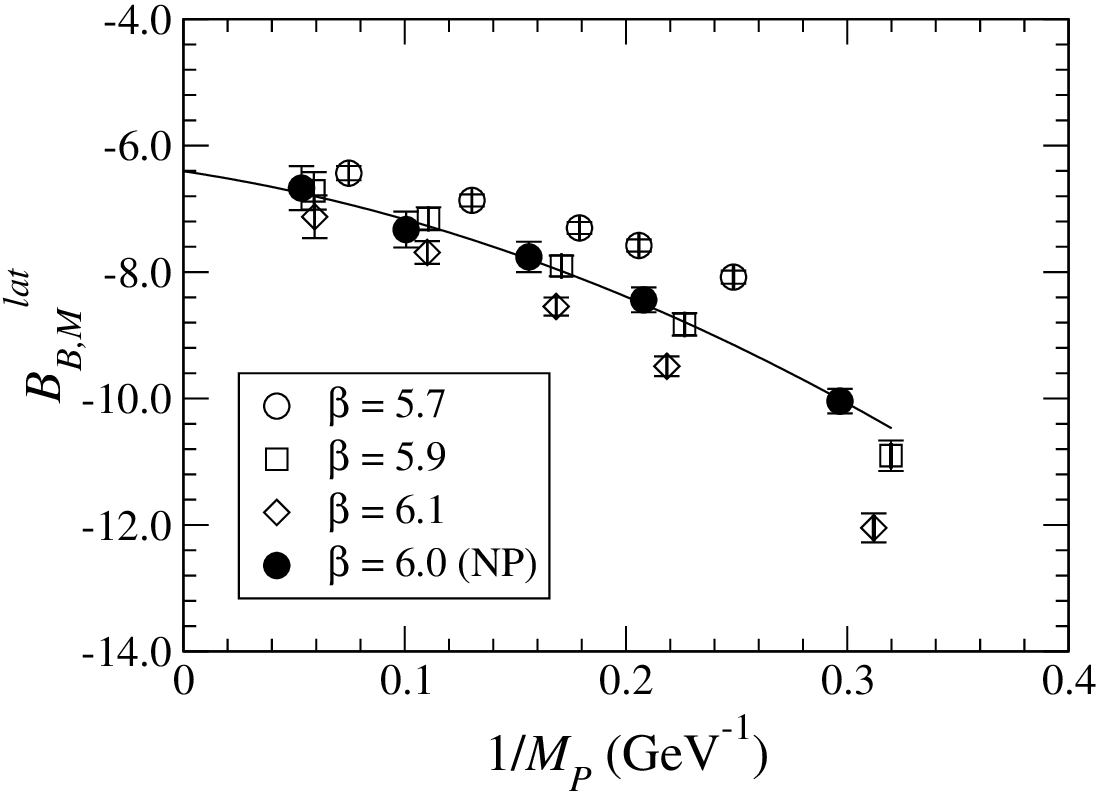}
  \caption{
    $1/M_P$ dependence of the lattice $B$ parameters $B_{B,X}$
    ($X$ = $LR$, $S$, $N$ and $M$).
    A quadratic fit is plotted for the data at $\beta$ = 6.0.
  }
  \label{fig:BB_HQMd}
\end{figure}

\begin{figure}[p]
  \centering
  \includegraphics*[width=8cm,clip=true]{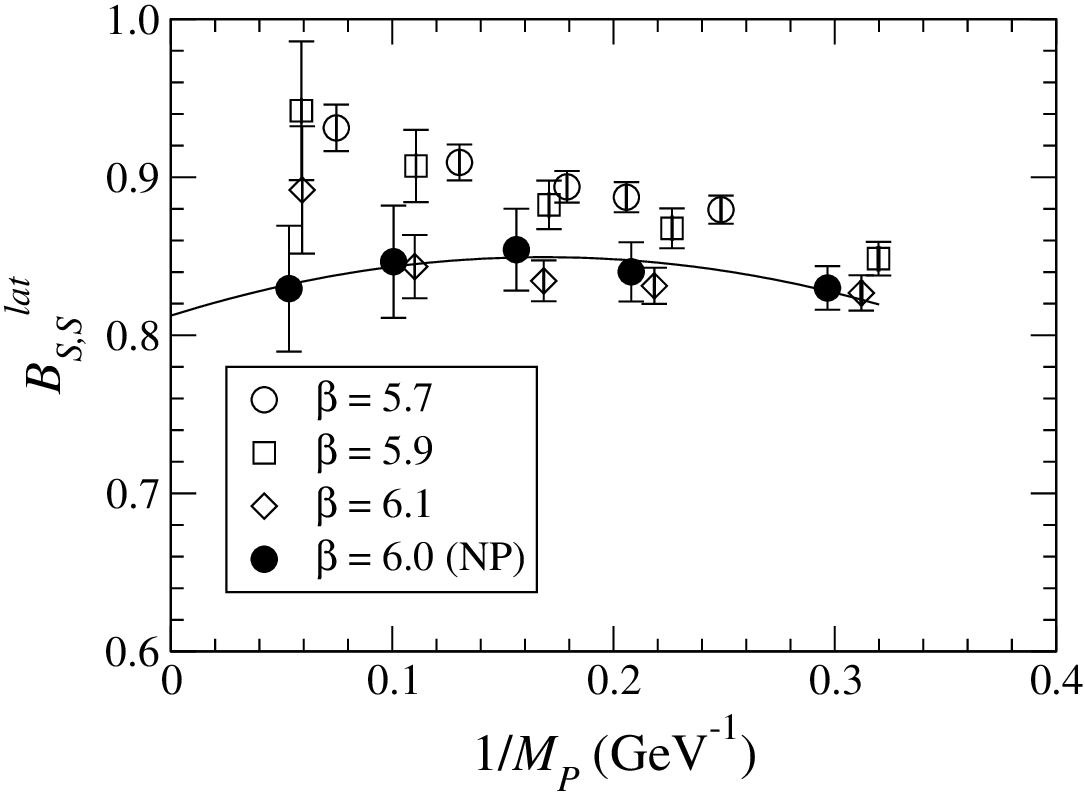}
  \includegraphics*[width=8cm,clip=true]{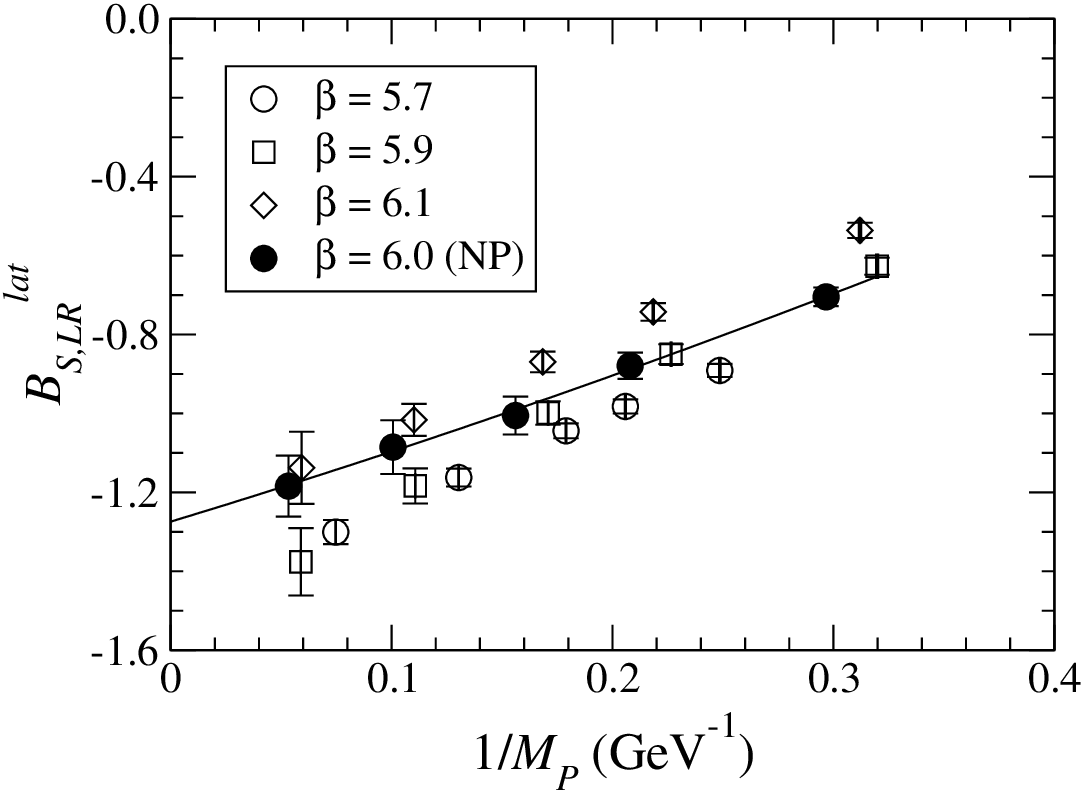}\\
  \includegraphics*[width=8cm,clip=true]{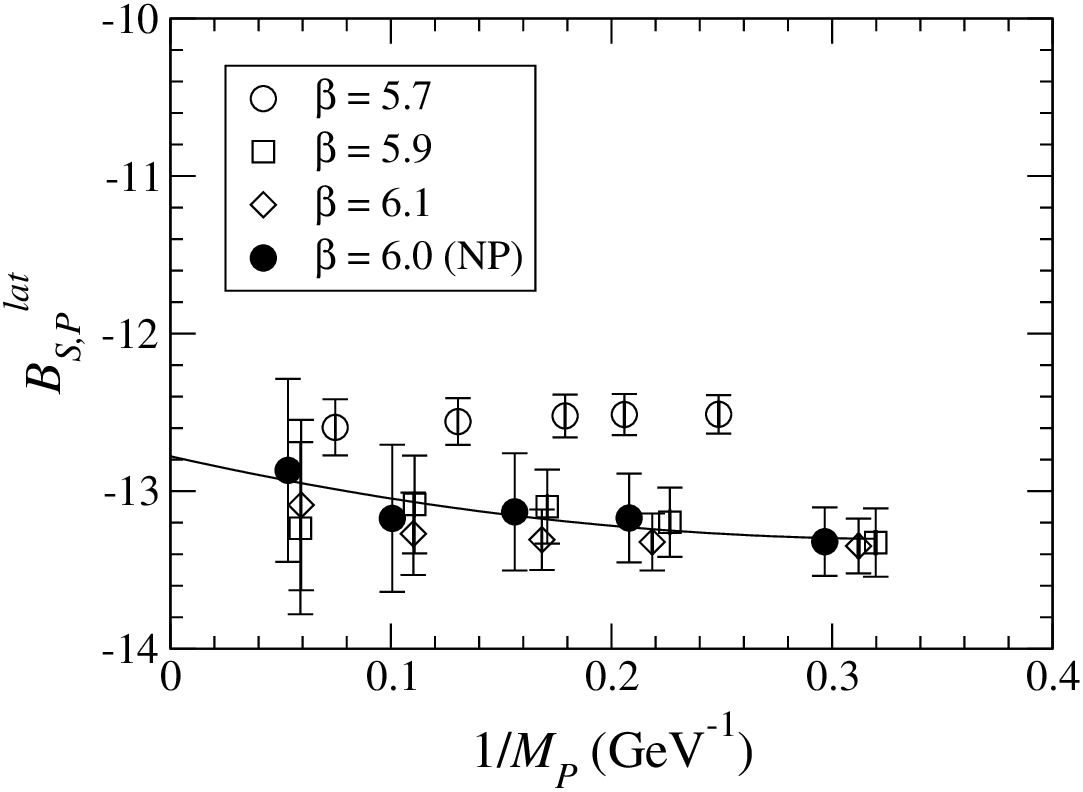}
  \includegraphics*[width=8cm,clip=true]{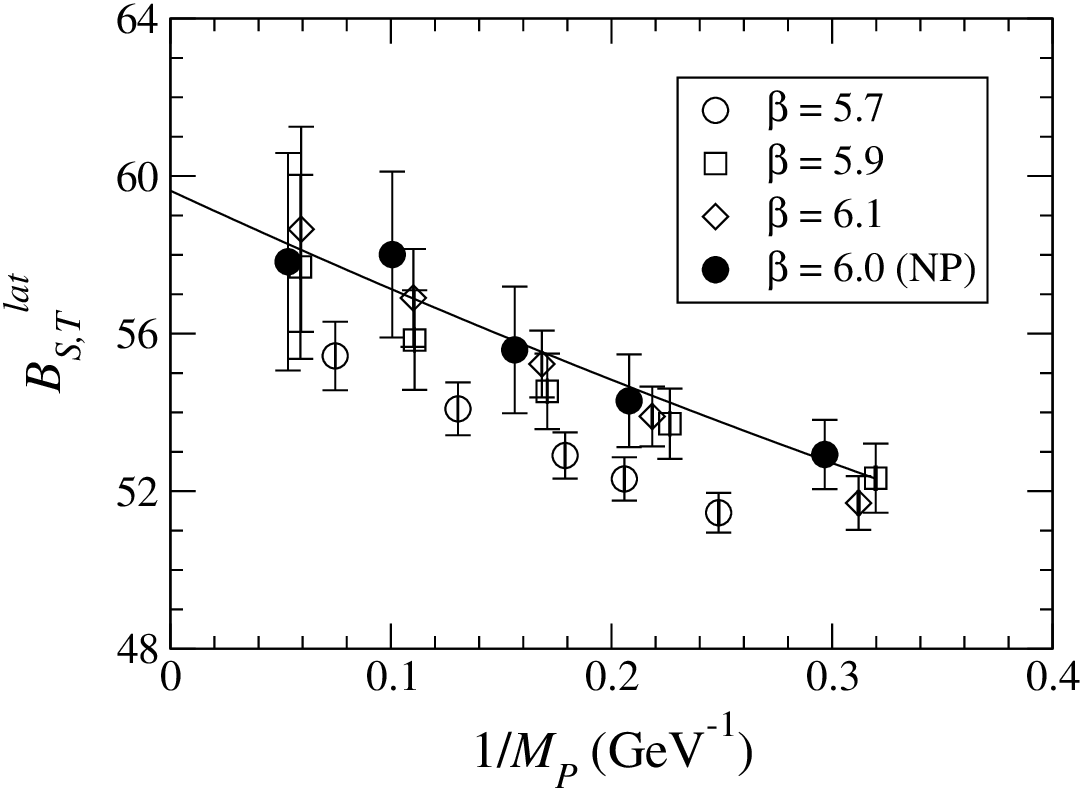}
  \caption{
    $1/M_P$ dependence of the lattice $B$ parameters $B_{S,X}$
    ($X$ = $S$, $LR$, $P$ and $T$).
    A quadratic fit is plotted for the data at $\beta$ = 6.0.
  }
  \label{fig:BS_HQMd}
\end{figure}

The $1/M_P$ dependence of the lattice $B$ parameters $B_{B,X}^{lat}$
($B_{S,X}^{lat}$) is plotted in Figure~\ref{fig:BB_HQMd}
(Figure~\ref{fig:BS_HQMd}). 
The light quark is extrapolated to the chiral limit.
Although the data at different $\beta$ values are overlaid, 
they do not necessarily agree because the operators are not
renormalized. 
Comparison with an adequate definition (the $B$ parameter in the
continuum renormalization scheme) will be discussed in detail in
Section~\ref{subsec:Results_at_different_lattice_spacings}.

We find that the mass dependence is small for $B_{B,LR}^{lat}$, while
it is significant for others.
This behavior can be mostly understood using the vacuum saturation
approximation (VSA) \cite{Hashimoto:1999ck}.
In VSA the matrix element in the numerator of the $B$
parameter is generally decomposed into $|\langle 0|A_\mu|P\rangle|^2$
and $|\langle 0|P|P\rangle|^2$.
For $B_{B,LR}^{lat}$, however, it is written by 
$|\langle 0|A_\mu|P\rangle|^2$ only and no 
$|\langle 0|P|P\rangle|^2$ term appears by definition, 
so that $B_{B,LR}^{\mathrm{(VSA)}}$ = 1 is independent of $1/M_P$. 
For others, the term $|\langle 0|P|P\rangle|^2$ gives a strong mass
dependence proportional to $(M_P/M)^2=(1+\bar{\Lambda}/M)^2$, where
$\bar{\Lambda}$ represents the binding energy produced by the light
degrees of freedom.
Comparison of the lattice data with VSA is made in Ref.~
\cite{Hashimoto:1999ck}.

\subsection{Renormalized $B$ parameters}
\label{subsec:Renormalized_B_parameters}
The $B$ parameters for the continuum operators are obtained from the
lattice $B$ parameters using (\ref{eq:BB_from_ratio_I}) for $B_B$
and (\ref{eq:BS_from_ratio_I}) for $B_S$.
We consider the truncation method I in this subsection.
The results of other truncations are discussed in the next
subsection. 

In order to see the effect of $1/M$ corrections we consider the
quantity
\begin{equation}
  \label{eq:Phi_BB}
  \Phi_{B_B}(\mu_b) \equiv
  \left(\frac{\alpha_s(M_P)}{\alpha_s(M_B)}\right)^{2/\beta_0}
  B_B(\mu_b)
\end{equation}
as a function of $1/M_P$.
The factor $(\alpha_s(M_P)/\alpha_s(M_B))^{2/\beta_0}$ is introduced
to cancel the logarithmic dependence on $M$ coming from the
continuum one-loop integral, so that the heavy quark expansion in
$1/M_P$ is explicit.
Up to two-loop corrections $\Phi_{B_B}(\mu_b)$ is
equivalent to $B_B(\mu_b)$ obtained with a replacement of 
$M$ in $\rho_{L,L}$ (\ref{eq:rho_LL}) and in $\rho_A^{(0)}$
(\ref{eq:rho0_A4}) by the physical $b$ quark mass $m_b$, which can be
confirmed by expanding the factor
$(\alpha_s(M_P)/\alpha_s(M_B))^{2/\beta_0}$ in $\alpha_s(M_P)$
explicitly.
With the replacement the static limit ($1/M_P\rightarrow 0$)
of NRQCD simply becomes the conventional static approximation.
Therefore, in the calculation of $\Phi_{B_B}(\mu_b)$ we
explicitly set the physical $b$ quark mass $m_b$ in the
matching coefficients (\ref{eq:rho_LL}) and (\ref{eq:rho0_A4}).
It should be also noted that at the physical $B$ meson mass, namely
$M_P$=$M_B$, our definition of $\Phi_{B_B}(\mu_b)$ exactly agrees with
the definition (\ref{eq:Phi_BB}).

\begin{figure}[p]
  \centering
  \includegraphics*[width=8cm,clip=true]{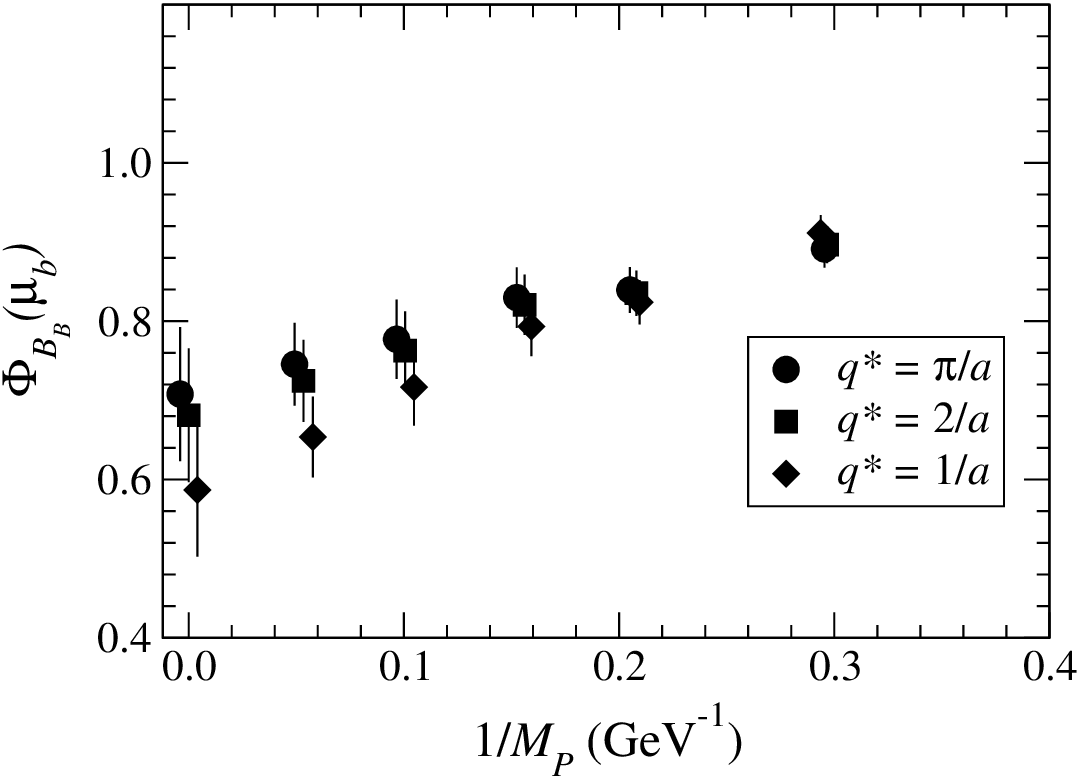}
  \\[2mm]
  \includegraphics*[width=8cm,clip=true]{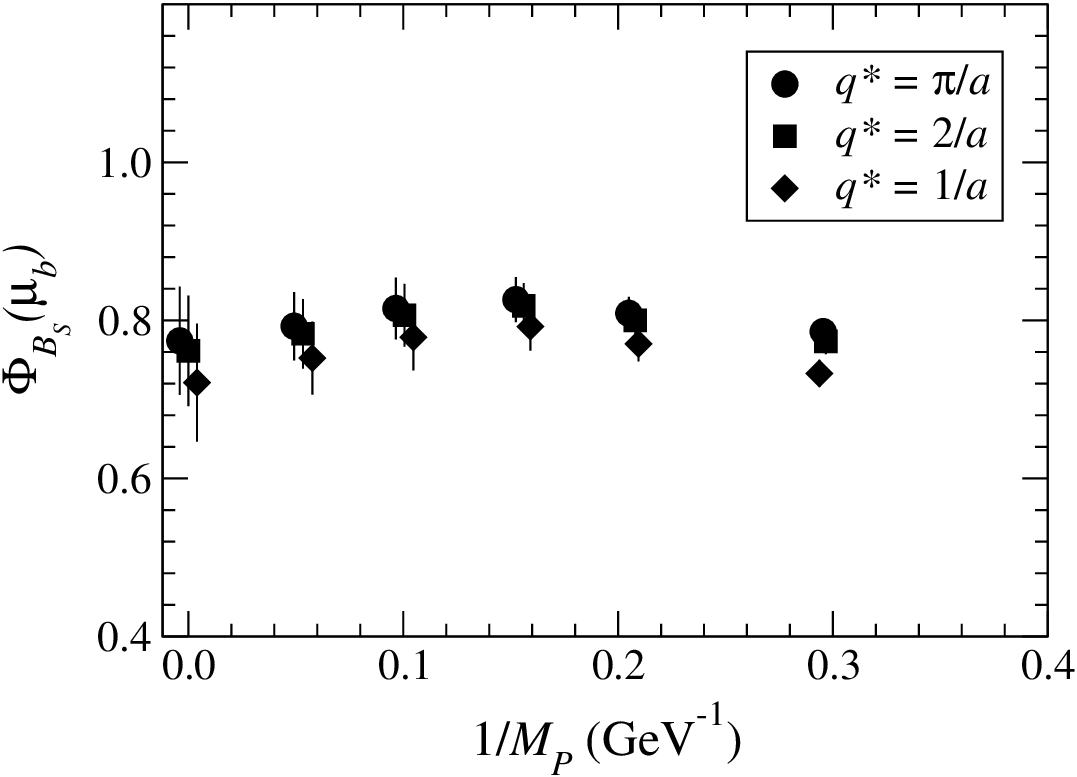}
  \caption{
    $\Phi_{B_B}(\mu_b)$ (top panel) and
    $\Phi_{B_S}(\mu_b)$ (bottom panel) for $\mu_b=m_b$.
    Data show the result at $\beta=6.0$ in the chiral limit of the
    light quark.
    The truncation method I is chosen as a demonstration. 
    Different symbols correspond to different scales of the coupling
    constant in the perturbative matching, and data points are
    slightly shifted in the x-direction for clarity.
    The data at the static limit ($1/M_P=0$) is obtained by an
    extrapolation in $1/M_P$ with a quadratic function.
  }
  \label{fig:B_HQMd_phys_I}
\end{figure}

Figure~\ref{fig:B_HQMd_phys_I} (top panel) shows 
$\Phi_{B_B}(\mu_b)$ at $\beta=6.0$.
The light quark mass is extrapolated to the chiral limit, and
the renormalization scale $\mu_b$ is set to $m_b$.
In the one-loop matching (\ref{eq:BB_from_ratio_I})
we use the renormalized coupling $\alpha_V(q^*)$ defined through the
heavy quark potential \cite{Lepage:1993xa}.
The scale $q^*$ represents the momentum region where the relevant
one-loop integral dominates.
Since it is not known, we use three typical values $\pi/a$, $2/a$ and
$1/a$ and consider their variation as an indication of systematic
uncertainty from higher order perturbative corrections.
We find that the variation among different coupling constants becomes
substantial as one goes to the static limit, while it is relatively
unimportant in the physical mass region $1/M_P\sim$ 0.2~GeV$^{-1}$.
This is because the one-loop coefficients in the matching
(\ref{eq:BB_from_ratio_I}) grows toward the static limit.
For the $1/M_P$ dependence of $\Phi_{B_B}(\mu_b)$ we observe a slight
positive slope and curvature, but the large systematic uncertainty
implies that the mass dependence is insignificant.

We obtain a similar plot for $\Phi_{B_S}(\mu_b)$ in
Figure~\ref{fig:B_HQMd_phys_I} (bottom panel), which is an analog
of $\Phi_{B_B}(\mu_b)$ but for $B_S(\mu_b)$.
The definition of $\Phi_{B_S}(\mu_b)$ with the renormalization group
improvement as in (\ref{eq:Phi_BB}) is more complicated, since the
logarithmic dependence appears in more than one coefficients,
so that we have to consider a mixing of operators. 
In this work, however, we avoid this problem by replacing the heavy
quark mass in $\rho_{S,S}$ and $\rho_{S,L}$ by the physical value as
we did for $\Phi_{B_B}(\mu_b)$.

For $B_S(\mu_b)$ the one-loop coefficients are relatively small and
their dependence on the heavy quark mass is mild.
Hence, we obtain a smaller variation due to different scale settings in
the coupling constant.

\subsection{Effect of truncation of expansions}
\label{subsec:Effect_of_truncation_of_expansions}
As we discussed in Section~\ref{subsec:Truncation_of_expansions} there
are several method to truncate the perturbative and $1/M$ expansions.
We consider the following four methods.
In the methods I and I' the perturbative matching
is truncated in the numerator and denominator separately as in 
(\ref{eq:BB_from_ratio_I}) and (\ref{eq:BS_from_ratio_I}), while in
the methods II and II' the denominator is linearized as 
(\ref{eq:BB_from_ratio_II}) and (\ref{eq:BS_from_ratio_II}).
In the primed methods the heavy quark field without the FWT rotation 
(\ref{eq:FWT_transformation}) is used for one-loop correction terms.

\begin{figure}[p]
  \centering
  \includegraphics*[width=8cm,clip=true]{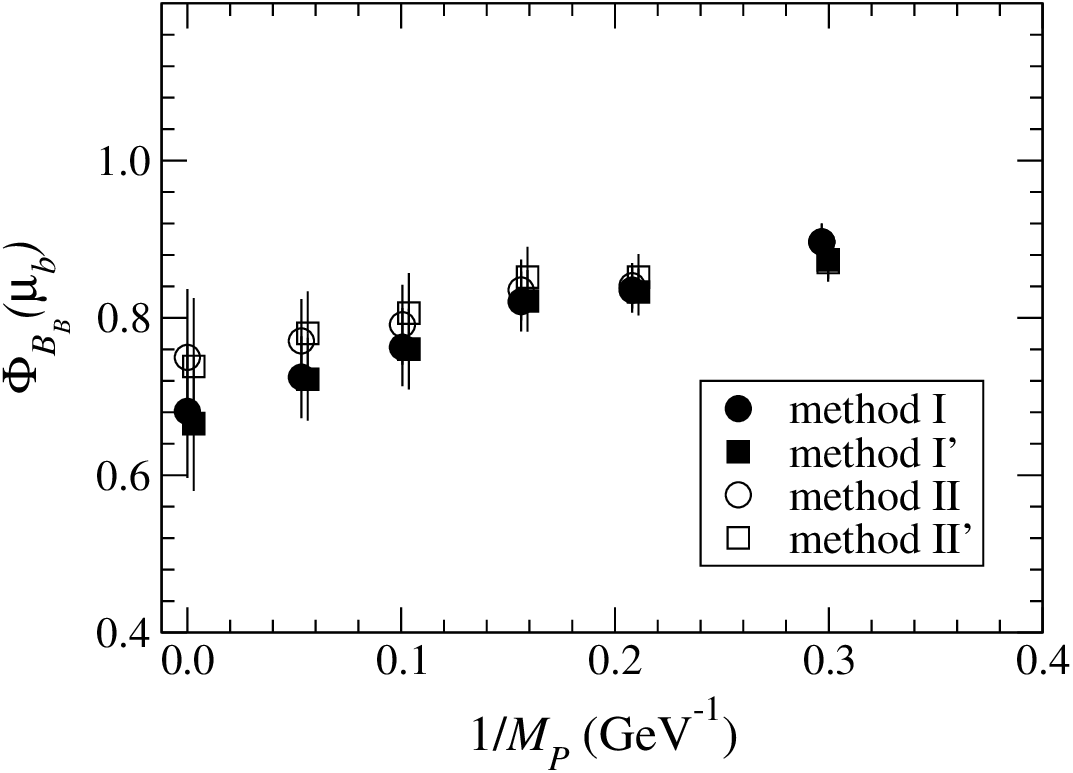}
  \\[2mm]
  \includegraphics*[width=8cm,clip=true]{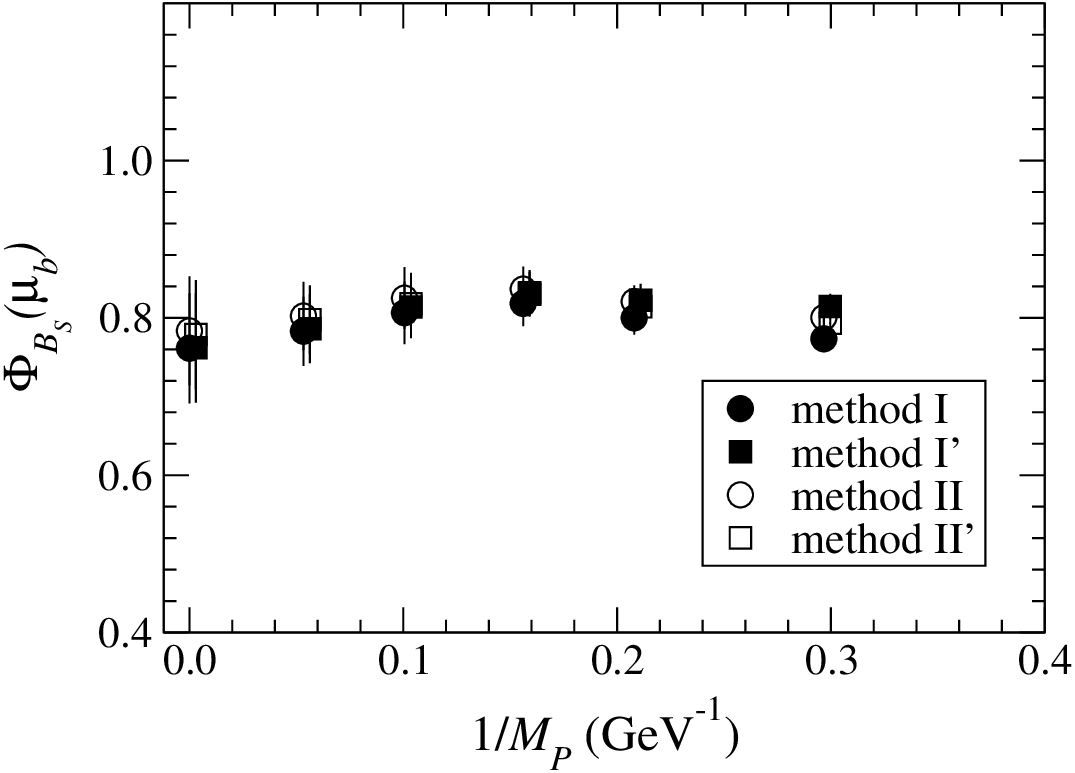}
  \caption{
    $\Phi_{B_B}(\mu_b)$ (top panel) and
    $\Phi_{B_S}(\mu_b)$ (bottom panel) for $\mu_b=m_b$.
    Data show the result at $\beta=6.0$ in the chiral limit for the
    light quark and renormalized with the coupling $\alpha_V(2/a)$.
    Different symbols correspond to different truncations of
    perturbative and heavy quark expansions.
    The data at the static limit ($1/M_P=0$) is obtained by an
    extrapolation in $1/M_P$ with a quadratic function.
  }
  \label{fig:B_HQMd_phys_all}
\end{figure}

In Figure~\ref{fig:B_HQMd_phys_all} we plot $\Phi_{B_B}(\mu_b)$ (top
panel) and $\Phi_{B_S}(\mu_b)$ (bottom panel) for four different
truncation methods.
As can be seen from the figure,
the methods I and I' (or II and II') agree in the
static limit, since their difference is only in the FWT rotation.
On the other hand, the difference between the methods I and II
(or I' and II') is smaller for lighter heavy quarks because the
one-loop correction in the denominator $\rho_A^{(0)}$ becomes small. 

We consider the variation among different truncations as an indication
of systematic uncertainties from higher orders of perturbative and
$1/M$ expansions.
The error estimation is given in the next section.

\subsection{Results at different lattice spacings}
\label{subsec:Results_at_different_lattice_spacings}

Since NRQCD is formulated by an expansion in $1/M$ and not a
renormalizable field theory, it does not allow a continuum limit;
instead it has to be considered as an effective theory valid in the 
region where $1/(aM_0)$ is small enough.
The dependence of systematic errors on the lattice spacing $a$ is not
just a simple power series in $a$, but contains its inverse powers.
Therefore, the question is how one can find a region of $a$ where 
the discretization error is small while the errors scaling as $1/a^n$
($n$ is a positive integer) is under control.
Although the order counting argument as discussed in
\cite{Hashimoto:1999ck,Hashimoto:2000eh} provides a rough
estimate of errors, it is essential to confirm it using
actual simulation data.

\begin{figure}[p]
  \centering
  \includegraphics*[width=8cm,clip=true]{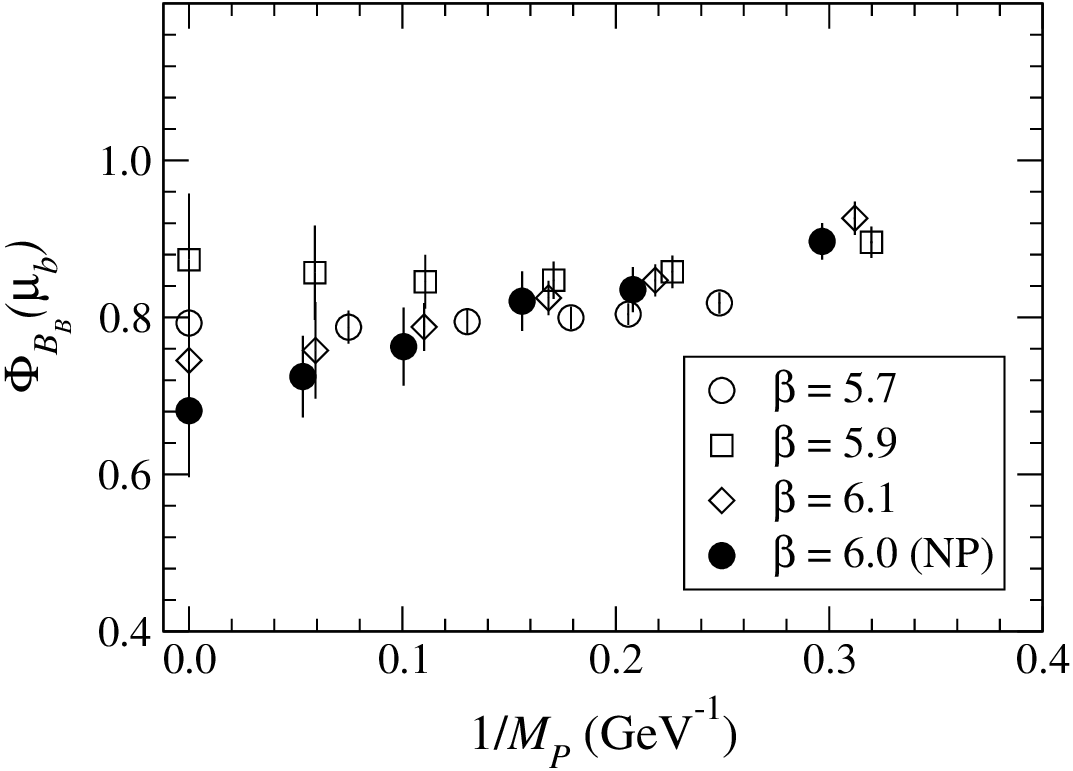}
  \\[2mm]
  \includegraphics*[width=8cm,clip=true]{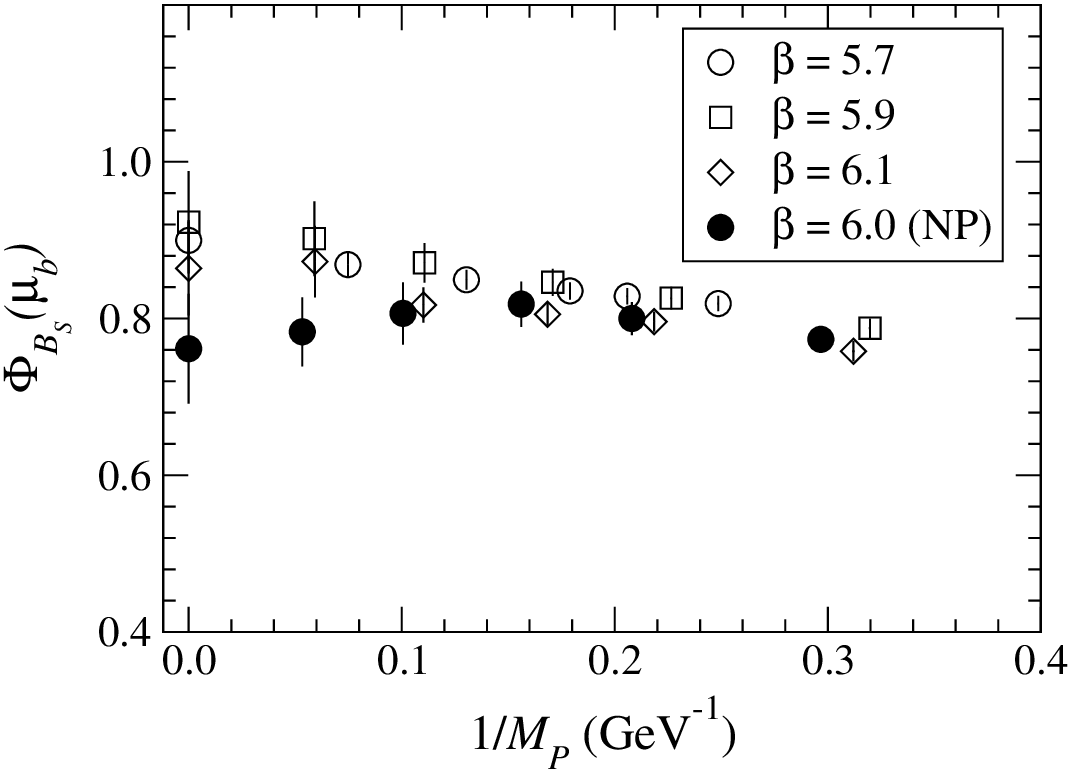}
  \caption{
    $\Phi_{B_B}(\mu_b)$ (top panel) and
    $\Phi_{B_S}(\mu_b)$ (bottom panel) for $\mu_b=m_b$.
    Results at different lattice spacings are compared.
    Data show the results in the chiral limit for the
    light quark and renormalized with the coupling $\alpha_V(2/a)$.
    The truncation method I is chosen as a demonstration. 
    The data at the static limit ($1/M_P=0$) is obtained by an
    extrapolation in $1/M_P$ with a quadratic function.
  }
  \label{fig:B_HQMd_phys_ad}
\end{figure}

In Figure~\ref{fig:B_HQMd_phys_ad} we plot $\Phi_{B_B}(\mu_b)$ (top
panel) and $\Phi_{B_S}(\mu_b)$ (bottom panel) obtained with the
truncation method I for four different lattice spacings.
The largest (smallest) inverse lattice spacing is 2.3~GeV at 
$\beta = 6.1$ (1.1~GeV at $\beta = 5.7$).
We find that around the physical $B$ meson mass 
($1/M_B\sim$ 0.2 GeV$^{-1}$) the results agree within order 10\% for 
$\Phi_{B_B}(\mu_b)$ or even better for $\Phi_{B_S}(\mu_b)$.
The agreement becomes marginal toward the static limit especially for
the nonperturbatively improved lattice ($\beta = 6.0$), but it is not
statistically significant.

\begin{figure}[p]
  \centering
  \includegraphics*[width=8cm,clip=true]{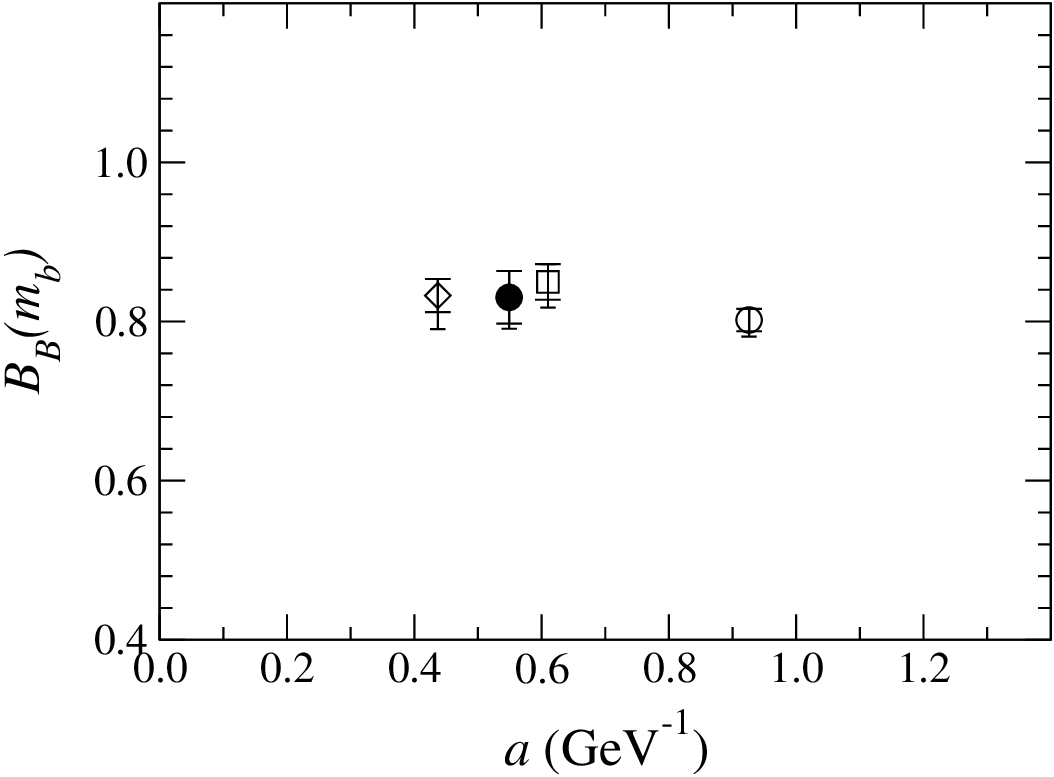}
  \\[2mm]
  \includegraphics*[width=8cm,clip=true]{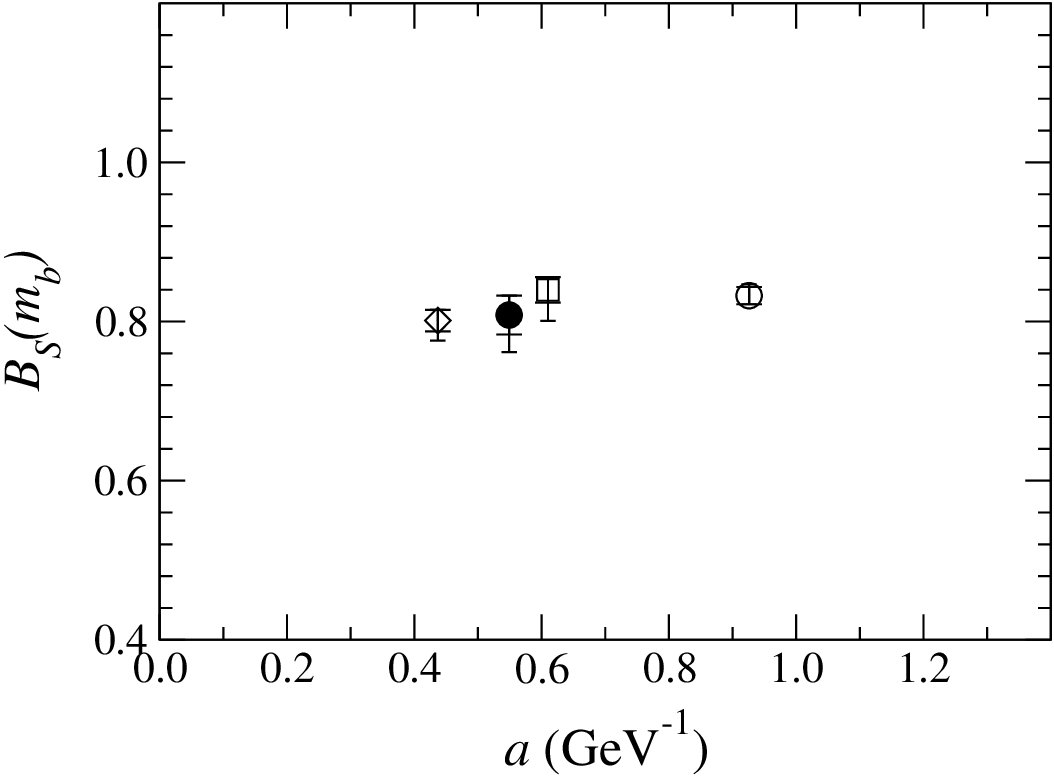}
  \caption{
    Dependence of $B_B(m_b)$ (top panel) and $B_S(m_b)$ (bottom
    panel) on the lattice spacing $a$.
    Data show the results in the chiral limit for the
    light quark and renormalized with the coupling $\alpha_V(2/a)$.
    The truncation method I is chosen as a demonstration. 
    The variation due to the different choice of fit range is
    added to the error bar at each $\beta$.
  }
  \label{fig:B_ad}
\end{figure}

Results of physical $B_B(m_b)$ (top panel) and $B_S(m_b)$ (bottom
panel) are plotted in Fig.~\ref{fig:B_ad} as a function of the
lattice spacing, where the variation due to the different choice
of fit range is added to the error bar at each $\beta$.
The $a$ dependence is a mixture of the discretization error scaling as
$a^m$ and the truncation error containing a form like $1/a^n$.
In addition, the truncation of perturbative expansion gives a
functional dependence like $1/\ln a$.
It is, therefore, difficult to determine the shape of the $a$
dependence, but the data imply that none of these errors is diverging
in the region we measured.

\section{Physics results}
\label{sec:Physics_results}

\subsection{Analysis of systematic errors}
\label{subsec:Analysis_of_systematic_errors}

As discussed in the previous section, we have performed the
calculation of the $B$ parameters with four different truncations
of $1/M$ and $\alpha_s$ expansions.
Furthermore, the calculations are made at four different lattice
spacings.
All of these calculations have different amount of various systematic
errors, and thus they allow us to estimate the uncertainty in
our final results.
In this subsection we first list possible sources of systematic errors
and estimate their size using a naive order counting.
Then, their results are compared with the actual lattice data.

One of the possible systematic errors arises from the discretization of
derivatives, which scales as a power of the lattice spacing $a$.
Because our actions and operators are $O(a)$-improved at tree level,
the leading error is of order $a^2$ and of order $\alpha_s a$.
Since we are using an effective theory for heavy quark, the truncation
of the $1/M$ expansion leads to a systematic error.
For our choice of actions and operators the leading contribution is of
order $1/M_b^2$.
Again, since the matching of the $1/M$ terms is done at the tree level
only, we also expect an error of order $\alpha_s/M_b$.
The perturbative matching of operators are truncated at the one-loop
level, so that there is an uncertainty of order $\alpha_s^2$.
In Table~\ref{tab:systematic_error} we list their typical size at each
$\beta$ value using a naive order counting.
Where the scale is needed we assume the typical spatial
momentum inside a heavy-light meson to be
$p \sim \Lambda_{\mathrm{QCD}}\sim$ 300~MeV.
For the strong coupling constant we use a typical value
$\alpha_V(2/a)$ as listed in Table~\ref{tab:simulation_parameters}.

The contribution of order $p^2/M_b^2$ is not
investigated in this paper, as we just neglect the $1/M^2$ terms in
the action and operators.
In Table~\ref{tab:systematic_error} we assign 2\% as the
corresponding uncertainty rather than a naive order counting 0.4\%,
taking the estimate from explicit lattice study in
\cite{Hashimoto:1999ck}.

Since we have removed errors of $\alpha_s/(aM)^m$ ($m\ge$0) by
perturbative matching,
the leading contribution which prevents us from the continuum
extrapolation with the NRQCD action has the form $\alpha_s^2/(aM)$.
Although its size in the naive order counting is smaller than the pure
two-loop correction $\alpha_s^2$, we include it in our error analysis
(thus in Table~\ref{tab:systematic_error}), as it gives the
leading contribution growing toward the continuum limit.

\begin{figure}[p]
  \centering
  \includegraphics*[width=12cm,clip=true]{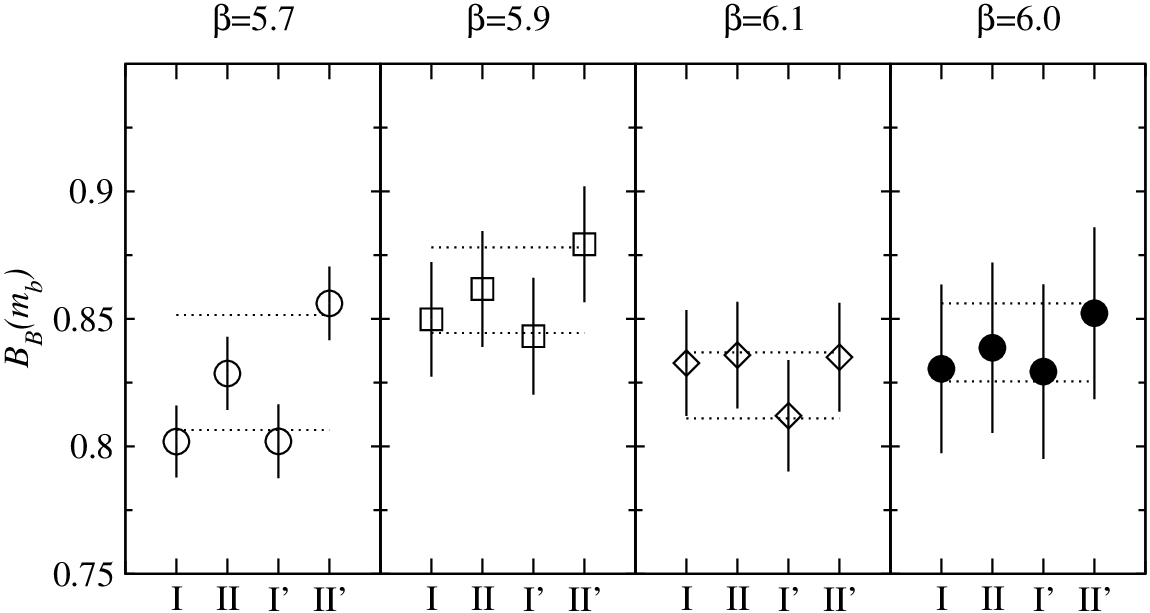}
  \caption{
    Results for $B_B$ at each $\beta$ with four different truncation
    methods (see text).
    Comparison is made with the estimate using the naive order
    counting (a band given by the dotted lines).
  }
  \label{fig:BB_syserr}
\end{figure}

\begin{figure}[p]
  \centering
  \includegraphics*[width=12cm,clip=true]{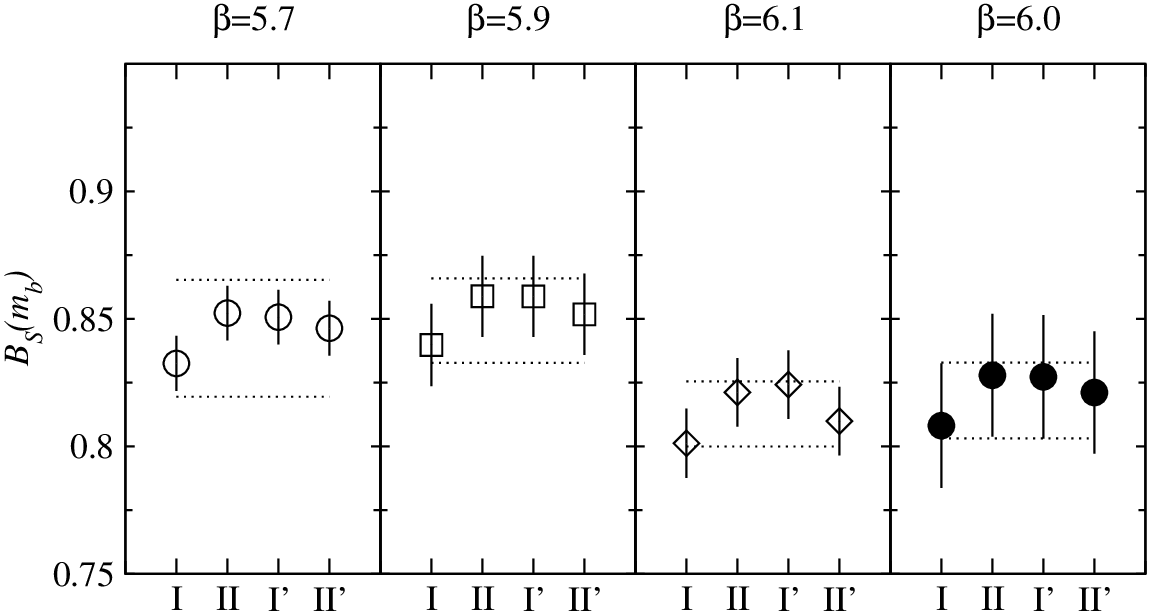}
  \caption{
    Same as Figure~\ref{fig:BB_syserr}, but for $B_S$.
  }
  \label{fig:BS_syserr}
\end{figure}

As mentioned in Section~\ref{subsec:Truncation_of_expansions},
the results from the different truncation methods (I, II, I' and II') 
are expected to differ from each other by $O(\alpha_s^2)$ or
$O(\alpha_s p/M_b)$.
We compare their results in Figure~\ref{fig:BB_syserr} for $B_B$ and
in Figure~\ref{fig:BS_syserr} for $B_S$.
The results of the four truncation methods when $\alpha_s(2/a)$ is used
in the one-loop matching are plotted.
In these Figures we also show the size of the systematic errors of
order $O(\alpha_s^2)$ and $O(\alpha_s p/M_b)$ estimated with the naive
order counting (first two lines of Table~\ref{tab:systematic_error}
added in quadrature). 
Although the statistical error of the data points makes the comparison
somewhat ambiguous, we conclude that the naive order counting
reasonably explains the scatter of the lattice data.
The same conclusion is reached when we use different values of $q^*$
in the perturbative matching.

An alternative way to estimate the $O(\alpha_s^2)$ error is to see the
variation of results with $q^*$=$\pi/a$, $2/a$ and $1/a$. 
From Figure~\ref{fig:B_HQMd_phys_I}, where the data at $\beta$ =  6.0
are plotted, we find that the variation with the choice of scale in
the coupling constant is consistent with our order counting 
($\sim$ 3\%) for the data points around the $B$ meson mass.
In the static limit, on the other hand, the one-loop coefficient is
uncomfortably large, and the variation among the results with
different $q^*$ is much larger than our naive estimate.

Other sources of systematic errors to be tested are 
$O(a^2p^2)$, $O(ap\alpha_s)$ and $O(\alpha_s^2/(aM_b))$.
Although it is difficult to disentangle various $a$ dependent
systematic errors solely from the data, our results are stable against
the change of lattice spacing suggesting that the systematic error is
well estimated by the naive order counting.

Finally we also investigated the systematic error from the
contamination of excited states by changing the fit range for the
ratio $R_{B,X}(t_1,t_2)$ and $R_{S,X}(t_1,t_2)$.
We find that the effect is at most 3\%.
In particular, at $\beta$=6.0 it is found to be 1\% for $B_B$ and 3\%
for $B_S$.
These variations are taken into account in the final results.

\subsection{$B$ parameter results}
\label{subsec:B_parameter_results}

As we discussed above, the systematic errors estimated with the naive
order counting actually describe the differences among different
calculations.
We, therefore, use the order counting argument to quote our estimate
of systematic uncertainties.

We take the central values from the data at $\beta$ = 6.0 (set $B$)
with $q^*$=$2/a$, and obtain the following results in the quenched
approximation. 
\begin{eqnarray}
  B_B(m_b)     &=& 0.84(3)(5), \\
  B_{B_s}(m_b) &=& 0.86(2)(5)(0), \\
  B_S(m_b)     &=& 0.82(2)(5), \\
  B_{S_s}(m_b) &=& 0.85(1)(5)(^{+1}_{-0}),
  \label{eq:results}
\end{eqnarray}
where the first and second errors represent statistical and systematic
ones respectively.
In the systematic error the contamination from excited states is added
in quadrature with other sources estimated by the order counting.
The third error is from the uncertainty of $m_s$ arising from the
different input physical quantities, \textit{i.e.} $m_K$ or $m_\phi$. 

The corresponding renormalization scale independent $B$ parameter is
obtained from (\ref{eq:para_BBhat})
\begin{eqnarray}
  \hat{B}_B=1.29(5)(8),
  \label{eq:Bhat_NRQCD}
\end{eqnarray}
which may be compared with the previous calculations using the
relativistic actions for heavy quark,
1.38(11)$(^{+0}_{-9})$ \cite{Becirevic:2001nv} and
1.40(5)$(^{+6}_{-1})$ \cite{Lellouch:2001tw}.
These two results are slightly higher than our result, although 
\cite{Becirevic:2001nv} is still consistent within the large error.
A possible reason for the high values in the relativistic approach
is in the extrapolation in the heavy quark mass from the charm quark
mass region to the bottom.
In fact, the combined analysis of the HQET and relativistic heavy
quark, in which the interpolation in $1/M$ can be made, gives
1.34(6)$(^{+8}_{-6})$ \cite{Becirevic:2001xt}, \textit{i.e.} closer to
our result.

\subsection{Applications}
\label{sec:Applications}

In this subsection we present a few examples of physics applications
of our results.
It should be noted, however, that our calculation is still in the
quenched approximation and there is no rigorous estimate for the
associated uncertainty.
For the following applications we assume that the quenching effect is
\textit{negligible} for the $B$ parameters as suggested by our
preliminary calculations
\cite{Hashimoto:2001zq,Yamada:2001ik,Yamada:2001xp}.

For the $B$ meson decay constant, on the other hand, the large effect
of quenching has been found
\cite{Ryan:2001ej,Kronfeld:2001ss,Yamada:2002_rev}.
Furthermore, a large uncertainty due to the presence of chiral
logarithm is suggested for $f_B$
\cite{Yamada:2002_rev,Kronfeld:2002ab}, while the effect is not too
large for $f_{B_s}$.
In the following analysis we therefore consider the quantities for
which only $f_{B_s}$ is needed, and use the recent world average of
unquenched lattice calculations of $f_{B_s}$ = 230(30)~MeV
\cite{Ryan:2001ej,Yamada:2002_rev} when needed.

Assuming the three generation unitarity $|V_{ts}|\simeq|V_{cb}|$, we 
obtain the mass difference in $B_s^0-\bar{B}_s^0$ mixing using 
(\ref{eq:dm_formula}) as
\begin{equation}
  \Delta M_s = 19.4(5.5)~\mbox{ps}^{-1},
\end{equation}
where the statistical and systematic errors in theoretical and
experimental quantities are added in quadrature, but the final error
is dominated by the uncertainty of $f_{B_s}$.
The value is consistent with the current lower bound
$\Delta M_s > 13.1$~ps$^{-1}$ at a 95\% CL \cite{PDG2002}.
Tevatron Run II is expected to measure the mass difference very
precisely in a few years.

The width difference in the $B_s$ meson system could also be measured
at Tevatron Run II, if it is large enough.
Using (\ref{eq:width_difference_formula_3}) and (\ref{eq:results}),
we obtain
\begin{equation}
  \left(\frac{\Delta\Gamma}{\Gamma}\right)_{B_s}
  = 
  0.106 \pm 0.020 \pm 0.028 \pm 0.037 \pm 0.024,
\end{equation}
where the first through third errors are from 
$m_b$ = 4.8(3)~GeV, $f_{B_s}$ = 230(30)~MeV, 
and
$\mathcal{R}$ (or $\bar{m}_b(m_b)$ = 4.25(25) GeV and
$\bar{m}_s(m_b)$ = 0.10(3) GeV), respectively.
The last error comes from the uncertainty in the estimation of the
$1/m$ correction, for which we assign 30\%.
The error from the $B$ parameters is much smaller than the others
listed above.

The uncertainty in the calculation of $(\Delta\Gamma/\Gamma)_{B_s}$ is
still very large ($\sim$ 50\% if added in quadrature).
In order to improve it one has to calculate the $1/m$ corrections
reliably, as it largely cancels the leading contribution from $B_S$
as seen in (\ref{eq:width_difference_formula_3}).
Currently, only an upper bound is obtained for this quantity 
experimentally.
Our prediction is consistent with the bound
$(\frac{\Delta\Gamma}{\Gamma})_{B_s}<0.31$ at a 95 \% CL
\cite{PDG2002}.

\section{SU(3) breaking ratio $\xi$}
\label{sec:SU(3)_breaking_ratio_xi}

Since we expect that the bulk of systematic uncertainty in the
calculation of $f_B$ and $B_B$ cancels in their SU(3) breaking ratio 
$f_{B_s}/f_B$ and $B_{B_s}/B_B$, they could be useful to reduce
the errors in the determination of $|V_{td}|$ through the relation 
(\ref{eq:dm_ratio}).

In the lattice calculation, the deviation of their ratio from unity is
the quantity to be calculated and the errors scale as $B_{B_s}/B_B-1$
rather than $B_{B_s}/B_B$ itself.
In the present case, the naive estimate of SU(3) breaking is 
$O(m_K^2/\Lambda_{\chi}^2)\sim$ 25\%, where $\Lambda_{\chi}$ is
a scale of the chiral symmetry breaking $\sim$ 1~GeV, and the order
counting of uncertainties for the ratio are starting from this order.

\begin{figure}[p]
  \centering
  \includegraphics*[width=12cm,clip=true]{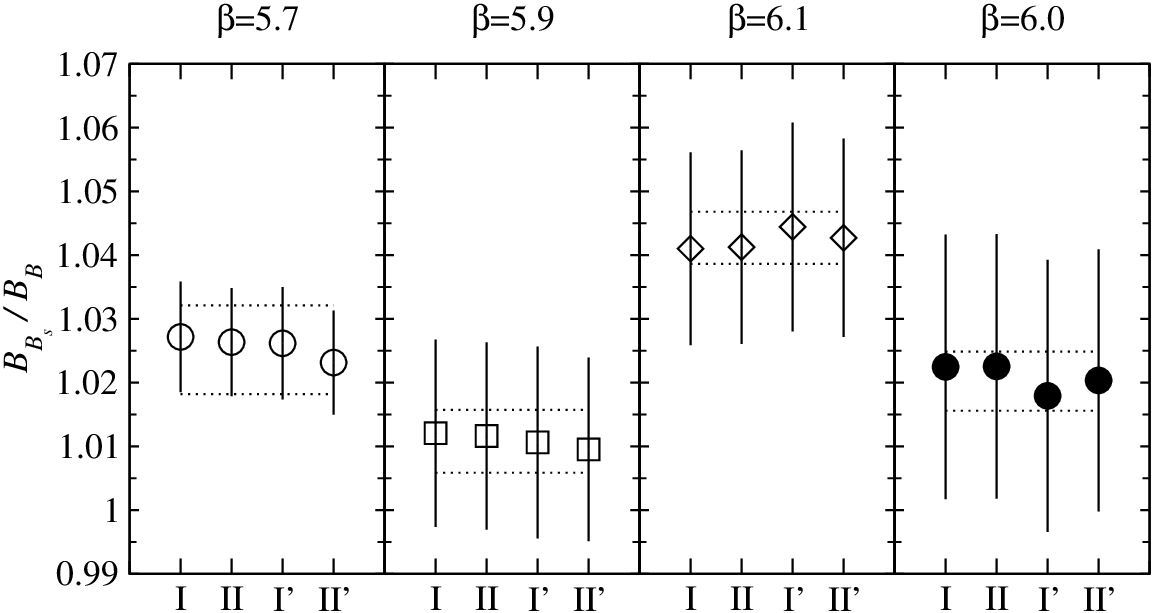}
  \caption{
    Same as Figure~\ref{fig:BB_syserr}, but for $B_{B_s}/B_B$.
  }
  \label{fig:BBratio_syserr}
\end{figure}

As done in Section~\ref{subsec:Analysis_of_systematic_errors},
we compare our order counting with data in 
Figure~\ref{fig:BBratio_syserr}.
As we expected, the variation with different truncations is much
smaller than the expected systematic errors (dotted line).
The dependence on $\beta$ is sizable, but not significant compared to
the relatively large statistical error.

Our result is
\begin{equation}
  \label{eq:SU(3)_breaking_ratios_results}
  \frac{B_{B_s}}{B_B} = 1.020(21)(^{+15}_{-16})(^{+5}_{-0}),
\end{equation}
where the central value is taken from $\beta$ = 6.0 (set \textit{B}),
and the errors are statistical, systematic and the uncertainty of
$m_s$ in the order given.

For the calculation of $\xi$ (\ref{eq:xi}) we have to combine
$B_{B_s}/B_B$ with $f_{B_s}/f_B$, which is called the indirect
method.
On the other hand, it is also possible to directly obtain $\xi$ from a 
ratio of
$\langle O_{L_s}\rangle$ and $\langle O_{L_d}\rangle$
(direct method).
It has been discussed that one may obtain rather large value of $\xi$
if one uses the direct method \cite{Bernard:1998dg,Lellouch:2001tw}.
Therefore, in the following we check if we could obtain
consistent results from both methods using our data.

\begin{figure}[p]
  \centering
  \includegraphics*[width=12cm,clip=true]{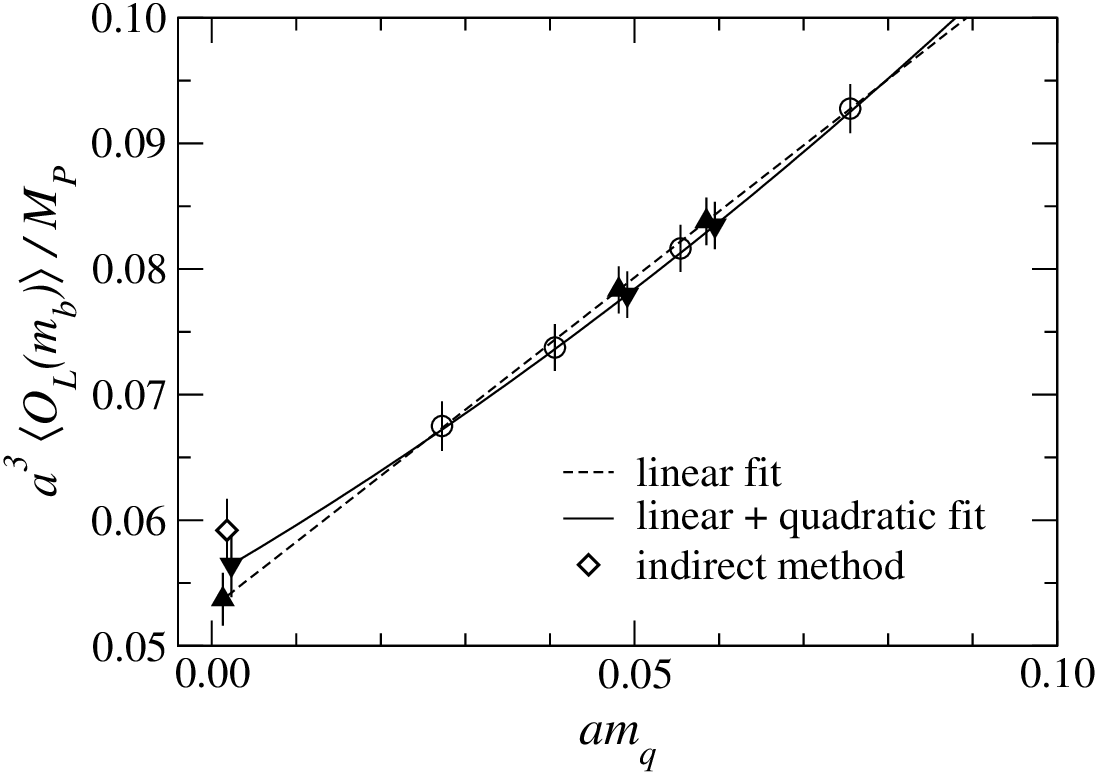}
  \caption{
    Comparison of direct and indirect methods. 
    The dashed line is a linear fit and the solid curve is obtained
    with a fit with linear and quadratic terms. 
    An open diamond at the chiral limit is obtained through
    the indirect method.
  }
  \label{fig:xi}
\end{figure}

Figure \ref{fig:xi} shows the chiral extrapolation of 
$\langle O_{L}\rangle$, as required in the direct method.
The data are obtained at $\beta$ = 6.0 (set \textit{B}) for a heavy
quark mass closest to the $b$ quark mass.
The dashed line is obtained by a linear fit to the data (open
circles), while the solid curve represents a fit with a linear plus
quadratic term in $am_q$.
Although the data look consistent with the linear fit, the chiral
limit with the quadratic fit is higher by about one standard
deviation.

An open diamond at the chiral limit, on the other hand, is obtained
through the indirect method, \textit{i.e.} the decay constant and $B$
parameter are separately extrapolated to the chiral limit with a
linear fit.
Although we have not presented a calculation of the decay constant in
this paper, they are done on the same set of gauge configurations at
$\beta$ = 6.0 and the lattice axial current is renormalized as
described in Section~\ref{subsec:Bilinear_operators}.
The result is completely consistent with the quadratic fit in the
direct method.
It implies that in the direct method the chiral extrapolation is more
difficult and needs enough statistics to control, since $\langle
O_{L}\rangle$ is effectively the decay constant squared so that the
finite $am_q$ correction is amplified\footnote{
  The similar discussion may be found in \cite{Lellouch:2001tw}.
}.

\section{Conclusions}
\label{sec:Conclusions}

In this work, we calculate the $B$ meson $B$ parameters on the lattice
in the quenched approximation.
The calculation is an extension of our previous works
\cite{Hashimoto:1999ck,Hashimoto:2000yh}, in which $B_B$ and $B_S$
were calculated for the first time with the lattice NRQCD action.

In the present work we include a detailed study of systematic 
uncertainties.
Using the lattice simulations at four different $\beta$ values with
the $O(a)$-improved actions, we find that the $B$ parameters are
essentially insensitive to the discretization error.
We also investigate the systematic errors associated with the
truncation of heavy quark and perturbative expansions, which are
necessary in the effective theory approach such as NRQCD.
By comparing four different truncations of these expansions, we
are able to confirm that the naive order counting argument of the
systematic errors could actually give a reasonable estimate.

In our final results for the $B$ parameters the systematic error is
$\sim$ 6\%, which is already smaller than that in the equivalent
calculations of $f_B$ (10--20\%), owing to the fact that it is
defined as a ratio to the vacuum saturation approximation.
Further reduction of systematic errors, if it is necessary, requires
higher order calculation of perturbation theory and the $O(a^2)$
improvement. 
Approaching to the continuum limit will not help to reduce the errors
in the NRQCD approach.

For a precise extraction of the important CKM element $|V_{td}|$
through the SU(3) breaking ratio of the $B-\bar{B}$ mixing one needs
$\xi^2$. 
It is preferable to take the chiral limit separately for $f_B$
and $B_B$, as they have milder light quark mass dependence.
The SU(3) breaking ratio of $B_B$ is obtained with accuracy of order a
few percent, since the $B$ parameter is extremely insensitive to the
light quark mass and the large cancellation of systematic errors is
expected.

The largest remaining uncertainty in our calculation is in the
quenching approximation, though it is not explicitly discussed in the
paper. 
We are currently performing an unquenched simulation with the same
lattice action at similar lattice spacing, which will allow us to
directly study the quenching effect.

\begin{acknowledgments}
This work is supported by the Supercomputer Project No.~79 (FY2002)
of High Energy Accelerator Research Organization (KEK), and also
in part by the Grant-in-Aid of the Ministry of Education 
(Nos.
11640294,
12640253,
12740133,
13135204,
13640259,
13640260,
14046202,
14740173).
N.Y. is supported by the JSPS Research Fellowship.
\end{acknowledgments}

\clearpage

\clearpage
\begin{table}
  \begin{tabular}{|c|ccc|c|}
    \hline
    set & \multicolumn{3}{c|}{\textit{A}} & \textit{B} \\
    \hline\hline
    $\beta$  & 6.1     & 5.9     & 5.7     & 6.0\\
    \hline
    size & $24^3\times 64$ & $16^3\times 48$ & 
           $12^3\times 32$ & $20^3\times 48$ 
    \\ \hline
    \#conf   & 518     & 419     & 420     & 655\\
    \hline
    $c_{\mathrm{SW}}$ 
             & 1.525   & 1.580   & 1.674   & 1.769\\
    \hline
    $1/a$ (GeV)    & 2.29    & 1.64    & 1.08 & 1.82 \\
    \hline\hline
    $\kappa$ & 0.13586 & 0.13630 & 0.13690 & 0.13260\\
             & 0.13642 & 0.13711 & 0.13760 & 0.13331\\
             & 0.13684 & 0.13769 & 0.13840 & 0.13384\\
             & 0.13716 & 0.13816 & 0.13920 & 0.13432\\ 
    \hline
    $\kappa_{s1} $ & 0.13635 & 0.13702 & 0.13800 & 0.13355\\
    $\kappa_{s2} $ & 0.13609 & 0.13657 & 0.13707 & 0.13318\\ 
    \hline
    $\kappa_c$ 
                   & 0.13767 & 0.13901 & 0.14157 & 0.13531\\ 
    \hline\hline
    $(aM_0,n)$
             & (7.0,2) & (10.0,2)& (12.0,2) & (10.0,2)\\
             & (3.5,2) &  (5.0,2)&  (6.5,2) &  (5.0,2)\\
             & (2.1,2) &  (3.0,2)&  (4.5,2) &  (3.0,2)\\
             & (1.5,3) &  (2.1,3)&  (3.8,2) &  (2.1,3)\\
             & (0.9,4) &  (1.3,3)&  (3.0,2) &  (1.3,3)\\ 
    \hline\hline
    $[t_{1min},t_{1max}]$
             & [8,26]  & [6,17]  & [4,13]  & [ 7,18] for $aM_0$ = 10.0\\
             &         &         &         & [ 9,18] for $aM_0$ =  5.0\\
             &         &         &         & [10,18] for $aM_0$ =  3.0, 2.1, 1.3\\
    \hline\hline
    $u_{0}$  & 0.8816  & 0.8734  & 0.86087 & 0.87603\\ 
    \hline
    $\alpha_{V}(\pi/a)$ & 0.149 & 0.164 & 0.188 & 0.159\\
    $\alpha_{V}(  2/a)$ & 0.172 & 0.193 & 0.229 & 0.186\\
    $\alpha_{V}(  1/a)$ & 0.229 & 0.270 & 0.355 & 0.256\\ 
    \hline
  \end{tabular}
  \caption{
    Simulation parameters.
    For the simulation set \textit{A}, the $O(a)$-improvement
    coefficient $c_{\mathrm{SW}}$ is determined at the
    tadpole-improved one-loop level.
    For the set \textit{B}, the nonperturbatively tuned value 
    \cite{Luscher:1997ug} is used.
    }
  \label{tab:simulation_parameters}
\end{table}

\begin{table}
  \begin{tabular}{|c|ccccc|}
\hline
   $aM_0$ & 12.0 & 6.5 & 4.5 & 3.8 & 3.0 \\
\hline\hline
   $B^{lat}_{B,LR}$ &&&&&\\
   $\kappa_{ud}$
   & 0.916(21) & 0.906(17) & 0.894(15) & 0.889(14) & 0.882(14) \\
   $\kappa_{s1}$
   & 0.931(13) & 0.923(11) & 0.915(09) & 0.911(09) & 0.905(09) \\
   $\kappa_{s2}$
   & 0.934(12) & 0.927(09) & 0.920(08) & 0.916(08) & 0.911(08) \\   
   \hline
   $B^{lat}_{B,S}$ &&&&&\\
   $\kappa_{ud}$
   & $-$0.656(11) & $-$0.708(10) & $-$0.765(10) & $-$0.802(11) & $-$0.870(12) \\
   $\kappa_{s1}$
   & $-$0.659(07) & $-$0.713(06) & $-$0.770(07) & $-$0.806(07) & $-$0.872(08) \\
   $\kappa_{s2}$
   & $-$0.659(06) & $-$0.714(06) & $-$0.772(06) & $-$0.808(06) & $-$0.872(08) \\
   \hline
   $B^{lat}_{B,N}$ &&&&&\\
   $\kappa_{ud}$ 
   & 1.220(36) & 1.459(31) & 1.707(32) & 1.864(34) & 2.144(40) \\
   $\kappa_{s1}$ 
   & 1.212(23) & 1.444(20) & 1.683(21) & 1.831(23) & 2.095(27) \\
   $\kappa_{s2}$ 
   & 1.210(20) & 1.440(17) & 1.676(19) & 1.822(20) & 2.081(24) \\
   \hline
   $B^{lat}_{B,M}$ &&&&&\\
   $\kappa_{ud}$ 
   & $-$6.44(11) & $-$6.87(10) & $-$7.30(09) & $-$7.58(10) & $-$8.08(10) \\
   $\kappa_{s1}$ 
   & $-$6.44(07) & $-$6.87(06) & $-$7.31(06) & $-$7.58(07) & $-$8.06(07) \\
   $\kappa_{s2}$ 
   & $-$6.44(07) & $-$6.87(06) & $-$7.31(06) & $-$7.58(06) & $-$8.06(07) \\
\hline
  \end{tabular}
  \caption{Numerical values for lattice $B$ parameters
           $B^{lat}_{B,X}$ at $\beta$=5.7.}
  \label{tab:lattice_B_parameters_57B}
\end{table}
\begin{table}
  \begin{tabular}{|c|ccccc|}
\hline
   $aM_0$ & 12.0 & 6.5 & 4.5 & 3.8 & 3.0 \\
\hline\hline
   $B^{lat}_{S,S}$ &&&&&\\
   $\kappa_{ud}$ 
   & 0.931(15) & 0.909(11) & 0.894(10) & 0.887(10) & 0.879(09) \\
   $\kappa_{s1}$ 
   & 0.941(10) & 0.927(07) & 0.916(07) & 0.910(06) & 0.904(06) \\
   $\kappa_{s2}$ 
   & 0.944(09) & 0.931(07) & 0.922(06) & 0.917(06) & 0.910(05) \\
   \hline
   $B^{lat}_{S,LR}$ &&&&&\\
   $\kappa_{ud}$ 
   & $-$1.301(31) & $-$1.163(23) & $-$1.044(19) & $-$0.982(18) & $-$0.891(17) \\
   $\kappa_{s1}$ 
   & $-$1.330(19) & $-$1.200(14) & $-$1.088(12) & $-$1.028(12) & $-$0.938(11) \\
   $\kappa_{s2}$ 
   & $-$1.338(17) & $-$1.209(13) & $-$1.099(11) & $-$1.040(10) & $-$0.951(10) \\
   \hline
   $B^{lat}_{S,P}$ &&&&&\\
   $\kappa_{ud}$ 
   & $-$12.59(18) & $-$12.56(15) & $-$12.52(14) & $-$12.51(13) & $-$12.51(12) \\
   $\kappa_{s1}$
   & $-$12.66(12) & $-$12.68(10) & $-$12.69(09) & $-$12.69(09) & $-$12.70(08) \\
   $\kappa_{s2}$ 
   & $-$12.68(10) & $-$12.71(09) & $-$12.73(08) & $-$12.74(08) & $-$12.76(08) \\
   \hline
   $B^{lat}_{S,T}$ &&&&&\\
   $\kappa_{ud}$ 
   & 55.43(87) & 54.09(67) & 52.91(58) & 52.31(55) & 51.45(50) \\
   $\kappa_{s1}$ 
   & 55.79(57) & 54.72(45) & 53.77(39) & 53.27(37) & 52.50(35) \\
   $\kappa_{s2}$ 
   & 55.89(51) & 54.90(40) & 54.01(36) & 53.53(34) & 52.78(32) \\
\hline
  \end{tabular}
  \caption{Numerical values for lattice $B$ parameters
           $B^{lat}_{S,X}$ at $\beta$=5.7.}
  \label{tab:lattice_B_parameters_57S}
\end{table}
\begin{table}
  \begin{tabular}{|c|ccccc|}
\hline
   $aM_0$ & 10.0 & 5.0 & 3.0 & 2.1 & 1.3 \\
\hline\hline
   $B^{lat}_{B,LR}$ &&&&&\\
   $\kappa_{ud}$
   &  0.977(61) &  0.936(35) &  0.904(25) &  0.884(22) &  0.848(22) \\
   $\kappa_{s1}$
   &  0.956(34) &  0.931(19) &  0.911(13) &  0.897(12) &  0.871(12) \\
   $\kappa_{s2}$
   &  0.952(30) &  0.930(16) &  0.913(11) &  0.899(10) &  0.877(11) \\
   \hline
   $B^{lat}_{B,S}$ &&&&&\\
   $\kappa_{ud}$
   &$-$0.669(32) &$-$0.718(19) &$-$0.800(16) &$-$0.904(17) &$-$1.148(25) \\
   $\kappa_{s1}$
   &$-$0.650(17) &$-$0.715(10) &$-$0.805(09) &$-$0.911(10) &$-$1.150(17) \\
   $\kappa_{s2}$
   &$-$0.646(15) &$-$0.714(09) &$-$0.806(08) &$-$0.912(09) &$-$1.150(15) \\
   \hline
   $B^{lat}_{B,N}$ &&&&&\\
   $\kappa_{ud}$ 
   &  1.346(122) &  1.595(74) &  1.979(60) &  2.453(67) &  3.541(102) \\
   $\kappa_{s1}$ 
   &  1.288(070) &  1.560(41) &  1.935(35) &  2.381(41) &  3.395(066) \\
   $\kappa_{s2}$ 
   &  1.275(060) &  1.552(35) &  1.924(31) &  2.364(37) &  3.361(059) \\
   \hline
   $B^{lat}_{B,M}$ &&&&&\\
   $\kappa_{ud}$ 
   &$-$6.715(297) &$-$7.156(182) &$-$7.903(162) &$-$8.828(177) &$-$10.904(241) \\
   $\kappa_{s1}$ 
   &$-$6.521(176) &$-$7.087(104) &$-$7.862(093) &$-$8.742(102) &$-$10.684(145) \\
   $\kappa_{s2}$ 
   &$-$6.476(157) &$-$7.071(091) &$-$7.852(082) &$-$8.723(090) &$-$10.634(128) \\
\hline
  \end{tabular}
  \caption{Numerical values for lattice $B$ parameters
           $B^{lat}_{B,X}$ at $\beta$=5.9.}
  \label{tab:lattice_B_parameters_59B}
\end{table}
\begin{table}
  \begin{tabular}{|c|ccccc|}
\hline
   $aM_0$ & 10.0 & 5.0 & 3.0 & 2.1 & 1.3 \\
\hline\hline
   $B^{lat}_{S,S}$ &&&&&\\
   $\kappa_{ud}$ 
   &  0.942(44) &  0.907(23) &  0.882(15) &  0.868(13) &  0.848(11) \\
   $\kappa_{s1}$ 
   &  0.923(24) &  0.912(12) &  0.899(08) &  0.888(07) &  0.873(06) \\
   $\kappa_{s2}$ 
   &  0.918(21) &  0.913(11) &  0.902(07) &  0.893(06) &  0.878(06) \\
   \hline
   $B^{lat}_{S,LR}$ &&&&&\\
   $\kappa_{ud}$ 
   &$-$1.376(85) &$-$1.184(45) &$-$0.999(30) &$-$0.850(25) &$-$0.627(22) \\
   $\kappa_{s1}$ 
   &$-$1.357(48) &$-$1.188(24) &$-$1.018(16) &$-$0.875(15) &$-$0.662(14) \\
   $\kappa_{s2}$ 
   &$-$1.353(42) &$-$1.188(21) &$-$1.022(14) &$-$0.880(13) &$-$0.670(12) \\
   \hline
   $B^{lat}_{S,P}$ &&&&&\\
   $\kappa_{ud}$ 
   &$-$13.24(55) &$-$13.08(31) &$-$13.10(24) &$-$13.20(22) &$-$13.33(22) \\
   $\kappa_{s1}$
   &$-$12.90(31) &$-$13.02(17) &$-$13.10(13) &$-$13.17(12) &$-$13.27(12) \\
   $\kappa_{s2}$ 
   &$-$12.82(27) &$-$13.01(15) &$-$13.10(12) &$-$13.17(11) &$-$13.25(10) \\
   \hline
   $B^{lat}_{S,T}$ &&&&&\\
   $\kappa_{ud}$ 
   & 57.69(233) & 55.84(126) & 54.53(96) & 53.71(89) & 52.33(88) \\
   $\kappa_{s1}$ 
   & 56.36(138) & 55.71(072) & 54.75(55) & 53.86(50) & 52.36(49) \\
   $\kappa_{s2}$ 
   & 56.05(122) & 55.68(063) & 54.80(48) & 53.90(43) & 52.36(42) \\
\hline
  \end{tabular}
  \caption{Numerical values for lattice $B$ parameters
           $B^{lat}_{S,X}$ at $\beta$=5.9.}
  \label{tab:lattice_B_parameters_59S}
\end{table}

\begin{table}
  \begin{tabular}{|c|ccccc|}
\hline
   $aM_0$ & 7.0 & 3.5 & 2.1 & 1.5 & 0.9 \\
\hline\hline
   $B^{lat}_{B,LR}$ &&&&&\\
   $\kappa_{ud}$
   &  0.833(66) &  0.846(32) &  0.845(23) &  0.834(22) &  0.819(24) \\
   $\kappa_{s1}$
   &  0.892(37) &  0.892(18) &  0.882(12) &  0.871(12) &  0.857(13) \\
   $\kappa_{s2}$
   &  0.904(33) &  0.901(16) &  0.889(11) &  0.878(10) &  0.865(12) \\
   \hline
   $B^{lat}_{B,S}$ &&&&&\\
   $\kappa_{ud}$
   &$-$0.653(30) &$-$0.701(17) &$-$0.808(14) &$-$0.929(16) &$-$1.255(28) \\
   $\kappa_{s1}$
   &$-$0.658(18) &$-$0.725(10) &$-$0.836(09) &$-$0.958(11) &$-$1.282(20) \\
   $\kappa_{s2}$
   &$-$0.659(16) &$-$0.730(09) &$-$0.842(08) &$-$0.964(10) &$-$1.287(19) \\
   \hline
   $B^{lat}_{B,N}$ &&&&&\\
   $\kappa_{ud}$ 
   &  1.227(112) &  1.674(61) &  2.179(55) &  2.715(66) &  4.130(115) \\
   $\kappa_{s1}$ 
   &  1.294(064) &  1.689(37) &  2.181(35) &  2.700(44) &  4.047(078) \\
   $\kappa_{s2}$ 
   &  1.307(057) &  1.692(34) &  2.181(33) &  2.697(41) &  4.031(073) \\
   \hline
   $B^{lat}_{B,M}$ &&&&&\\
   $\kappa_{ud}$ 
   &$-$7.125(338) &$-$7.690(180) &$-$8.545(143) &$-$9.490(155) &$-$12.046(230) \\
   $\kappa_{s1}$ 
   &$-$6.857(220) &$-$7.575(112) &$-$8.452(089) &$-$9.389(099) &$-$11.877(156) \\
   $\kappa_{s2}$ 
   &$-$6.804(204) &$-$7.552(102) &$-$8.433(082) &$-$9.369(092) &$-$11.843(146) \\
\hline
  \end{tabular}
  \caption{Numerical values for lattice $B$ parameters
           $B^{lat}_{B,X}$ at $\beta$=6.1.}
  \label{tab:lattice_B_parameters_61B}
\end{table}
\begin{table}
  \begin{tabular}{|c|ccccc|}
\hline
   $aM_0$ & 7.0 & 3.5 & 2.1 & 1.5 & 0.9 \\
\hline\hline
   $B^{lat}_{S,S}$ &&&&&\\
   $\kappa_{ud}$ 
   & 0.892(40) & 0.843(20) & 0.834(13) & 0.831(12) & 0.827(11) \\
   $\kappa_{s1}$ 
   & 0.907(24) & 0.879(11) & 0.869(7) & 0.863(7) & 0.855(7) \\
   $\kappa_{s2}$ 
   & 0.910(21) & 0.886(10) & 0.876(7) & 0.870(6) & 0.861(6) \\
   \hline
   $B^{lat}_{S,LR}$ &&&&&\\
   $\kappa_{ud}$ 
   & $-$1.138(92) & $-$1.016(41) & $-$0.870(26) & $-$0.741(22) & $-$0.536(19) \\
   $\kappa_{s1}$ 
   & $-$1.229(52) & $-$1.080(23) & $-$0.916(15) & $-$0.784(13) & $-$0.571(12) \\
   $\kappa_{s2}$ 
   & $-$1.247(46) & $-$1.094(21) & $-$0.925(14) & $-$0.792(12) & $-$0.578(12) \\
   \hline
   $B^{lat}_{S,P}$ &&&&&\\
   $\kappa_{ud}$ 
   & $-$13.09(54) & $-$13.27(26) & $-$13.31(19) & $-$13.32(18) & $-$13.35(17) \\
   $\kappa_{s1}$
   & $-$13.01(35) & $-$13.28(17) & $-$13.31(12) & $-$13.31(11) & $-$13.31(11) \\
   $\kappa_{s2}$ 
   & $-$12.99(33) & $-$13.28(15) & $-$13.31(11) & $-$13.31(10) & $-$13.31(10) \\
   \hline
   $B^{lat}_{S,T}$ &&&&&\\
   $\kappa_{ud}$ 
   & 58.65(260) & 56.91(124) & 55.23(85) & 53.90(76) & 51.70(68) \\
   $\kappa_{s1}$ 
   & 57.32(171) & 56.66(078) & 55.11(53) & 53.78(47) & 51.65(42) \\
   $\kappa_{s2}$ 
   & 57.06(158) & 56.61(072) & 55.08(48) & 53.76(43) & 51.63(38) \\
\hline
  \end{tabular}
  \caption{Numerical values for lattice $B$ parameters
           $B^{lat}_{S,X}$ at $\beta$=6.1.}
  \label{tab:lattice_B_parameters_61S}
\end{table}
\begin{table}
  \begin{tabular}{|c|ccccc|}
\hline
   $aM_0$ & 10.0 & 5.0 & 3.0 & 2.1 & 1.3 \\
\hline\hline
   $B^{lat}_{B,LR}$ &&&&&\\
   $\kappa_{ud}$
   &  0.820(53) &  0.832(51) &  0.868(39) &  0.857(30) &  0.858(25) \\
   $\kappa_{s1}$
   &  0.864(37) &  0.869(29) &  0.888(22) &  0.877(17) &  0.872(15) \\
   $\kappa_{s2}$
   &  0.874(34) &  0.877(25) &  0.892(18) &  0.881(14) &  0.875(13) \\
   \hline
   $B^{lat}_{B,S}$ &&&&&\\
   $\kappa_{ud}$
   &$-$0.574(28) &$-$0.649(27) &$-$0.738(23) &$-$0.820(20)
   &$-$1.011(22) \\
   $\kappa_{s1}$
   &$-$0.601(21) &$-$0.673(17) &$-$0.757(14) &$-$0.840(12)
   &$-$1.029(14) \\
   $\kappa_{s2}$
   &$-$0.607(19) &$-$0.678(15) &$-$0.761(12) &$-$0.845(11)
   &$-$1.033(13) \\
   \hline
   $B^{lat}_{B,N}$ &&&&&\\
   $\kappa_{ud}$ 
   &  1.128(107) &  1.376(97) &  1.787(84) &  2.207(75) &  3.095(81) \\
   $\kappa_{s1}$ 
   &  1.148(073) &  1.393(57) &  1.785(48) &  2.184(45) &  3.023(53) \\
   $\kappa_{s2}$ 
   &  1.152(066) &  1.397(49) &  1.784(41) &  2.178(39) &  3.007(48) \\
   \hline
   $B^{lat}_{B,M}$ &&&&&\\
   $\kappa_{ud}$ 
   &$-$6.67(36) &$-$7.33(28) &$-$7.76(24) &$-$8.44(20) &$-$10.04(20) \\
   $\kappa_{s1}$ 
   &$-$6.68(24) &$-$7.15(17) &$-$7.69(14) &$-$8.35(12) &$-$ 9.88(13) \\
   $\kappa_{s2}$ 
   &$-$6.68(23) &$-$7.11(15) &$-$7.67(12) &$-$8.33(11) &$-$ 9.84(12) \\
\hline
  \end{tabular}
  \caption{Numerical values for lattice $B$ parameters
           $B^{lat}_{B,X}$ at $\beta$=6.0.}
  \label{tab:lattice_B_parameters_60B}
\end{table}
\begin{table}
  \begin{tabular}{|c|ccccc|}
\hline
   $aM_0$ & 10.0 & 5.0 & 3.0 & 2.1 & 1.3 \\
\hline\hline
   $B^{lat}_{S,S}$ &&&&&\\
   $\kappa_{ud}$ 
   &  0.829(40) &  0.847(35) &  0.854(26) &  0.840(19) &  0.830(14) \\
   $\kappa_{s1}$ 
   &  0.868(30) &  0.881(22) &  0.881(15) &  0.869(11) &  0.857(08) \\
   $\kappa_{s2}$ 
   &  0.877(28) &  0.888(20) &  0.887(13) &  0.875(10) &  0.863(07) \\
   \hline
   $B^{lat}_{S,LR}$ &&&&&\\
   $\kappa_{ud}$ 
   &$-$1.184(77) &$-$1.085(68) &$-$1.005(48) &$-$0.879(33)
   &$-$0.705(23) \\
   $\kappa_{s1}$ 
   &$-$1.249(53) &$-$1.138(39) &$-$1.034(27) &$-$0.907(19)
   &$-$0.726(14) \\
   $\kappa_{s2}$ 
   &$-$1.264(50) &$-$1.149(34) &$-$1.040(23) &$-$0.913(16)
   &$-$0.730(13) \\
   \hline
   $B^{lat}_{S,P}$ &&&&&\\
   $\kappa_{ud}$ 
   &$-$12.87(58) &$-$13.17(47) &$-$13.13(37) &$-$13.17(28)
   &$-$13.32(22) \\
   $\kappa_{s1}$
   &$-$12.95(41) &$-$13.02(27) &$-$13.11(22) &$-$13.14(17)
   &$-$13.26(14) \\
   $\kappa_{s2}$ 
   &$-$12.97(38) &$-$12.99(24) &$-$13.11(19) &$-$13.14(15)
   &$-$13.24(12) \\
   \hline
   $B^{lat}_{S,T}$ &&&&&\\
   $\kappa_{ud}$ 
   & 57.82(276) & 58.01(211) & 55.59(161) & 54.30(118) & 52.93(88) \\
   $\kappa_{s1}$ 
   & 57.99(194) & 56.99(126) & 55.44(094) & 54.21(072) & 52.78(57) \\
   $\kappa_{s2}$ 
   & 58.02(180) & 56.76(111) & 55.41(082) & 54.19(063) & 52.74(51) \\
\hline
  \end{tabular}
  \caption{Numerical values for lattice $B$ parameters
           $B^{lat}_{S,X}$ at $\beta$=6.0.}
  \label{tab:lattice_B_parameters_60S}
\end{table}
\clearpage
\begin{table}
  \begin{tabular}{|c|cccc|}
    \hline
    $\beta$ & 5.7 & 5.9 & 6.1 & 6.0 \\
    \hline\hline
    $O(\alpha_s^2)$
    & 5\% & 4\% & 3\% & 3\% \\
    $O(\alpha_sp/M_b)$
    & 1\% & 1\% & 1\% & 1\% \\
    $O(a^2p^2)$
    & 8\% & 3\% & 2\% & 3\% \\
    $O(ap\alpha_s)$
    & 6\% & 4\% & 2\% & 3\% \\
    $O(\alpha_s^2/(aM_b))$
    & 1\% & 1\% & 1\% & 1\% \\
    $O(p^2/M_b^2)$
    & \multicolumn{4}{c|}{ 2 \%} \\
    \hline
    Total (added in quadrature)
    & 11 \% & 7\% & 5 \% & 6 \% \\
    \hline
  \end{tabular}
  \caption{
    Estimate of systematic uncertainties by a naive dimensional
    counting at each $\beta$ value. 
  }
  \label{tab:systematic_error}
\end{table}

\clearpage
\clearpage
\clearpage
\clearpage

\end{document}